\documentclass[12pt]{article} 
\usepackage[dvipsnames]{xcolor}
\usepackage{amsmath}
\usepackage{amssymb}
\usepackage[colorlinks=false, bookmarksnumbered,hypertexnames=false]{hyperref}  
\usepackage{cleveref}

\newcommand{\boxedeq}[2]{\begin{empheq}[box={\fboxsep=6pt\fbox}]{align}\label{#1}#2\end{empheq}}
\newcommand{\eqfig}[2]{\vcenter{\hbox{\includegraphics[width=#1]{#2}}}}

\usepackage{empheq} 
\usepackage{cite}
\usepackage{mciteplus} 
\usepackage{graphicx}
\usepackage{physics}
\usepackage{float}
\usepackage{comment}
\usepackage{bbold}
\graphicspath{ {./figs_couplings/} }
\usepackage[normalem]{ulem}
\usepackage[utf8]{inputenc}
\usepackage{bbm} 

  \newlength{\abstractwidth}
  \newcommand{\be}{\begin{equation}}
  \newcommand{\bea}{\begin{eqnarray}}
  \newcommand{\eea}{\end{eqnarray}}
  \newcommand{\beq}{\begin{equation}}
  \newcommand{\ee}{\end{equation}}
  \newcommand{\eeq}{\end{equation}}

    \newcommand\pcr{{Poincar\'e}}
\def\lll{{\mathsf{LL}}}
\def\lr{{\mathsf{LR}}}
\def\lll{{\mathsf{LL}}}
\def\rr{{\mathsf{RR}}}
\def\rl{{\mathsf{RL}}}
\def\glr{g_{\lr} }
\def\gll{g_{\lll} }
\def\betax{\beta_\mathrm{aux}}

\newcommand{\AAl}[1]{{\textbf{\textcolor{blue}{#1}}}}

\newcommand{\lan}{\langle}
\newcommand{\ran}{\rangle}

	\newcommand{\HL}[1]{{\color{red}[HL\@: #1]}}
	
	\newcommand{\eqn}[1]{\begin{equation}\begin{split} #1 \end{split}\end{equation}}
	
	\newcommand{\lp}{\left (}
	\newcommand{\rp}{\right )}

	\newcommand{\R}{\mathbb{R}}
	
	\newcommand{\bz}{\bar{z}}

	\newcommand{\pd}{\partial}
	\newcommand{\inv}{^{-1}}
	
	\newcommand{\hf}{\frac{1}{2}}

	\newcommand{\lb}{\left [}
	\newcommand{\rb}{\right ]}

    \def\tfd{\mathsf{TFD}}
    
    \def\zb{\bar{z}}

\def\la{\label}

\def\sgn{\mathrm{sgn}}

  \def\ba{\begin{eqnarray}}
  \def\ea{\end{eqnarray}}

 \def\simleq{\; \raise0.3ex\hbox{$<$\kern-0.75em
      \raise-1.1ex\hbox{$\sim$}}\; }
 \def\simgeq{\; \raise0.3ex\hbox{$>$\kern-0.75em
	\raise-1.1ex\hbox{$\sim$}}\; }

\def\nref#1{(\ref{#1})}

\begin{document}

\begin{titlepage}
  \bigskip

  \bigskip\bigskip

  \bigskip

\begin{center}
 
\centerline
{\Large \bf {The Entanglement Wedge of Unknown Couplings }}
 \bigskip

 \bigskip
{\Large \bf { }} 
    \bigskip
\bigskip
\end{center}

  \begin{center}

 \bf {Ahmed Almheiri$^{\ket{i}}$ and Henry W. Lin$^{\ket{j}}$}
  \bigskip 
  
  \rm
\bigskip
 $^{\ket{i}}$Institute for Advanced Study,  Princeton, NJ 08540, USA\\

 \rm 
 \bigskip
 $^{\ket{j}}$Jadwin Hall, Princeton University, Princeton, NJ 08540, USA  
\rm
 \bigskip

  \bigskip \rm
\bigskip
 
\rm

\bigskip
\bigskip

  \end{center}

 \bigskip\bigskip
  \begin{abstract}

	The black hole interior is a mysterious region of spacetime where non-perturbative effects are sometimes important. These non-perturbative effects are believed to be highly theory-dependent. We sharpen these statements by considering a setup where the state of the black hole is in a superposition of states corresponding to boundary theories with different couplings, entangled with a reference which keeps track of those couplings. 
    The entanglement wedge of the reference can then be interpreted as the bulk region most sensitive to the values of the couplings.
    In simple bulk models, e.g., JT gravity $+$ a matter BCFT, the QES formula implies that the reference contains the black hole interior at late times. 
	We also analyze the Renyi-2 entropy $\tr \rho^2$ of the reference, which can be viewed as a diagnostic of chaos via the Loschmidt echo. 
	We find explicitly the replica wormhole that diagnoses the island and restores unitarity. Numerical and analytical evidence of these statements in the SYK model is presented. Similar considerations are expected to apply in higher dimensional AdS/CFT, for marginal and even irrelevant couplings. 
	  
 \medskip
  \noindent
  \end{abstract}
\bigskip \bigskip \bigskip

\vspace{1cm}

\begin{figure}[H]
\hspace*{-1in}
\begin{raggedleft}
	\includegraphics[scale=1, trim = -210 0 0 0]{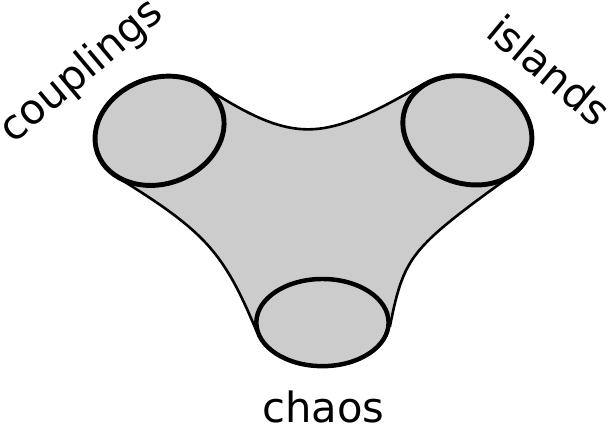}
\end{raggedleft}
\end{figure}

\vspace{2cm}

  \end{titlepage}

   \tableofcontents

\def\cj{\mathcal{J}}
\def\sch{\mathsf{Sch}}
\def\slt{\mathrm{SL}(2,\R)}
\def\bulk{\mathsf{bulk}}
\def\jour{\mathsf{journal}}
\def\disk{\mathsf{disk}}

\def\toc{\mathsf{toc}}
\def\otoc{\mathsf{otoc}}
\def\ads{\mathsf{AdS}}
\def\cft{\mathsf{CFT}}
\def\syk{\mathsf{SYK}}
\section{Introduction}

The black hole interior is a mysterious region of spacetime where non-perturbative quantum gravity effects are sometimes important. 
Despite the importance of such non-perturbative effects, work from the past few decades supports the idea that ``plain'' gravity (e.g. a sum over metrics and possibly a few light fields) knows a lot about fine-grained quantum information.
The paradigmatic examples of this include the geometrization of von Neumann entropy in holographic systems, including its suitable generalization to generate the Page curve of Hawking radiation of an evaporating black hole \cite{Penington:2019npb,Almheiri:2019psf}.
Such features were thought to require a UV complete theory of gravity, such as string theory, and obtaining them from gravity came as a pleasant surprise.

Gravity, however, doesn't know everything. It appears to know of the underlying random unitary dynamics, but it fails to pin down a particular realization of those dynamics. In particular, while gravity is able to reproduce arbitrary moments of the signal drawn from an ensemble of random dynamics, it fails to capture the large fluctuations that come with a given realization. This is a pretty big miss since the size of those fluctuations is of order the signal.

A related point is that whereas the gravity calculations seem reliable for some universal quantities, it seems to know only statistical properties about the non-perturbative effects that are highly theory-dependent. For example, the spectral form factor, which at large times probes the detailed energy spectrum of the black hole, is a highly erratic function that depends sensitively on the couplings\cite{Cotler:2016fpe,Saad:2018bqo}. Similarly, the black hole $S$-matrix that governs the formation and evaporation of a black hole is suspected to be an erratic and possibly pseudo-random matrix  \cite{Polchinski:2015cea}. Presumably a precise computation of such quantities from the bulk point of view will involve strings, branes, half-wormholes \cite{Saad:2021rcu}, etc, and the answers would depend on the particular string vacua. 

The dependence of these non-perturbative effects on the couplings of the theory suggests that the interior of the black hole is in some sense highly theory dependent. The goal of our work will be to make this more precise in the context of bulk reconstruction.  We will consider models which admit an ensemble of boundary theories parameterized by a set of boundary couplings, and analyze the sensitivity of bulk reconstruction on the level of precision in specifying those couplings. This means we will look for instances where the reconstruction fails, and map out which bulk regions are most sensitive to this. Hence, those bulk regions require exquisite knowledge of the couplings to reconstruct.

{\it Note added}: as we were finishing this work, we became aware of work by Qi, Shangnan, and Yang \cite{Qi:2021oni}. We have arranged to coordinate our preprints. See also \cite{Renner:2021qbe}.

\subsection{Knowing the couplings -- an operational definition}

\def\sys{{\mathsf{sys}}}
\def\re{{\mathsf{journal}}}
\def\po{\mathsf{ptr}}
\def\env{\mathsf{env}}
\def\un{\, \cup \, }
\def\en{\mathcal{N}=4}

\def\blam{{\boldsymbol{\lambda}}}
\def\blam{{\mathbf{J}}}
The first point we make precise is the notion of ``knowing the couplings." Suppose we have a system whose Hamiltonian depends on a set of parameters $\blam = \{\lambda_i \}$. For example, in SYK it is natural to choose $\blam = \{ J_{ijkl} \}$ to be the set of random couplings. However, in general the couplings could be non-random, for example we could consider $\en$ and take $\blam = \{ g^2_{YM} \} $.
We denote the state of the system prepared with those couplings as
\begin{align}
   \ket{ \psi; \blam }_\sys. 
\end{align}
As a concrete example, we could take $ \ket{ \psi; \blam }_\sys = \ket{\beta+i2T; \blam}_\sys$ to be the thermofield double state associated to the Hamiltonian $H(\blam)$ at some temperature $\beta$ and some Lorentzian time $T$.
Let's imagine that these couplings are drawn form some distribution $P(\blam)$. We can keep track of this by using a standard method in quantum information theory of entangling the system with a reference system labelled by the couplings
\begin{align}
  \ket{  \Psi}_{\sys  \un \re} = \sum_{\blam} \ \sqrt{P\lp \blam \rp } \, | \psi; \blam \rangle_\sys | \blam \rangle_{\re}. \la{global:pure}
\end{align}
We call the reference system $\re$ since in the SYK context it records the $J$'s.
Tracing over the $\re$, we get a density matrix 
\eqn{\rho_\sys = \sum_{\blam} P(\blam) \rho_\psi(\blam)  = \ev{\rho_\psi}_\blam, \quad \rho_\psi(\blam) = \ketbra{\psi; \blam}{\psi; \blam}_\sys. \la{eq:entangle}}
This state represents the situation where we have no information about the couplings. This ignorance represented in equation \nref{eq:entangle} captured by the non-zero von Neumann entropy between $\sys$ and $\re$. If we think of $\blam$ as parameters in a disorder average, we may use condensed matter jargon and refer to the von-Neumann entropy as the {\it annealed} entropy $S\lp \ev{\rho_\psi}_\lambda\rp $. In the above discussion we started with a pure state density matrix $\rho_\psi$ but after averaging over $\blam$ we get a mixed state. As a slight generalization, we could consider any subsytem of $\sys$, and get a similar formula, where $\rho_\psi$ is replaced by some partial trace of $\ketbra{\psi; \blam}{\psi; \blam}_\sys$. 

The main question we'd like to answer is the following: How much of the bulk can be reconstructed given the density matrix of the system after tracing out the reference? Or in other words, how much of the bulk is contained within the entanglement wedge of $\sys$? In a theory with a semiclassical holographic dual, the answer is given by the quantum extremal surface (QES) formula, which states that the boundary of the entanglement wedge is given by the QES responsible for the von Neumann entropy $S(\sys)$.
An alternative version of this question is: how much of the bulk is contained within the entanglement wedge of the reference? Since the $S(\sys) = S(\re)$ for a pure state, the entanglement wedge is simply the complement of the entanglement wedge of $\sys$. 

Let us make some preliminary comments on what we mean by bulk reconstruction.
First notice that semi-classically, the entropy $S(\sys) > 0$. The bulk matter is not pure, since it is entangled with $\re$. 
This semi-classical entropy can ``pollute'' the bulk and lead to problems with reconstructing any operator using traditional methods like HKLL\cite{Hamilton:2005ju}.
Any matter in the bulk will generically interact (at least weakly) with the fields that are sourced by the couplings we turn on at the boundary of AdS; this will cause problems with reconstruction methods such as HKLL. We will not be discussing  such semiclassical problems in this paper. Instead, we will ask for bulk reconstruction in the modern sense of entanglement wedge reconstruction \cite{Jafferis:2015del, Dong:2016eik}. The failure of our ability to do reconstruction will be a non-perturbative effect, signaled by replica wormholes and the appearance of an island.

In the above discussion, we entangled $\sys$ to $\re$ in such a way that the global state is pure. 
However, one could also consider a setup where instead of entangling $\sys$ to a $\re$, we instead classically correlate $\sys$ to a ``pointer'' system $\po$:
\eqn{\rho_{\sys{} \un \po{}} = \sum_{\blam} P(\blam) \rho_\psi(\blam)  \otimes \ketbra{\blam}{\blam}_\po. \la{densitypo} }
This mixed state would be the appropriate description of a setup where Alice flips some coins and uses the outcome to decide which couplings to prepare the system in.
In fact, following the standard discussion of decoherence/measurement theory, this state can be purified by adding an auxiliary system $\env$. Often one adopts the interpretation that the system $\po$ is a pointer or measurement device, and $\env$ is the environment. The purification of this state is simply \eqref{global:pure}, where $\re = \po \cup \env$ and $\ket{\blam}_\re = \ket{\blam}_{\po{}} \ket{\blam}_\env$. Tracing over $\env$ gives \eqref{densitypo}. 
Alternatively, we can go the other direction: we start from the global pure state of \eqref{global:pure} and perform a complete measurement in the $\lambda$ basis:
\eqn{\ketbra{  \Psi}{\Psi}_{\sys  \un \re} \to \sum_\blam \Pi_\blam^{\re} \ketbra{  \Psi}{\Psi}_{\sys  \un \re}\Pi_\blam^{\re}.}
Then relabelling $\re \to \po$, we obtain \eqref{densitypo}. Thus including the $\po$ system makes precise the idea of having classical (e.g. decohered) knowledge of the couplings.

To ask about bulk reconstruction given perfect classical knowledge of the couplings is to ask for the entanglement wedge of $\sys \cup \po$.
Following the QES rules, we should compute the von Neumann entropy of $\sys \cup \po$:
\eqn{S(\sys \un \po) = \sum_\blam P(\blam) S(\rho_\psi(\blam) )-\sum_\blam P(\blam) \log P(\blam).  \la{union} }
The first term is the von Neumann entropy, averaged over the couplings. 
If $\rho_\blam$ is pure, this vanishes. But this formula also applies to an initially mixed state, or a subsystem, e.g., we could take $\sys$ to be the left side of the thermofield double. Then the first term would be the disorder-averaged thermal entropy, known as the \emph{quenched} entropy, $\ev{ S\lp \rho_\psi\rp}_\lambda $. 

The second term in $\eqref{union}$ is the Shannon entropy of the couplings $S(\po)$. It is an entirely classical entropy and does not play a role in determining the QES. To see this more explicitly, note that \eqref{union} is a formula for the exact entropy of a boundary subsystem, but in a holographic system we could also consider the reduced density matrix $\rho_{A \cup \po}$ of the semiclassical quantum fields in some bulk region $A$. This density matrix will have the same form as \nref{densitypo}, so a similar formula holds for the semi-classical bulk matter entropy:
\eqn{ S_\text{matter}(A \cup \po) = \sum_\blam P(\blam) S(\rho^A_\blam)-\sum_\blam P(\blam) \log P(\blam), \quad \rho_\blam^A = \Pi_\blam \rho_A^\text{semi} \Pi_\blam. \la{eq:matterent}}
The second term is independent of the choice of $A$, so it will never contribute to the derivative of the matter entropy (relevant for the QES). 
The conclusion is that the entanglement wedge of $\sys \cup \po$ can be diagnosed by computing $S(\sys|\po) = S(\sys \cup \po) - S(\po)$, which is simply the quenched entropy. In this context, we have a QES formula for the conditional entropy:
\def\mat{\mathsf{matter}}
\eqn{S(\sys|\po) = \mathrm{min} \; \mathrm{ext}_I  \lb  \mathrm{Area}(\partial I)/(4G) + S_\mat(I|\po) \rb  \la{eq:matterent2} .}
To summarize, if we are interested in the entanglement wedge of $\sys$ with no knowledge of the couplings whatsoever, we should compute the annealed entropy $S(\ev{\rho_\psi}_\blam)$. If we are interested in bulk reconstruction when we have perfect classical knowledge of the couplings $\sys \cup \po$, we should compute the quenched entropy $\ev { S(\rho_\psi) }_\blam$. 

Note that whether we choose to entangle $\sys$ with a reference or classically correlate $\sys$ to a pointer makes no difference for the density matrix $\rho_\sys$. A crucial difference is for the complement of $\sys$: the density matrix of $\rho_\po$ is diagonal in the $\lambda$ basis whereas $\rho_\re$ has off-diagonal elements. Furthermore, the entanglement wedge of $\re$ can contain an island (as we will show), whereas a quick argument \cite{Qi:2021sxb} rules out any possible island for the entanglement wedge of $\po$: for the pointer system, the matter conditional entropy is positive
$S(A |\po) = S(A \cup \po) - S(\po) \ge 0$ for any bulk region $A$, so adding any the bulk region $A$ can only increase the entropy. Therefore, any island QES must be non-minimal.

\subsection{Both known and unknown couplings}
\def\unk{\mathsf{unknown}}
\def\kno{\mathsf{known}}
\def\bkap{\kappa}
\def\bmu{{\mu}}

More generally, we can imagine that all the couplings in $\re$ can be divided into $\kno$ and $\unk$. We write $\blam = \{ \bkap, \bmu\}$ with $\bkap$ the known parameter(s) and $\bmu$ the unknown, $\re = \kno \un \unk$. The global state is a density matrix on $\sys \cup \re$:
\eqn{
\rho = \sum_{\bkap, \bmu, \bmu'} \sqrt{P(\bmu, \bkap) P(\bmu', \bkap)}\ketbra{\psi; \bkap, \bmu}{\psi; \bkap, \bmu'}_\sys \otimes \ketbra{\bkap}{\bkap}_\kno \otimes \ketbra{\bmu}{\bmu'}_\unk 
\la{eq:joint}}
Here we are treating the known couplings as ``classical'' pointer states $\po$, whereas the unknown parameters are entangled. 
The reduced density matrices
\eqn{
    \rho_{ {\re}} = \sum_{ \bmu, \bmu',\bkap } \ketbra{  \bmu}{\bmu'}_{\unk} \ketbra{\bkap}{\bkap}_\kno \sqrt{P\lp \bmu, \bkap \rp   P\lp \bmu', \bkap \rp } \, \langle \psi ; \bmu, \bkap | \psi ; \bmu', \bkap \rangle_\sys \la{eq:rho_unk}
}
\eqn{
    \rho_{ {\kno}} = \sum_{\bkap }  \ketbra{\bkap}{\bkap}_\kno P(\kappa), \quad P(\kappa) = \sum_\mu P(\mu, \kappa) \la{eq:rho_k}
}
We see that the density matrix of $\kno$ is completely classical as before. 
Once we include ``known'' couplings, the global state is not pure. Therefore, $S(\re) \ne S(\sys)$. Nevertheless, a close analog is given by the conditional entropies:
\boxedeq{eq:cond}{S(\sys|\kno) = S(\unk|\kno).}
This follows from \eqref{eq:rho_unk}. 
As we have already argued, the entanglement wedge of $\sys \un \kno$ may be diagnosed by computing the conditional entropy $S(\sys \cup \kno)$, and similarly for the entanglement wedge of $\kno \cup \unk$. The equality of the above conditional entropies essentially shows that the entanglement wedges are in fact complementary even though the global state is not pure.


The fact that we are including the known couplings when computing the entanglement wedges reflects the fact that the known couplings are completely classical, so there is no obstruction to cloning. The experimentalist simply publishes in a journal the values of the couplings in which she prepared her system.

\pagebreak 
\section{The entanglement wedge with uncertain couplings \la{matterentropy}}

In this section we analyze the candidate entanglement wedges of the reference keeping track of the unknown couplings. This will be done in a bulk model of JT gravity \cite{Almheiri:2014cka,Jensen:2016pah,Maldacena:2016upp,Engelsoy:2016xyb} coupled to general conformal matter:
\eqn{-I[g] =- S_0 \chi +\int_{\Sigma_{2}} \frac{\phi}{4 \pi}(R+2)+\frac{\phi_{b}}{4 \pi} \int_{\partial \Sigma_{2}} 2 K+\log Z_{\cft}[g],}
where $\chi$ is the Euler characteristic.
We will specialize to a concrete BCFT when needed.

The couplings in this model will be the choice of CFT boundary conditions along the AdS boundary, which we label $\ket{J}$ and interpret as arising from a holographic boundary Hamiltonian $H_J$. We require the state $\ket{J}$ to be a conformally invariant Cardy state.
Here $J$ could be a discrete variable (e.g., if the CFT is a minimal model) or continuous. We will study the state given by entangling the thermofield double to the $\re$ as in \eqref{global:pure}:
\begin{align}
    | \Psi \rangle_{\sys \un \re } &= \sum_{J} \sqrt{P(J)} | \beta+2iT, J \ran_{\sys} | J \ran_{\re} \la{eq:comp}
\end{align}
where the time-evolved thermofield double $\ket{\beta+ 2iT ,J}_\sys = e^{-iH_L T}\ket{\beta,J}_\sys $. We will frequently drop the $2iT$ and think of $\beta$ as a complex number. The reduced density matrix
 \begin{align}
     \rho_\re = \sum_{J, J'} \sqrt{P(J) P(J')} \lan \beta, J' | \beta, J \ran \ | J \ran \lan J' |_\re. \label{rho_unk}
 \end{align}
To compute the entanglement wedge of the journal, we will use the QES formula \cite{Faulkner:2013ana, Engelhardt:2014gca}:
\eqn{S(\re) =\operatorname{min} \left\{ \operatorname{ext}_{I} \left[ \sum_{\partial I} {\phi(\partial I)}{}+ S_{m}(I \cup \re)\right] \right\} \la{eq:qes}.}
Here we are instructed to compute the generalized entropy of extremal islands, and then pick the smallest extremized entropy. Note that for marginal deformations of the boundary theory, we expect the dilaton to be independent of the boundary state $J$. Hence the non-trivial calculation is just the matter entropy in \eqref{eq:qes}. We will compute this in some special cases in this section.

Before getting into the weeds, let us clarify somewhat the interpretation of the calculation.
We are thinking of the above theory as an effective theory of the bulk. It is not UV complete due to divergences when wormholes get narrow \cite{Saad:2019lba}. 
We do not know what the precise boundary dual of this bulk model is (or even if it exists). 
The simplest possibility is that the boundary dual is a theory with a small number of couplings. 
For example, one might be able to embed such a setup in a traditional higher dimensional example of AdS/CFT by considering near extremal black holes in AdS.
In this case, we assume that all of the couplings besides $J$ are fixed, and the state of the combined system is given by \eqref{eq:comp}.

However, it is also possible that the gravity description only arises after a disorder average over other random couplings, like in pure JT gravity \cite{Saad:2019lba} or in SYK \cite{Maldacena:2016hyu}. 
In this case, the above \eqref{eq:comp} and \eqref{rho_unk} are not quite right; instead, we assume that the additional couplings are ``known'' while the $J$ couplings are unknown, and use \eqref{eq:joint}.
 The density matrix of $\sys \cup \unk$ would be
\eqn{
\rho_{\sys \un \unk} = \sum_{\bkap, J, J'}P(\bkap) \sqrt{P( J ) P(J')}\ketbra{\beta; \bkap,J }{\beta; \bkap, J'}_\sys  \otimes \ketbra{J}{J'}_\unk 
\la{eq:jointt}}
Then following \eqref{eq:matterent} we would interpret the QES computation as giving the conditional entropy $S(\unk| \kno)$:
\eqn{S(\unk|\kno) =\operatorname{min} \left\{ \operatorname{ext}_{I} \left[ \sum_{\partial I} {\phi(\partial I)}{}+\ev{ S_{\text {m}}(I \cup \unk )}\right] \right\} \la{eqn:qes2}.}
Here we have used  $S_{\text {matter }}(I \un \unk | \kno) = \ev{S_{\text {matter }}(I \cup \unk)}$, where $\ev{ \cdots} $ is a disorder average with respect to the known couplings, see \eqref{eq:matterent} and \eqref{eq:matterent2}.

\subsection{Inconsistency of the trivial surface \la{trivialsurface}}

Here we analyze the contribution from the trivial surface, namely where the entanglement wedge of the reference doesn't include any part of the gravitational system. The entire bulk is encoded on the boundary. We will find that this contribution to the entanglement between the boundary and the reference leads to an ever growing entropy, producing to a Hawking-like information paradox. This signals that at late times, the couplings should contain an island, to prevent the entropy from growing.

To compute the entropy of the trivial surface, all we need is the bulk matter entropy. The most straightforward way of getting this is by computing the entropy of the density matrix \eqref{rho_unk} using the replica trick while freezing the gravitational saddle to be the product of Euclidean cigars on the $n$ copies. The $n$-th Renyi entropy\footnote{In the standard quantum information literature, the Renyi's usually refer to $S_n = \frac{1}{1-n} \log R_n$; here we will refer to both $S_n$ and $R_n$ as Renyi entropy.} of the journal is therefore given by the path integral with the boundary conditions $Z_n = \tr \rho^n_\unk$

\eqn{Z_n  \quad   &= \quad \eqfig{0.7\columnwidth}{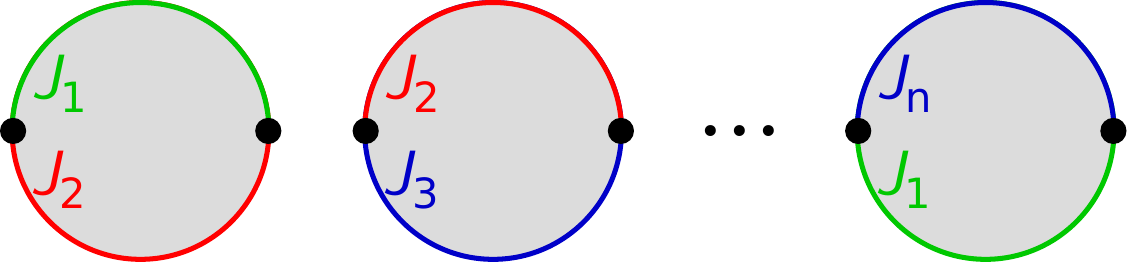} \\
            &= \quad \int \prod_{i = 1}^n P(J_i) \, d J_i  \ \prod_{i = 1}^n \lan \beta, J_i | \beta, J_{i+1} \ran,
}
where $J_{n + 1} = J_1$. 

In the above drawing, we are supposed to sum over all the indices, weighted by the probability distribution $P(J_i)$. 
The overlaps can be represented as the BCFT disk partition function in the presence of boundary changing operators.
\begin{align}
    \lan \beta , J_i | \beta, J_{i+1}  \ran  = \ev{O_{J_i,J_{i+1}}(0) O_{J_{i+1},J_i}({\tau}) }_\disk, \quad \tau = \beta/2. 
\end{align}
where $O_{J_i,J_{i + 1}}(\tau)$ changes the boundary condition by $J_{i + 1} - J_i$ as $\tau$ is crossed along the path integral contour. They satisfy $\left( O_{J_i,J_{i + 1}}(\tau)\right)^\dagger = O_{J_{i+1},J_{i }}({-\tau})$, and behave as primary operators with a dimension that depends on the boundary conditions $\Delta[{J_{i + 1}, J_{i}}]$, and their two point function is
\begin{align}
    \ev{O_{J_i,J_{i+1}}(0) O_{J_{i+1},J_i}(\tau) }_\disk  = \left[ {\pi \epsilon \over \beta \sin (\pi \tau / \beta)} \right]^{2 \Delta[{J_{i + 1} , J_{i}}]}
\end{align}
Here the $\epsilon$ comes from the Weyl factor evaluated on the boundary.
Instead of consisting a product of $n$ disks, we may  equivalently consider the quotient picture where we instead consider the $n$-fold tensor product of the CFT on a single disk, with only 2 boundary condition changing operators:
\eqn{Z_n  \quad   &= \quad \eqfig{0.3\columnwidth}{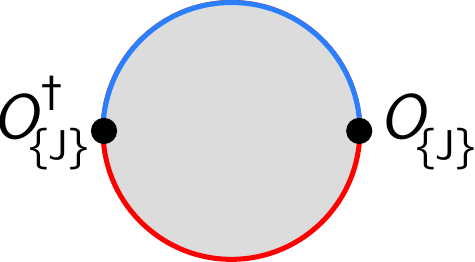}
}
\def\oj{O_{\{J\}}}
\noindent In this picture, the black dot represents a composite boundary condition changing operator $\oj = O_{J_1, J_2} \otimes O_{J_2, J_3} \otimes \cdots \otimes O_{J_n, J_1}$ which shifts the boundary condition $\{J_1, J_2 \cdots J_n\} \to \{J_2, J_3 \cdots J_1\}  $.
This quotient picture is a bit overkill for the no-island computation, but we are introducing it now since it will be crucial when the QES is non-trivial.

Since we are interested in the time evolution of the entropy, we need to consider the Renyi computation in the time evolved state 
$    | \beta + 2 i T,  J \ran = e^{-i H_{ J}^L T} | \beta,  J \ran$.
We can achieve this by setting $\tau = \beta/2 +  i T$. The overlaps become
\begin{align}
    \lan \beta + 2 i T , J_i | \beta + 2i T , J_{i+1}  \ran  =\left[ {\pi \epsilon  \over \beta \cosh (\pi T / \beta)} \right]^{2 \Delta[{J_{i + 1} , J_{i}}]}
\end{align}
Notice that all off-diagonal matrix elements are decaying to zero at large $T$; only the diagonal terms where $\Delta = 0$, e.g., $J_i = J_{i+1}$ do not decay. This implies that at late times the matter is close to maximally mixed.

Putting this back into the Renyi entropy, we get
\begin{align}
      Z_n = \int \prod_{i = 1}^n d J_i \, p(J_i) \ \left[ {\pi \epsilon  \over \beta \cosh ( \pi T / \beta)} \right]^{\sum_{i=1}^n 2 \Delta[{J_{i + 1} , J_{i}}]} \label{eq:renyi}
\end{align}
So the only dynamical input we need from the particular CFT to evaluate these Renyi entropies is the boundary dimensions $\Delta[{J_{i + 1}, J_{i}}]$. 
In Appendix \ref{app:2bd} we consider a generic BCFT with 2 boundary states. Here we will consider a non-compact boson, with action
\eqn{S = {1 \over 2 \pi \alpha'} \int d^2 z \pd X \bar{\pd} X \la{eq:freebosonaction}}
The boundary conditions we consider are simply Dirichlet conditions on the free field, labeled by $X$. The boundary condition changing operator\footnote{A quick way to see that changing Dirichlet conditions behaves as a local boundary primary is to use T-duality, which relates this setup to the insertion of a vertex operator with momentum $k$ on the boundary (with standard Neumann condtions.) We thank Juan Maldacena for pointing this out.} changes the value of the field from $X_1$ to $X_2$. Its dimension is equal to the energy on the theory on the strip, which is given by
\eqn{\Delta_b = {1 \over 4 \pi \alpha'} \int_0^\pi (\pd_\sigma X)^2 d \sigma = {1 \over \alpha'}  \lp{X_1 - X_2 \over 2\pi}\rp^2 \la{eq:bcc} }
This is familiar from string theory. The mass of a bosonic open string of level $N=1$ which stretches between two D-branes is $M^2 = (X_1 - X_2)^2/(2\pi \alpha')^2$.
We will take a Gaussian measure over the boundary conditions $p(X) =  \frac{m}{\sqrt{2\pi}} e^{-m^2 X^2/2}$. 
In terms of the boundary dual, we are considering an ensemble of boundary Hamiltonians parameterized by a marginal coupling $J$ that changes the Dirichlet condition of the bulk field $X$; the ensemble for this boundary coupling is Gaussian distributed.

Then Equation \eqref{eq:renyi} becomes
\eqn{ Z_n &=  \lp \frac{1}{\sqrt{2\pi} } \rp^n \int  \prod_{i=1}^n {dx_i} \exp \lb - \hf \lp x_i^2 +a (x_i - x_{i+1})^2 \rp \rb   \\
&=\lp \det M_n \rp^{-1/2}, \quad a = {1 \over 2\pi^2 \alpha' m^2} \log \lb \frac{\beta}{\epsilon \pi} \cosh \frac{\pi T}{\beta} \rb, \quad x_i = m X_i,  \la{eq:noncompact}}
Here $M_n$ is an $n \times n$ matrix; to take $n=5$ for example:
\eqn{M_5 = \left(
\begin{array}{ccccc}
 2 a+1 & -a & 0 & 0 & -a \\
 -a & 2 a+1 & -a & 0 & 0 \\
 0 & -a & 2 a+1 & -a & 0 \\
 0 & 0 & -a & 2 a+1 & -a \\
 -a & 0 & 0 & -a & 2 a+1 \\
\end{array}
\right)}
Here the determinant is a polynomial in $a$:
\eqn{\det M_n = \sum_{k=0}^{n-1} {2n-k \choose k } \frac{n}{2n-k} a^k  \la{eq:detpoly} }
This sum can be analytically continued to complex $n$:
\eqn{\det M_n =  \lp d_+^{2}\rp^n + \lp d_-^2\rp^n-2 a^n, \quad d_\pm = {1\pm \sqrt{4 a+1} \over 2 } \la{eq:carlson} }
One can show that for integer $n$ the Taylor series of this expression in $a$ reproduces the polynomial \eqref{eq:detpoly}. Furthermore, this function satisfies Carlson's theorem, since it grows exponentially in $n$ on the real axis, but only oscillates along the imaginary axis.
Then differentiating and taking the $n\to 1$ limit of the normalized Renyi entropy we get
\boxedeq{eq:replica}{S_m &=-\left.\partial_{n}   Z_{n} \right|_{n=1} = \frac{\log (a)}{2}+\sqrt{4 a+1} \coth ^{-1}\left(\sqrt{4 a+1}\right)}
 At late times, $a \gg 1$ so\footnote{Note that we could derive the large time expansion by simply taking $\det M \approx n^2 a^{n-1}$.}
\eqn{S_m \approx \hf \log a  +2 \sim \hf \log \lp  \frac{T }{\beta \alpha' m^2} \rp .}
In general, the entropy monotonically increases. This growth results from  states with different boundary conditions becoming more orthogonal under time evolution. If we had $c$ independent free bosons, each with uncorrelated boundary conditions, the entropy would simply be $c$ times the above answer.

This entropy eventually competes with the thermal entropy of the wormhole, which at low temperatures is $S = 2 S_0 + O(1/\beta)$, and produces a unitarity problem when
\eqn{T/\beta \sim  4 \pi \alpha' m^2 e^{4 S_0/c -4}.}
This signals that the trivial surface eventually becomes subdominant in the full non-perturbative  calculation of the entropy of the unknown couplings entropy, or equivalently of the system.\footnote{Note that whether we choose to regulate the problem by compactifying the boson $X \sim X+ 2\pi R$ or by changing the measure so that $X^2 \sim 1/m$, the net result is quite similar. See Appendix \ref{windings}. } It suggests that some part of the bulk might become inaccessible to the boundary system, falling out of it's entanglement wedge as a result of the uncertainty in the couplings.

It is also interesting to note the dependence of this time scale on the uncertainty in the couplings. For that, we use the Shannon entropy of the dimensionless variable $(\alpha')^{1/2} X$, which for a Gaussian distribution is
\eqn{S_\mathrm{Sh}(  X) = \frac{c}{2} \lp 1+  \log\frac{2\pi}{\alpha' m^2}\rp }
Recall that $m$ is the inverse width of the distribution, and the increase of $S_\mathrm{Sh}(X)$ as $m$ decreases is consistent with there being more uncertainty. The time scale is
\eqn{T/\beta \sim    e^{{2 \over c} [2S_0 - S_\mathrm{Sh}(X)]}.}
Thus, the transition time is sooner for higher uncertainty. This can be used to make the transition time less than exponential in the entropy, and seemingly as small as desired by considering a wide enough distribution for the couplings.



\subsection{Matter entropy for semi-infinite interval \la{sec:peninsula}}
\def\yb{\bar{y}}
\def\wb{\bar{w}}
Before addressing the matter entropy of an island, let us consider the matter entropy of a semi-infinite interval (a ``peninsula''). This may be viewed as a warmup to the island calculation, which we will see reduces to the peninsula calculation at late times. It is also the relevant computation for the entanglement wedge of the left side of the boundary system, tracing out the right side and the $\re$. If the journal is empty, the QES lives at the bifurcate horizon; with a journal, we expect the QES to shift to the left.

To compute the matter entropy, we will follow the same strategy as above. First note that the left density matrix
\begin{align}
    \rho_L = \sum_J P(J) \frac{e^{-\beta H_J}}{Z_J(\beta)}
\end{align}
For a free boson, the disk partition function $Z_J(\beta)$ is independent of $J$, so we can ignore the denominator (it will just contribute to an overall normalization of the density matrix). More generally, for a BCFT on a disk one can we write $Z_J(\beta) = g(J) Z(\beta)$, where $\log g(J)$ is the ``boundary entropy'' which is the boundary analog of the central charge \cite{Cardy:2004hm}.\footnote{ $Z(\beta)$ is generated from the conformal anomaly, which only depends on the central charge.} Therefore we can absorb $g(J)$ into the definition of $P(J)$. 
Then the computation of the unnormalized density matrix $\tr \rho_L^n$ is given by
\eqn{Z_n  \quad   &= \quad \raisebox{-0.3\totalheight}{\includegraphics[width = 0.75\columnwidth]{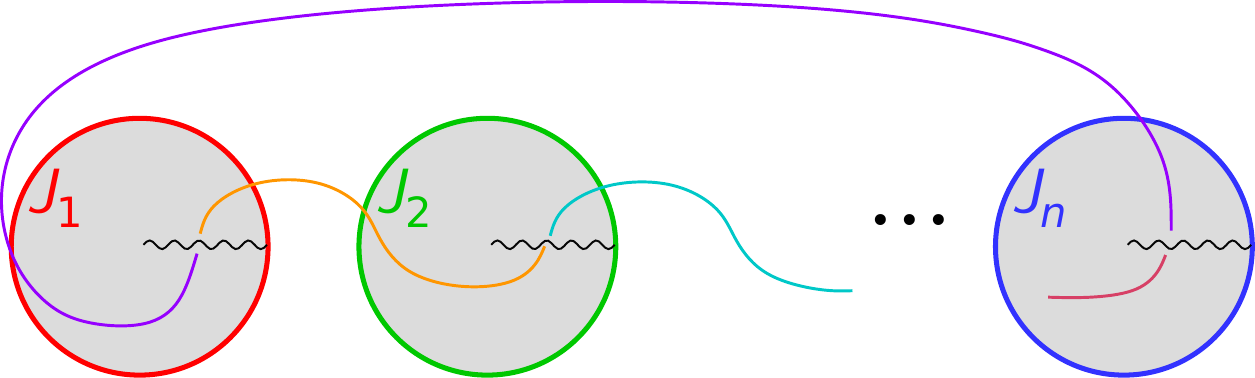}}
}
\noindent This is a path integral on a cone with boundary total length $\beta n $. Each segment of length $\beta$ has some (generically different) boundary conditions $J_i$. So there are $n$ boundary condition changing operators on the disk. We can quotient the picture to obtain:
\eqn{Z_n  \quad   &= \quad \eqfig{0.25\columnwidth}{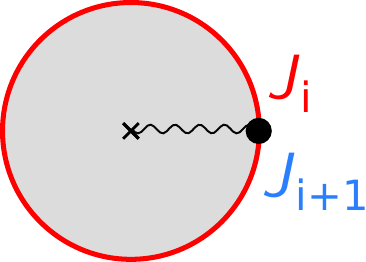} 
}
\noindent In the quotient picture, the CFT is again the $n$-fold tensor product of the seed theory, now with only a single boundary condition changing operator $\oj$ and a twist operator $\sigma$. We will compute the matter entropy as a function of the position of the twist operator. The virtue of the quotient picture is that whereas before we had an $n$-pt function on the disk, now we only have a bulk-to-boundary 2-pt function, which is fixed by conformal symmetry:
\eqn{\ev{\sigma(z,\zb) O_b(y) }_\disk 
\propto \frac{ 1 }{(|1-|z|^2)^{2h-h_b}  |1 -z \yb|^{2h_b} }  }
We can obtain this by mapping the disk to the upper half plane, and then using the doubling trick to relate the correlator to a chiral three-pt function on the plane.
This is the 2-pt function on a flat disk. We are interested in the \pcr{} disk 
$ds^2 = 4 dz  d\bz /(1-|z|^2)^2 $ with a circular boundary of circumference $\beta/\epsilon$ at $(1-|z|^2)/2 = 2\pi \epsilon/ \beta$, so a Weyl transformation leaves us with
\eqn{\ev{\sigma(z,\zb) \oj(\theta') }_{\ads} = c_n(\{J\}) \lp \frac{2\pi \epsilon}{\beta } \frac{  (1-|z|^2) }{ |1 -z \yb|^{2} }\rp^{h_b}. \la{eq:twist1}  }
Note that this expression is invariant under rotational (boost) symmetry $z \to z e^{i\theta}, y \to y e^{i \theta}$.
In our problem, the boundary condition changing operator is on the left side at $\yb = -1$ and $z$ is real.
The dependence on $X$ and comes in via both $h_B$ and the BOE coefficient $c_n(\{ J \})$.
In principle one can evaluate this coefficient, which is related to an $n$-pt function of boundary vertex operators on a disk:
\eqn{ c_n(\{J\}) =  \prod_{i < j}^n \left| w_i-w_j\right|^{-2\lp \frac{X_i - X_j}{2\pi} \rp^2/\alpha'  }  }
where $w_k = e^{2\pi i (k-1)/n}$. Together $h_b \propto -2\lp \frac{X_i - X_j}{2\pi} \rp^2/\alpha'$,  \eqref{eq:twist1} reduces to ``just'' a Gaussian integral. Nevertheless, the $c_n(\{J\})$ factors lead to a sufficiently complicated determinant that analytically continuing the answer in $n$ is not easy. However, defining
\def\ta{\tilde{a}}
\eqn{\ta&=  {1 \over 4\pi^2 \alpha' m^2} \log \lp  \frac{\beta}{2\pi  \epsilon } \frac{ 1+z }{1-z } \rp  ={d_\mathsf{twist}  \over 2\pi^2 \alpha' m^2} ,}
where $d_\mathsf{twist} $ is the distance from the boundary to the twist operator.2 
we expect that when $\ta \gg 1$, the integral will be dominated by the second factor in \eqref{eq:twist1}, which is much more sharply peaked in $X^2$. The entropy with then be of the form \eqref{eq:noncompact} where $a \to \tilde{a}$.

The entropy in the limit of large $\Tilde{a}$ is
\begin{align}
    S_m &\approx {1 \over 2} \ln \Tilde{a} + ...
\end{align}
At fixed $\alpha' m^2$, this answer is expected\footnote{The only possible complication is that the $n\to 1$ limit does not commute with $\tilde{a} \to \infty$} to be valid when $z+1 \gg \epsilon^{1/2}$, e.g., as long as the twist operator is far from the boundary cutoff.

\begin{figure}[H]
\begin{center}
\includegraphics[scale=.6]{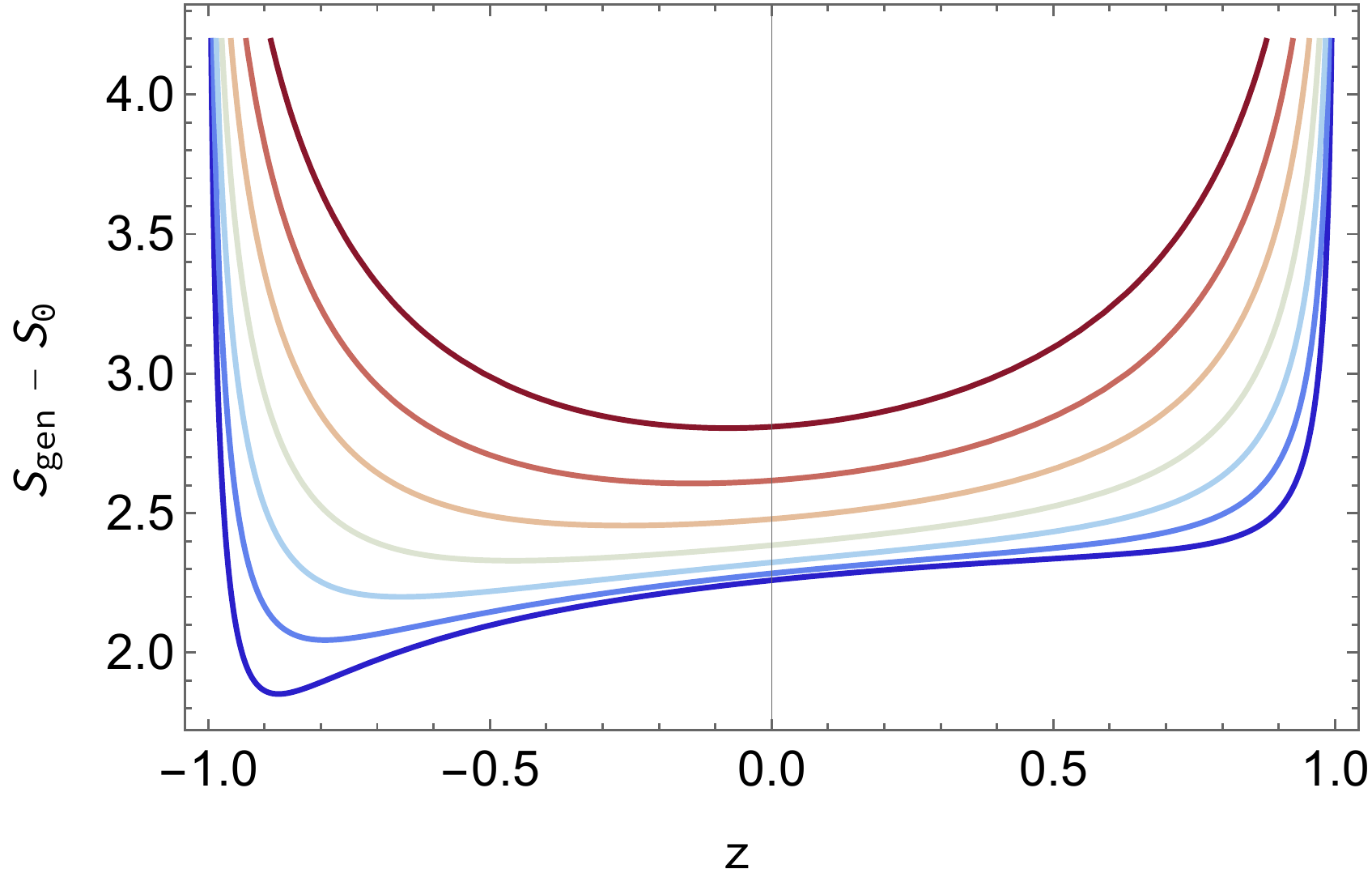}
\caption{ Generalized entropy $S_m + \phi - \phi_0$ as a function of the real coordinate $z$. Here $S_m$ is the matter entropy of an interval $[-1+\epsilon^{1/2},z]$. For large values of $\phi_h$, there is a QES near the horizon. As $\phi_r$ gets smaller, the QES shifts closer to the left side. 
\label{extqes}} 
\end{center}
\end{figure}

To find the QES, we need the profile of the dilaton, which in these coordinates is
\eqn{\phi =  \frac{2\pi \phi_r}{\beta} \frac{ 1 + |z|^2}{1-|z|^2}}
In general, there will be a QES to the left of the horizon as long as $\phi_r/\epsilon \gg 1$, e.g., as long as $\phi_r$ is fixed in the $\epsilon \to 0$ limit. This is shown in figure \ref{extqes}. Balancing out the derivatives of the $\phi$ and $S_m$ places the QES at
\begin{align}
    z_\mathsf{QES} = -{\beta \over 2 \pi \phi_r \ln \beta/(2\pi \epsilon)}
\end{align}
This is just outside the left horizon. To leading order in $1/\Tilde{a}$ we find no dependence of the location of the QES on the degree of uncertainty of the couplings $m^2$.


\def\wb{ {\bar{w}} }

\subsection{Entropy of the island \la{sec:island}}
\def\dbd{\Delta_{\partial}}
Let us finally consider the matter entropy of an island $S_m(I \un \re)$.
Again consider the $n$-th Renyi entropy of the semiclassical theory.
Like in the case of the trivial surface, there are 2 boundary condition changing operators; now, there are also 2 twist operators in the bulk: 
\eqn{Z_n  \quad   &= \quad \eqfig{0.25\columnwidth}{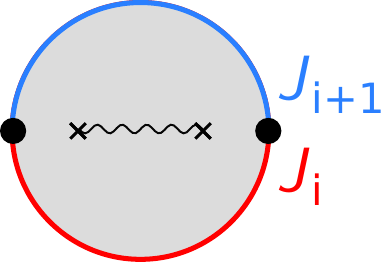} 
}
One can ask what happens when we bring one of the twist operators very close to a boundary operator. This is a bulk-boundary OPE (sometimes referred to as a BOE) limit. More explicitly, we can expand both twist operators in terms of boundary primaries and their descendants. Then we are left with a boundary 4-pt function. In the limit that two of the points are close, only the boundary identity will contribute:
\eqn{\lan \oj(y_1) \sigma(z_1, \zb_1) \sigma(z_2, \zb_2) \oj^\dagger(y_2)\ran   \quad   &= \quad \eqfig{0.35\columnwidth}{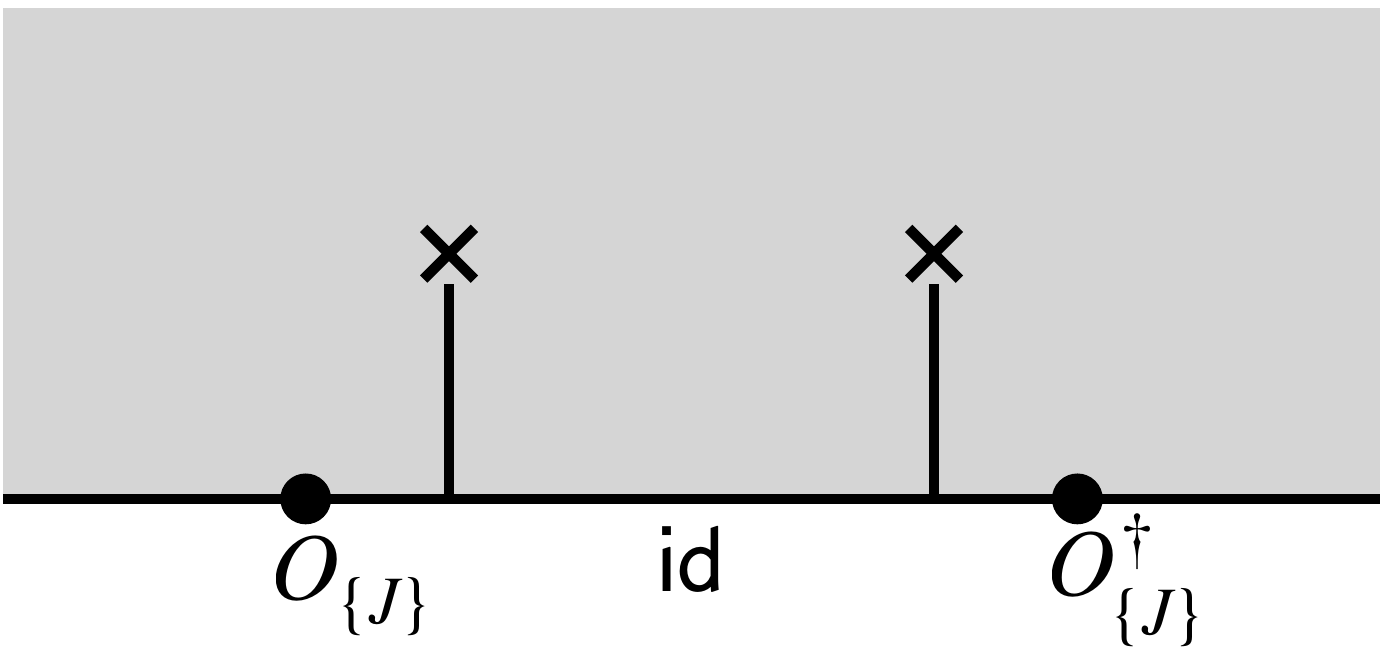} 
}
Let us consider the entropy of a finite interval with both endpoints near the horizon at very late times. This is a bulk-boundary OPE limit: the distance to the horizon is fixed whereas the length of the wormhole is growing linearly with time. So we expect the above four point function to factorize:
\eqn{\lan \oj(y_1) \sigma(z_1, \zb_1) \ran \lan \sigma(z_2, \zb_2) \oj^\dagger(y_2)\ran   \quad    }
The Renyi entropy, however, doesn't factorize, but is instead given by correlated sum of the above product:
\eqn{\int \prod_{i = 1}^n d J_i \, p(J_i) \ \lan \oj(y_1) \sigma(z_1, \zb_1) \ran \lan \sigma(z_2, \zb_2) \oj^\dagger(y_2)\ran   \quad    }
In the frame where we evolve in time symmetrically on the left and right, the problem has a symmetry that interchanges the left and right sides. This allows us to consider instead the quantity
\eqn{\int \prod_{i = 1}^n d J_i \, p(J_i) \  \lan \sigma(z_2, \zb_2) \oj^\dagger(y_2)\ran^2   \quad    }
This is identical to the single interval case, but with $\Tilde{a} \rightarrow 2 \Tilde{a}$. This minor modification does not change the location of the QES. See Figure \ref{fig:EC}.

This case demonstrates the strong sensitivity of the black hole interior to the values of the couplings. The Page-like transition gives a limit to the allowed uncertainty in the couplings, as measured by the entropy between the journal and the boundary, after which the entanglement wedge snaps and the interior falls outside the entanglement wedge of the boundary.

\pagebreak
\section{Renyi Entropies in gravity and SYK \la{sec:renyi}}

In the previous section, we discussed a model where the reconstructability absent precise knowledge of the couplings could be directly probed by finding the QES of the unknown couplings. Not all models enjoy this level solvability. However, it is sometimes easier to compute the Renyi entropy in these more general models. Thankfully, there are two signatures of the presence of a QES from the Renyi entropy. The first is a unitarity paradox where the semi-classical saddle produces a Renyi entropy too small to be consistent with the dimensionality of the Hilbert space of the system. The second signature (and what fixes the first) is the presence of Replica wormholes. In solvable models, the QES prescription falls out of the $n \rightarrow 1$ limit of the $n$-th Renyi entropy computation.

In this section, we study the $n = 2$ Renyi entropy in both JT gravity and SYK. This will be done using various methods/regimes, including the low temperature Schwarzian regime, at large $q$, and also numerically via exact diagonalization of the SYK Hamiltonian. We will again find signatures of the failure of reconstruction due to insufficient knowledge of the couplings. We will also consider the case where the unknown couplings correspond to an irrelevant deformation of the system.

\subsection{Disk contribution}
As an intermediate step for the Renyi computation, we will compute elements of the journal density matrix. In any holographic theory, this is determined by a computation like
\eqn{\bra{J} \rho \ket{J'}   \quad   &\propto \quad \eqfig{0.25\columnwidth}{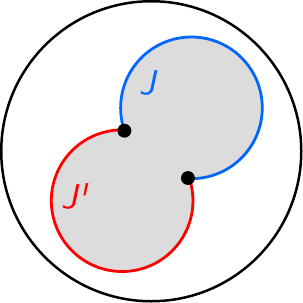} \la{eq:disk}
}
where we fill in some gravity solution (gray) with the appropriate boundary conditions that correspond to evolution by $H + \chi(u) O_\Delta$. $\chi(u)$ takes the value $J$ for $0<u < \tau$ (blue) and the value $J'$ (red) for the other part of the circle $\tau < u < \beta$, which determines the boundary conditions of the bulk scalar field $\chi$. By taking $\tau = \beta/2+iT$ one can study the matrix elements as a function of real time $T$.

To make further progress, we will now assume that field dual to $O_\Delta$ is a free field in AdS$_2$. We can then integrate out the field to get an effective matter action
\eqn{ -I_m = D\int du_1 \, du_2 \lb  \frac{t'(u_1) t'(u_2)}{2 \sin^2 \lp \frac{t(u_1) - t(u_2)}{2}\rp  } \rb^{\Delta} \chi_r(u) \chi_r(u')  \la{eq:matter}  }
where $D=\frac{\left(\Delta-\frac{1}{2}\right) \Gamma(\Delta)}{\sqrt{\pi} \Gamma\left(\Delta-\frac{1}{2}\right)}$.
In computing the overlap, we should divide by the norms of the states $\sqrt{\braket{J}{J} \braket{J'}{J'}}$; but these are not time dependent and do not play an important conceptual role.
Following the discussion of the free boson in the previous section, we will now specialize to the case of a marginal deformation $\Delta = 1$, where the computation simplifies significantly and the full gravitational backreaction can be computed. 
For one thing, we may shift the $J \to J + a$, $J' \to J' + a$ without changing the value of the overlap, so without loss of generality we may set $J'=0$. In Appendix \ref{app:matter} we evaluate this integral for the special case of a marginal deformation $\Delta = 1$, giving an answer that is remarkably bi-local in $u_1$ and $u_2$:
\eqn{e^{-I_m} &=  \lb \frac{\epsilon^2 t'(u_1) t'(u_2)}{ \sin^{2} \lp \frac{t(u) -t(0)}{2}\rp  }\rb^{\delta}, \quad \delta =  \frac{(J-J')^2}{2\pi} \la{eq:matt2} }
where $\epsilon$ is a UV regulator. This precisely agrees with the dimension of the boundary condition changing operator we found in \nref{eq:bcc} if we set $\alpha' = 1/(2\pi)$ so that the scalar is canonically normalized. Note that the final form is consistent with a picture where we have inserted a bulk ``domain wall'' of mass $\sim \delta$ that separates the $J$ and $J'$ vacua. The domain wall is where the gradient of the bulk field is appreciable; it is not a thin wall.

This matter action is valid for an off-shell $t(u)$. In other words, we can include the gravitational backreaction by integrating over $t(u)$ with the Schwarzian action appropriate for the disk. This is the same action we would have gotten from inserting two local operators of NCFT$_1$ dimension $\delta$ at $u=0$ and $u$. For a free scalar field in the bulk, these expressions combined with the results for the Schwarzian $n$-pt functions \cite{Mertens:2017mtv, Lam:2018pvp, Saad:2019pqd, Yang:2018gdb} give us the exact disk contribution, summing over all quantum fluctuations of the boundary mode. 

The fact that we have an exact quantum expression for the Renyi entropy gives us confidence that there is really an information paradox if we just focus on the disk. Although in the classical approximation, the disk contribution decays exponentially, without the quantum expressions, we would not be confident that the disk decays to a value smaller than what is required by unitarity $\sim e^{-S_0}$. Indeed, the quantum modifications show that the exponential decay is replaced by a power law decay $\sim T^{-3}$, with the exponent independent of $\delta$.

Here we would also like to comment that the above result also applies to SYK in the Schwarzian limit.  We will use the following conventions for SYK:
\eqn{H_\syk=i^{q / 2} \!\!\! \!\!\! \sum_{1 \leq i_{1} \ldots \leq i_{a} \leq N} \!\!\!\!\!\!\! J_{i_{1} \ldots i_{q}} \psi_{i_{1}} \ldots \psi_{i_{q}}, \quad\left\langle J_{i_{1} \ldots i_{q}}^{2}\right\rangle=\frac{N J^{2}}{q {N \choose q}}=\frac{N \mathcal{J}^{2}}{2 q^2 {N \choose q} \la{eq:sykdef} },
\quad\left\{\psi_{i}, \psi_{j}\right\}=2 \delta_{i j}}
To mimic the above expressions, we can imagine turning on a deformation by another SYK Hamiltonian,
\eqn{H=H_\syk(J_1,q_1) +  \chi(u) H_\syk(J_2,q_2) \la{syksum} }
Here $J_1, J_2$ are distributed like in \eqref{eq:sykdef}. We can introduce a $G,\Sigma$ action by following the usual steps, only integrating out both $J_1$ and $J_2$ to obtain an action $I_0 + I_m$, where
\eqn{-I_0/N &= \log \operatorname{Pf}\left(\partial_{t}-{\Sigma}\right)-\frac{1}{2} \int d \tau_{1} d \tau_{2}\left[{\Sigma}\left(\tau_{1}, \tau_{2}\right) {G}\left(\tau_{1}, \tau_{2}\right)-\frac{J^{2}}{q} {G}\left(\tau_{1}, \tau_{2}\right)^{q} \right]\\
&\approx \frac{ \alpha_{S}}{\mathcal{J}} \int d u \left\{\tan \frac{\pi t(u)}{\beta}, u\right\}\\
-I_m/N &=  \frac{J^2 }{2q'} \int du_1 \,  du_2 \, \lb  G(u_1, u_2) \rb ^{q'} \chi(u_1) \chi(u_2)\\
&\approx   \int du_1 \,  du_2 \,  \lb  \frac{t'(u_1) t'(u_2)}{2 \sin^2 \lp \frac{t(u_1) - t(u_2)}{2}\rp  } \rb^{q'/q} \chi_r(u_1) \chi_r(u_2), \quad \chi_r = \chi(u) b^{q'/2}    \la{eq:sykaction}
}
where $J^2 b^q \pi = \lp \hf - {1 \over q} \rp \tan (\pi/q)$.
Here $I_0$ comes from $J_1$ and $I_m$ comes from the $J_2$ couplings. The relative normalization of the two terms is controlled by the magnitude of $\chi$.
Now in the low temperature, large $N$ limit $I_0[G,\Sigma] \to \sch[t(u)]$ is replaced by the Schwarzian action, where we integrate over $t(u)$ instead of $G,\Sigma$. Similarly, the action $I_m$ can be re-written in terms of the Schwarzian mode.
Note that in writing these expressions, we also assume that $\chi$ is small so that we can simply integrate over the near-zero mode. This is a similar approximation to what is discussed in \cite{Maldacena:2018lmt}.

We expect that the disk partition function at early times to be self-averaging both in $J_1$ and $J_2$. Therefore, this computation has multiple interpretations. The first interpretation is that we draw some particular choice of $J_2$ and view the term $H_\syk(J_2)$ as a single operator that is deforming the original theory. The journal records not the values of $J_2$ but merely a single coupling $\chi_0$ which is the overall normalization of the deformation. Then the above action would govern the matrix element of the journal density matrix, e.g., an overlap between $\tfd$'s with different values of $\chi_0$. Since the disk answer (for early times) is self-averaging, we can average over $J_1, J_2$ in this computation and derive the above effective action in terms of $G,\Sigma$.

A second physically different setup is when $\chi_0$ is fixed, and the journal instead records the values of all ${N \choose q'}$ couplings $J_2$. 
Then we think of $J_1$ as ``known'' or fixed. We will see in the next section that \eqref{eq:sykaction} will still be relevant for computing Renyi entropies, or the overlaps averaged over $J_2$.

\subsection{Renyi-2 wormhole}
\def\taub{\bar{\tau}}
\def\cc{\mathcal{C}}
The Renyi-2 entropy of the journal is given by
\eqn{\tr \rho^2_\re = \int p(J)p(J') |\bra{J} \rho \ket{J'}|^2.}
As an intermediate step, we should compute the squares of density matrix elements. 
If we plug in our answer for the density matrix elements \eqref{eq:matter}, there will be a unitarity paradox at late times. In particular, $\tr \rho^2_\re$  will decay to zero, which is impossible since it is bounded by $2^{-N}$.

The resolution to this paradox is that the squares of the density matrix elements have an additional contribution given by wormholes. In a theory with other ``known'' couplings, this is an acceptable resolution, but such wormholes would violate factorization in a theory with fixed couplings.
To obtain the norm of the overlaps, we start with a Euclidean computation  real $\tau, \taub$ and then analytically continuing $\tau = \beta/2 + iT$, $\taub = \beta/2-iT$:
\eqn{|\braket{\beta + 2 i T, J}{\beta + 2i T, J'} |^2 &= \tr \lp e^{-\taub H_J}e^{- \tau H_{J'}} \rp \tr \lp e^{- \tau H_J}e^{-\taub H_{J'}} \rp \\
&\supset \quad  \eqfig{0.3\columnwidth}{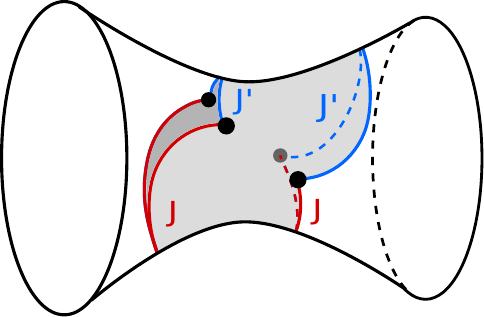} \la{eq:wormhole}}
In addition to the disk topology which we have already discussed, we have drawn a wormhole contribution. We can think of this wormhole as being supported by the bulk ``domain walls'' that separate the $J$ and $J'$ region. Such a wormhole is closely related to the one described by Douglas Stanford in Appendix B of \cite{StanfordMore}. 

Let us outline the steps to obtaining the wormhole, leaving a more thorough discussion to Appendix \ref{app:matter}. We will start with the double trumpet geometry, with a cutout parameterized by two boundary times $T_L(u_L)$ and $T_R(u_R)$. First, we turn on a source $\chi_L$ on the left side from times in $(0,\tau)$ and a source $\chi_R$ which is on from times in $(\tau,\beta)$. This leads to a matter action that has $LL$ correlators, $RR$ correlators, as well as $LR$ cross terms.
\def\trum{{\mathsf{wormhole}}}
\eqn{ -I_m^\trum &= D\int du_1 \, du_2  \Bigg\{ \lb \frac{T_L'(u_1) T_R'(u_2)}{\cosh^2 \lp \frac{T_L - T_R}{2}\rp } \rb^{\Delta} \chi_L(u_1) \chi_R(u_2) +\\
&  \lb \frac{T_L'(u_1) T_L'(u_2)}{\sinh^2 \lp \frac{T_L(u_1)- T_L(u_2)}{2}\rp } \rb^{\Delta} \chi_L(u_1) \chi_L(u_1)+  \lb \frac{T_R'(u_1) T_R'(u_2)}{\sinh^2 \lp \frac{T_R(u_1) - T_R(u_2)}{2}\rp } \rb^{\Delta} \chi_R(u_1) \chi_R(u_2) \Bigg\} }
In principle, one should evaluate this integral for off-shell $T_L,T_R$, and then find the classical solution.
For $\Delta = 1$ the integral can be performed  but the answer is a bit complicated. However, large Lorentzian times $T$, a sensible ansatz is that the distance between the quench sites on opposite sides of the wormhole and $u_2$ and $u_3$. In the above picture \eqref{eq:wormhole}, we are saying that the domain walls that cross the wormhole on the ``front'' and ``back'' sides have fixed length at large $T$. With this ansatz, one can show that the above matter integral becomes the insertion of two domain wall operators:
\eqn{e^{-I_m} = \cc^{LR}(u_1, u_4) \cc^{LR}(u_2, u_3), \quad \cc =  \lb \frac{T_L'(u_1) T_R'(u_2)}{\cosh^2 \lp \frac{T_L - T_R}{2}\rp } \rb^{\delta}.}
Since the overall effect of the matter is local, the on-shell solution must again be semi-circular arcs, with junctions at the quench sites. Therefore the wormhole can be obtained by starting with two copies of the disk geometry in \eqref{eq:disk}, one for the ``front side'' of the wormhole and the other for the ``back side.'' 
Then one cuts both of these disks and pastes them together along two geodesics so that there is no bulk discontinuity. 
The key equation we will need is the effective energy of the solution, or equivalently the circumference $\beta_E$ of the pieces of the disk that we use to make the solution. This is the same energy on all pieces of the solution away from the quench sites. 
It is determined by imposing $\slt$ charge conservation at the quench sites, see \ref{app:matter}. The result is that for any $T$, 
\eqn{\tan \left(\frac{\pi \beta}{2 \beta_{E}}\right)=\frac{\delta \beta_{E}}{2 \pi}.\la{eq:tanb} }
For small values of $\delta \beta$, we get $\beta_E^2 =  \pi^2 \beta/\delta$.
Several aspects of the wormhole geometry are discussed in \ref{app:matter}; here we will just check that the wormhole resolves the unitarity problem. To do so in a semi-classical approximation, we need to evaluate the on-shell value of 
\eqn{\exp \lp {-I^\trum} \rp \cc^{LR}(u_1, u_4) \cc^{LR}(u_2, u_3).}
Let us now evaluate $I^\trum$ in the limit of small $\delta \beta$ and show that this resolves the potential unitarity paradox at large $T$. There are two Schwarzian boundaries in the computation, both of which have the same action. The left boundary consists of two arcs:
\eqn{
-I_\sch &= 2  \int   du_L \{ \tan \frac{\pi t(u)}{\beta}, t\} = 2 \lb \tau + \taub \rb  \lp \frac{\pi }{\beta_E}\rp^2 \\
&= 2 \pi^2 \beta / \beta_E^2.
}
The correlators are essentially on opposite sides of an effective thermofield double $\beta_E$, so they take thermal values:
\eqn{Z_\text{wormhole} \sim \exp \lp {- 4 \pi^2 \beta / \beta_E^2} \rp \beta_E^{-4\delta}.  }
We see that the classical action approaches a value independent of $T$ at late times, which resolves the potential unitarity paradox for the Renyi-2 entropy. 
More generally, from the results in Appendix \ref{app:matter}, we can see that the solution at general $iT$ will give an action independent of $T$ as both the distance across the wormhole and the action, including corner contributions, will be time-independent.

\subsection{Numerics for finite $N$ SYK }
\def\ntrials{n_\mathrm{trials}}

In this section, we resort to the numerical analysis of the $n = 2$ Renyi entropy of the journal by exact diagonalization of the SYK Hamiltonian at finite $N$. We will present evidence of the wormholes. The ``known" couplings in all of our setups will be in a standard SYK Hamiltonian with ${N \choose q}$ couplings. The ``unknown'' couplings will be additional unknown couplings.

We argued around \eqref{eq:cond} that the relevant entropy that diagnoses the entanglement wedge given the ``known'' couplings is the relative entropy $S(\unk | \kno)$. The ``conditional'' Renyi entropy whose $n\rightarrow 1$ limit is equal to this conditional entropy is given by
\eqn{\tilde{R}_n &=  \sum_{\kappa} P(\bkap) \sum_{\bmu_1, \cdots, \bmu_n} \prod_{i=1}^n P(\bmu_i)  \braket{ \psi ; \bmu_i, \bkap }{ \psi ; \bmu_{i+1}, \bkap}_\sys.}
\eqn{ S(\unk | \kno) = -\pd_n \tilde{R}_n\big|_{n=1}.  }
An efficient numerical way to estimate the above quantity is to Monte Carlo sample the quantity $\prod_{i=1}^n \braket{ \psi ; \bmu_i, \bkap }{ \psi ; \bmu_{i+1}, \bkap}_\sys$ from the distribution $P(\mu_1) \cdots P(\mu_n) P(\kappa)$. The conditional Renyi $\tilde{R}_n$ has the appealing feature that it is linear in $P(\kappa)$, as opposed to the standard Renyi entropy of the full journal, which comes with a $P^n(\kappa)$ factor.


We will now employ this method in studying $\tilde{R}_2$ for various ``unknown'' deformations of the SYK model. The first case we will consider is the $q=4$ SYK, where the couplings are known with only limited precision. To model this, we decompose the usual SYK Hamiltonian into two terms:
\eqn{ H = \lp K_{i jkl } + J_{ijkl}\rp  \psi_i \psi_j \psi_k \psi_{l}, \quad  }
Here $K$ are the known couplings, and $J$ are the unknown couplings (the ones stored in the journal). We can think of this setup as modeling a situation where the usual SYK couplings are known but with some Gaussian errors. This is similar to \eqref{syksum}, except that there we were viewing the single coupling as an overall normalization of the second term; here, there are ${N \choose 4}$ couplings in the journal.  The conditional $n = 2$ Renyi entropy is given by
\eqn{\tilde{R}_2 &=  \int dJ_0 \,  p(J_0)  \int dJ dJ' \, p(J) p(J') \, |\ev{\beta/2+iT; J_0, J|\beta/2+iT; J_0,J'}|^2.}
We would like to test whether wormholes contribute to the above calculation. To do so, we consider the following ``disk'' approximation to $\tilde{R}_2$:
\eqn{\tilde{R}_2^\disk &=  \Bigg|\int dJ_0 dJ dJ' \,p(J_0) p(J) p(J') \, \ev{\beta/2+iT; J_0, J|\beta/2+iT; J_0,J'}\Bigg|^2.}
In this approximation, we have averaged first, before squaring. In the large $N$ analysis, only disconnected solutions can contribute to the above quantity. If $\tilde{R}_2 $ is significantly larger than $\tilde{R}_2^\disk $, we interpret this as evidence that a wormhole is dominating over the disconnected solution.

To Monte Carlo sample these integrals, we draw two independent SYK Hamiltonians $H_1$ and $H_2$ from Gaussian distributions. Then $H_\pm = (\cos \theta) H_1 \pm (\sin \theta) H_2$ are also normalized Hamiltonians such that $\ev{\tr H_\pm^2 } = J^2, \quad \ev{\tr H_+ H_-} =  \lambda^2 J^2$, where $\lambda^2 = \cos (2\theta)$. 
The results are displayed in Figure \ref{fig:2}. The disk approximation is a good one at small $JT$ but at large values, we see that it significantly underestimates $\tilde{R}_2$. In fact, at very late times, the underestimate is so bad that even without computing $\tilde{R}_2$ exactly, we can rule it out on the grounds that it would violate unitarity.

\begin{figure}[H]
\begin{center}
\includegraphics[scale=.7]{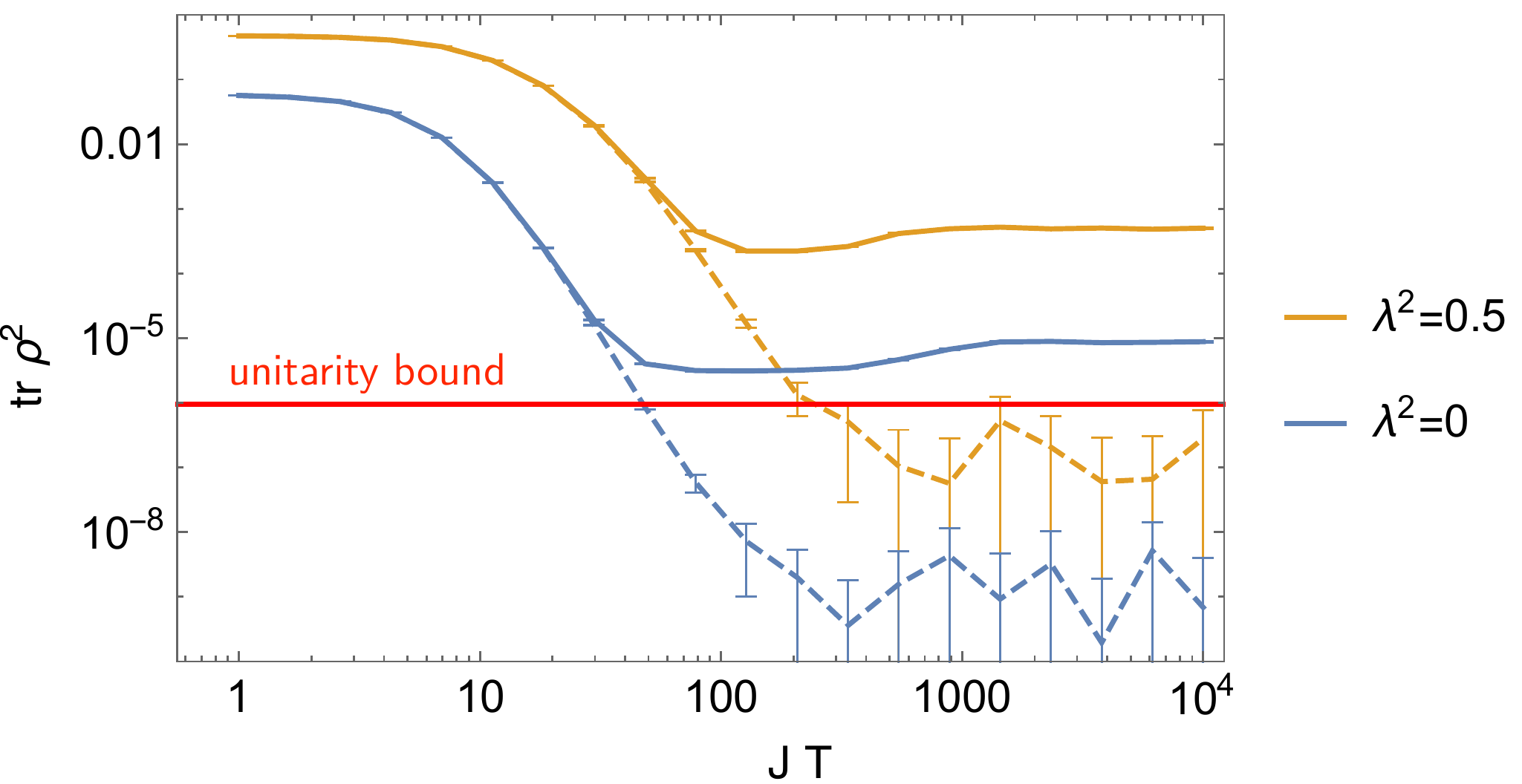}
\caption{ The conditional Renyi $\tilde{R}_2$ as a function of Lorentzian time evolution $J T$, obtained by numerically diagonalizing the SYK Hamiltonian for $N=20$, $\beta J = 16$ and $\ntrials = 3000$. $1 \sigma$ error bars are displayed. The solid curve is the full answer while the dashed curve is the ``disk'' approximation $\tilde{R}_2^\disk$. We see that the disk approximation is good for early times, but decays too rapidly at late times. Even without computing the full answer (in orange), the unitarity bound $\tr \rho^2 \ge 2^{-N}$ would rule out the disk answer at large times. \la{fig:2}} 
\end{center}
\end{figure}

We also consider deforming a standard $q = 4$ SYK Hamiltonian with a {\it single} unknown coupling $\chi$:
\eqn{ H = \lp K_{i jkl } \rp  \psi_i \psi_j \psi_k \psi_{l} + i^{q_2/2} \lp \psi_1 \cdots \psi_{q_2}\rp \chi . } 
Here $K_{ijkl}$ are drawn from a Gaussian as usual with a variance set by \eqref{eq:sykdef}; we also choose $\chi$ to be normally distributed, with a variance $\ev{\chi^2} = \epsilon^2 J^2$. 
This may be thought of as a variant of the problem in \eqref{syksum}. The results are shown in figure \ref{fig:q4q6}. 
We see clear evidence of a wormhole, even when $q_2 > 4$. In fact, at the moderate values of $N$ that we computed, there does not seem to be a large qualitative difference between $q_2=4$ and $q_2>4$. This is perhaps surprising since in the conformal regime, $q_2 > 4$ would correspond to an {\it irrelevant} coupling whereas $q_2 = 4$ would be marginal.
To further test this interpretation, we compute the $G_\lll$ correlators numerically and see that they qualitatively agree with the predictions given in the subsequent section on the large $q$ wormhole, see Figure \ref{fig:l0c} {for the case of many marginal couplings} and \ref{fig:q6c} {for the case of a single irrelevant coupling}. We also computed $G_\lr$ numerically and saw that it is nearly zero at early times $Jt \sim 1$ but becomes appreciable at times that correspond to the wormhole/disk transition. More specifically, in Figures \ref{fig:l0c} and \ref{fig:q6c} we compare 
\eqn{ |G_\lll^\disk| &=  \frac{1}{N'}\sum_i \frac{\overline{ \left| \bra{ \beta+2 iT, J} \psi_i(\beta/2) \psi_i(t) \ket{\beta+2i T,J'} \right|} }{\overline{|\braket{\beta+2iT,J}{\beta+2iT,J'}|} },\\
|G_\lll^\mathsf{WH}| &= \frac{  \overline{ \bra{ \beta+2 iT, J} \psi_i(\beta/2) \psi_i(t) \ket{\beta+2i T,J'}\braket{\beta+2iT,J'}{\beta+2iT,J} }}{ \overline{|\braket{ \beta+2 iT, J'}{\beta+2i T,J}|^2} },   }
where in the above expressions the overbar indicates averaging with respect to $J$ and $J'$. {For the single coupling case, the fermion index $i$ is summed over all $N' = N - q_2$ fermions except those appearing in the deformation, whereas in the case of many marginal couplings, the index $i$ is summed over all $N' = N$ fermions.}
Of course our limited numerics cannot conclusively test the conformal regime ($N \to \infty$, $\beta \cj \gg 1$) where the term ``irrelevant'' has a sharp  meaning. Nevertheless, we conjecture that in the conformal/JT regime, there exist wormhole solutions for irrelevant couplings; we hope to report progress in this direction in a future publication.

\begin{figure}[H]
\begin{center}
\includegraphics[scale=0.6]{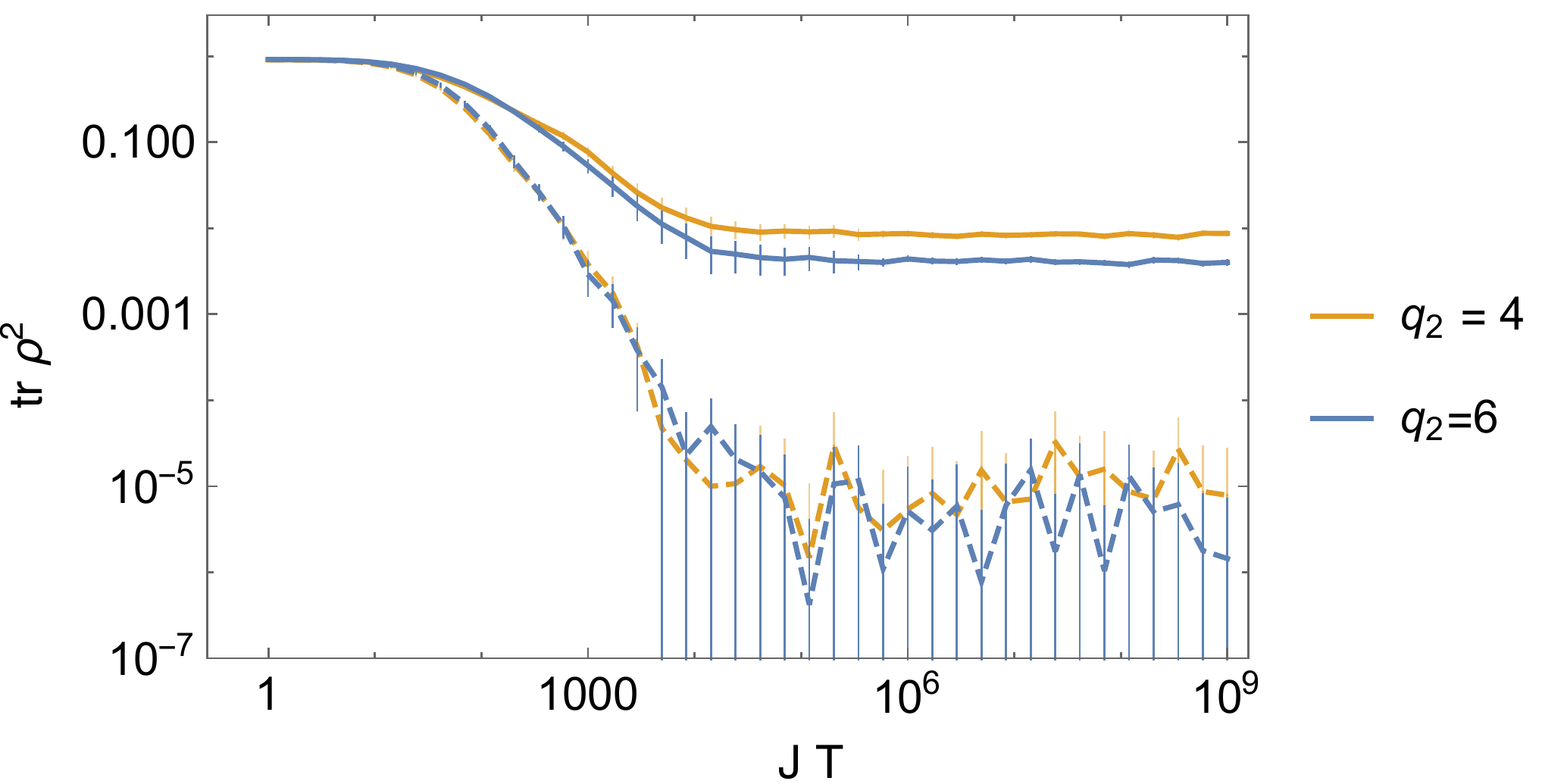}
\caption{$N = 20$, $\ntrials = 1000$, $\beta J = 16, \epsilon = 0.1$. We show $q_2 = 4,6$ deformations and $1\sigma$ error bars. Note that $q_2=6$ is an irrelevant deformation, but the curves seem qualitatively similar to the marginal case. \la{fig:q4q6}} 
\end{center}
\end{figure}

\begin{figure}[H]
\begin{center}
\includegraphics[scale=0.6]{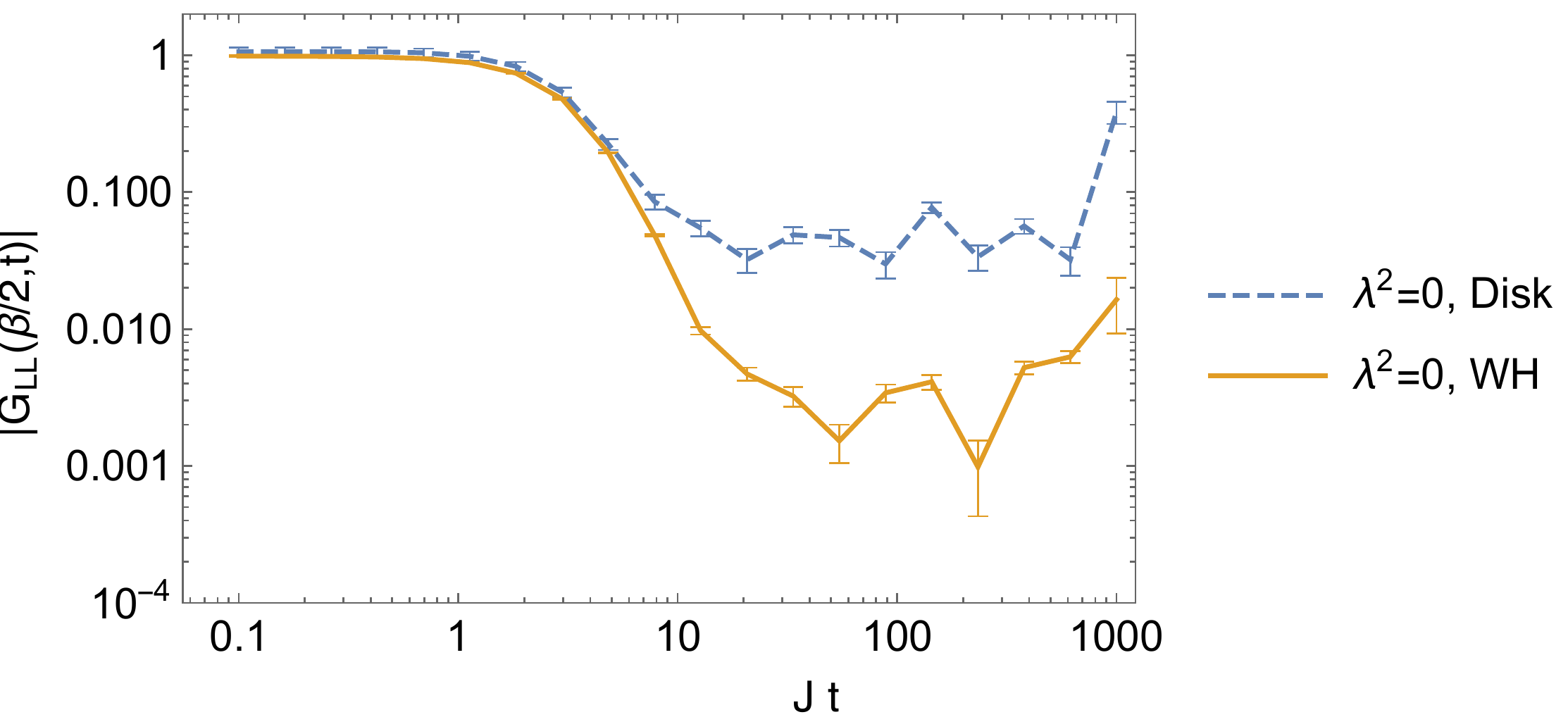}
\caption{The 2-pt correlator $|G_{LL}|$ as a function of Lorentzian time for the $\lambda^2=0$ disk and wormhole for the case of many marginal couplings.  We take $N = 20$, $\ntrials = 600, \beta \cj = 1, JT=10^3$. Notice that as $t \to T$ we reach the second quench site. For the disk solution, a prediction from the large q analysis is that the correlator becomes large on the disk, whereas it remains small on the wormhole.  This seems to be in rough agreement with the modest $N,q= 4$ numerics.  In computing the correlator, we divide by the average norm $\braket{\beta+2iT,J}{\beta+2iT,J'}$ for the disk and $|\braket{\beta+2iT,J}{\beta+2iT,J'}|^2$ for the wormhole. \la{fig:l0c}} 
\end{center}
\end{figure}

\begin{figure}[H]
\begin{center}
\includegraphics[scale=0.6]{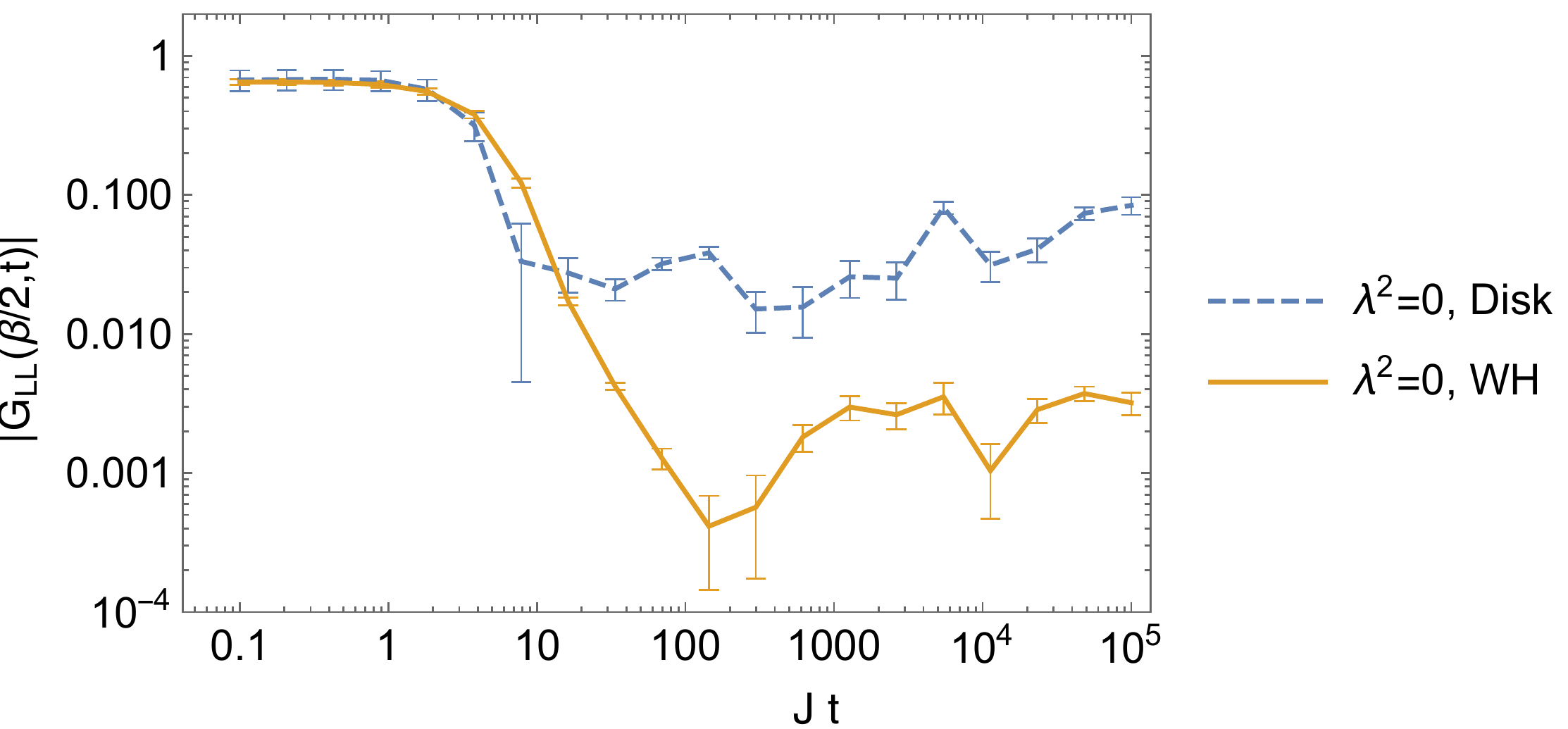}
\caption{
The 2-pt correlator $|G_{LL}|$ as a function of Lorentzian time, where a single irrelevant $q=6$ parameter is varied, as in Figure \ref{fig:q4q6}.  We take $N = 20$, $\ntrials = 2000, \beta \cj = 1, JT=10^5$. The correlators are qualitatively similar to the case with many marginal parameters in Figure  \ref{fig:l0c}.
\la{fig:q6c}} 
\end{center}
\end{figure}

\subsection{Large $q$ SYK \la{sec:largeq} }
\def\hm{\mu}
\def\tg{g}
\def\gh{ {\hat{g} }}
\def\omt{\Omega_{\tau}}
\def\betae{\beta_{E}}

In this section, we will consider the disk contribution to the Renyi-2 entropy in the $N\to \infty$ large $q$ SYK model. We will show that the disk contribution decays to zero at large Lorentzian times, in conflict with unitarity. This suggests that there should be a wormhole that dominates at late times. We show that the wormhole in the low temperature limit is closely related to the Schwarzian wormhole described in the gravity section. 

\begin{figure}[H]
\begin{center}
\includegraphics[width=0.5\columnwidth]{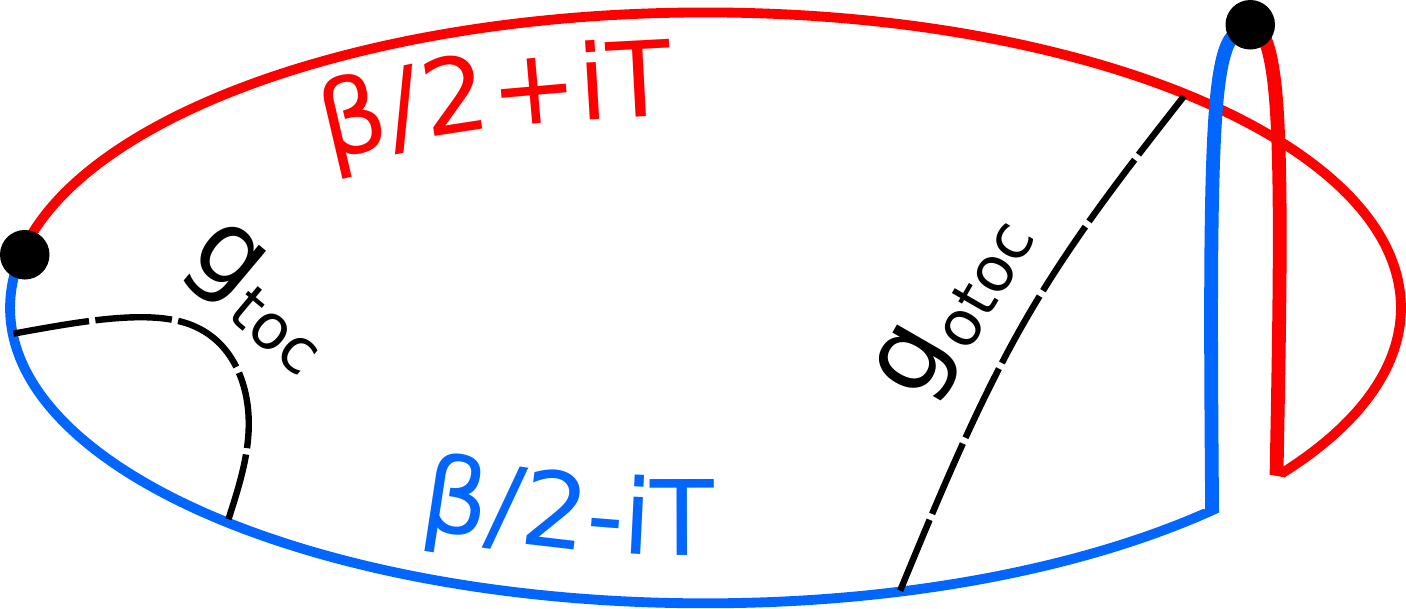}
\caption{ Contour relevant for matrix element of journal. This can be obtained by considering a red segment of length $\tau$ and a blue segment of length $\beta - \tau$, and then setting $\tau = \beta/2 + i T$. \la{fig:drawing2} }  
\end{center}
\end{figure}


We will consider the disk solution for the thermal circle of length $\beta$, with quench sites at times $0$ and $\tau$. The couplings are constant in the segments $(0,\tau)$ and $(\tau, \beta)$ and are partially correlated between these two segments.
Let us assume the correlation between the couplings on the two sides is $\lambda^2 = e^{- \mu}$. When $\lambda= 1$ there is no change in the coupling at the quench sites; when $\lambda = 0$ the two sides are uncorrelated. The large $q$ equations of motion are
\eqn{
	\pd_1 \pd_2 g_\toc + 2 \cj^2 e^{g_\toc} = 0, \quad &t_1, t_2 \text{ on same side}\\
\pd_1 \pd_2 \tg_\otoc + 2 e^{-\hat{\mu}} \cj^2 e^{\tg_\otoc} =0, \quad &t_1, t_2 \text{ on opposite sides}  \la{eq:verbose} }
We will sometimes use the subscripts $(\toc, \otoc)$ to emphasize when an equation applies when both times are on the same side or on opposite sides of the quench.
It is convenient to define 
\eqn{ 
    \Omega_\tau &=\Theta(\tau - t_1) \Theta(t_2 - \tau) + \Theta(\tau-t_2) \Theta(t_1 - \tau)\\
    \Omega_\toc &= 0,  \quad   \Omega_\otoc = 1.
    }  
    Then, we can summarize \eqref{eq:verbose} as
\eqn{
	\pd_1 \pd_2 g + 2 \cj^2 \exp \lp {g- \hm\omt } \rp = 0.}
In addition to the usual UV boundary condition $g(\tau, \tau) = 0$, we impose continuity at the quench site:
\eqn{g(t,-\epsilon) &= \tg(t,+\epsilon) \\
g(t,\tau-\epsilon) &= \tg(t,\tau+\epsilon) . \la{eq:cont}}
These boundary conditions follows from the fact that for very short times $\epsilon$, we can neglect the interactions from $\cj$.
A useful trick is the following. Consider the field redefinition 
\eqn{g =  \gh + \mu \Omega. \la{hatg} }
Then the equations of motion and the boundary conditions are
\eqn{
\pd_1 \pd_2  \gh &+ 2 \cj^2 e^\gh =0,\\
\gh_\toc &\to  \gh_\otoc - \mu .
}
By using the variable $\gh$ the equations of motion is uniform over the entire circle, at the price of a discontinuous boundary condition.
This discontinuous boundary condition is considered in \cite{Qi:2018bje, Streicher:2019wek, Eberlein:2017wah}. For general $\tau$, see Streicher \cite{Streicher:2019wek}. In our conventions,
\begin{align}
e^{g_\toc} &= \left( {\alpha_1 \over \cj \sin \left[  \alpha_1 t_{12} + \gamma_1  \right]} \right)^2 , \quad 0 < t_1 < t_2 < \tau \la{eqn:gtop} \\
e^{g_\toc} &= \left( {\alpha_2 \over \cj \sin \left[  \alpha_2 t_{12} + \gamma_2  \right]} \right)^2, \quad t_3 < t_1 < \taub   \la{eqn:gbot} \\
e^{g_\otoc} &= \lp { \alpha_1 \alpha_2 \cj^{-2} \over   \lambda^2 \sin \left[  \alpha_1 (t_1 -\tau)   \right]\sin \left[  \alpha_2 (t_2 -\taub)   \right]  -  \sin \left[  \alpha_1 (t_1 -\tau) + \gamma_1  \right] \sin \left[  \alpha_2 (t_2 -\taub) - \gamma_2  \right]} \rp^2. \label{eqn:gotoc} 
\end{align}
We also have the UV boundary conditions:
\begin{align}
&\alpha_i = {\cal J} \sin \gamma_i \\
&\sin \left(  \frac{ \alpha_1 \tau   \pm \alpha_2 \taub}{2}  + \gamma_1 \pm \gamma_2 \right) = \lambda^2 \sin \left(  \frac{ \alpha_1 \tau \pm \alpha_2 \taub    }{2} \right)\la{streich}
\end{align}
Here we are imagining solving the above equations for real $\tau, \taub$ and then analytically continuing and setting $\tau = \beta/2+iT, \taub = \beta/2-iT$.
The physical branch of the above equations is the one that is continuously connected to the $T=0, \lambda = 1$ solution.
In figure \ref{fig:gamma} we solve for $\gamma$ numerically at $T=0$ for varying $\beta \cj$. We also solve for $\gamma$ as a function of $T$ for fixed $\beta \cj$ in figure \ref{fig:gamma2}.

It is interesting to study these constraints for small $\gamma \approx \alpha/\cj$:
\eqn{ -\frac{\gamma_1 \pm \gamma_2 }{1-\lambda^2} \approx \tan \lp   \frac{\alpha_1 \tau \pm \alpha_2 \taub}{2} \rp. \la{eq:tana}}
This can be compared with a classical Schwarzian analysis for the insertion of a heavy operator of dimension $\delta$ on the disk, which gives (see \cite{Goel:2018ubv} equation 2.9):
\eqn{-\frac{k_1 \pm k_2}{\Delta} =  \tan \lp  \frac{k_1 \tau \pm k_2 \taub }{4 C} \rp. \la{eq:tanSch} } 
This means that we may identify $k_i/2C = \alpha_i $ and $(1-\lambda^2) 2 C \cj =  \delta$. Here $C$ is the coefficient of the Schwarzian action, which for large $q$ SYK is $C \sim N /\cj q^2$. In the Schwarzian theory, $k_i = 2\pi C/(\beta_E)_i$, where $\beta_E$ is the length of the circle that is pieced together to form the disk solution.

The left hand side of the above equation is of order $(\beta \Delta)\inv \sim (\beta \cj (1-\lambda^2))\inv$. So for large $\beta \Delta$, we have $\tan \lb \hf \lp \alpha_1 \tau \pm \alpha_2 \bar{\tau}\rp \rb  = 0$. The correct branch is to take $\alpha_1 \tau +\alpha_2 \taub = 2 \pi$ and $\alpha_1 \tau - \alpha_2 \taub = 0$, which yields
\eqn{\alpha_1 = \frac{ \pi}{\tau}, \quad \alpha_2 = \frac{\pi}{\taub} . \la{eq:pinch}}
This has a pleasing geometric interpretation for real $\tau, \taub$. As we have noted, $\alpha_i = \pi/(\betae)_i$ where $\betae$ is the effective inverse temperature  (e.g. correlators on the same side of the quench agree with thermal correlators at a temperature $\beta_E\inv$.) The solution tells us to set $(\betae)_1 = \tau$ and $(\betae) = \taub$. This means the geometry has the form of two circles which are completely ``pinched.'' Indeed, if we plug in this solution, we can compute the correlator \eqref{eqn:gtop} at $t_1 = 0$ and $t_2 = \tau$. We see that $e^g= 1$, its maximal value. From the Schwarzian point of view, the domain wall is so massive that it pinches the disk together. {This is depicted in Figure \ref{fig:infbeta}.}

\begin{figure}[H]
\begin{center}
\includegraphics[width=0.7 \columnwidth]{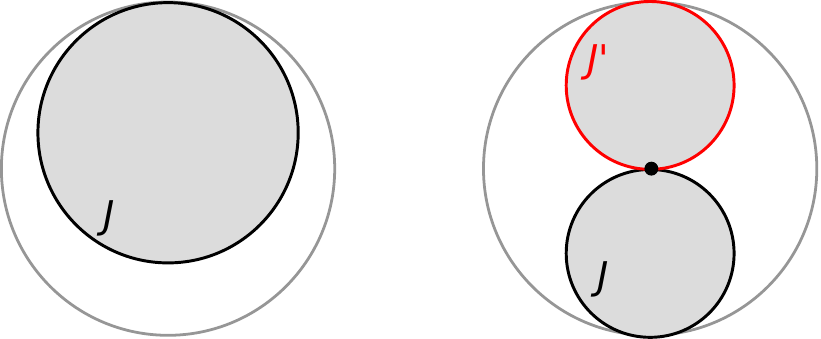}
\caption{ Left: the ordinary extremal or zero temperature solution. Since the boundary particle intercepts the asymptotic boundary, the total proper time is infinite. Right: if we take $\beta = \infty$ while holding the mass of the brane constant, the solution pinches. This means that the correlator between the quench sites is getting large. In the large $q$ theory, we showed that the correlator in fact becomes maximal $e^{g} = 1$. \la{fig:infbeta} }  
\end{center}
\end{figure}

This should be contrasted with the case when $\Delta \beta \ll 1$ in which case $\alpha_i = 2 \pi/\beta$. So for $\tau =\taub$ large but general $\lambda$, we expect that $\pi /(\beta \cj) \le \gamma \le 2\pi /(\beta \cj)$. These bounds are shown in figure \ref{fig:gamma}. 

\begin{figure}
\begin{center}
\includegraphics[scale=.6]{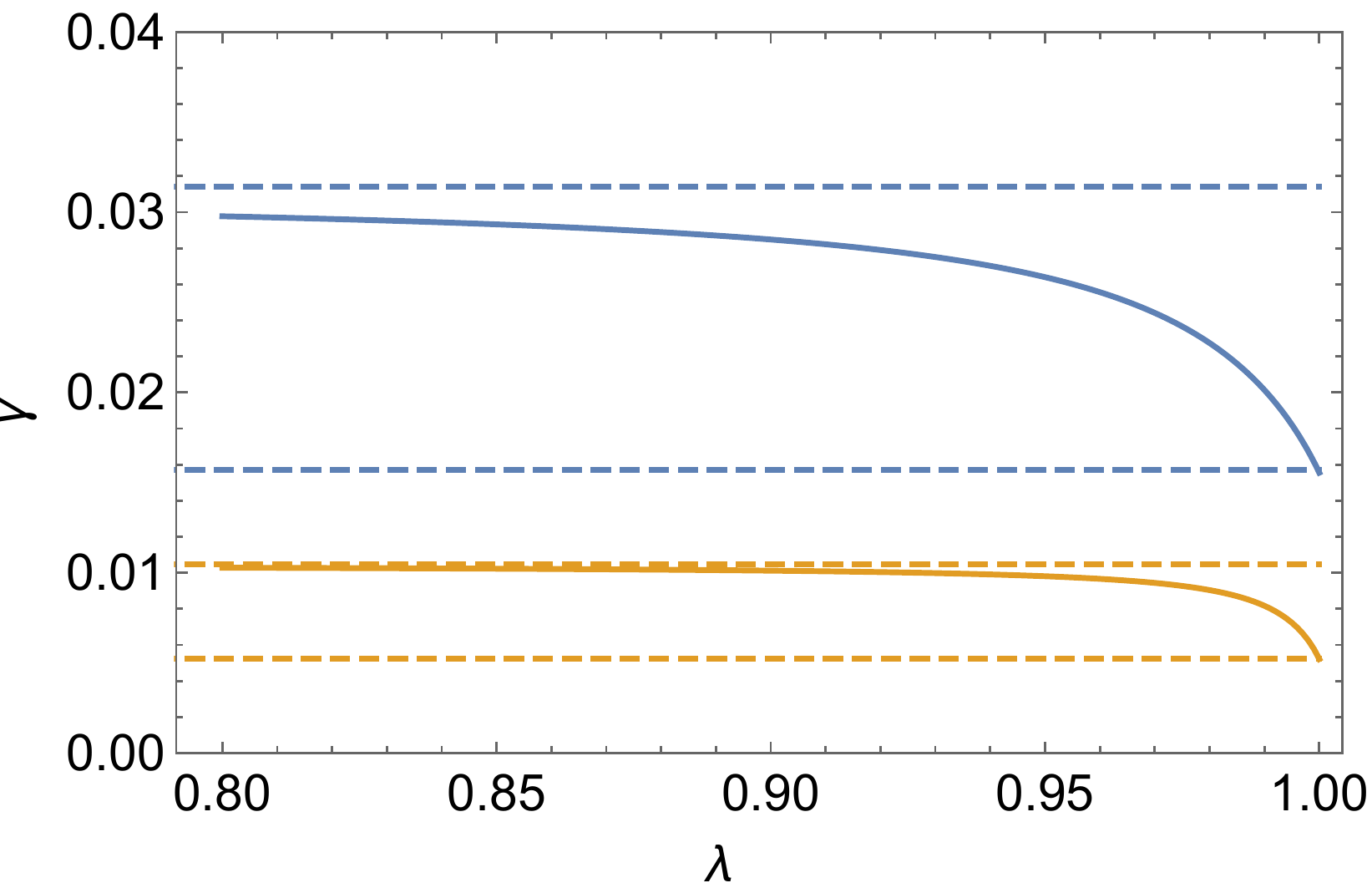}
\caption{ Here we show $\gamma$ as a function of $\lambda$ for $\beta\cj = 200$ and $\beta \cj = 800$ in blue and orange respectively. We also show in dashed lines the upper and lower bounds on $\gamma$ obtained analytically (in a $1/\beta \cj$ expansion.) Note that for larger $\beta \cj$, $\gamma$ quickly obtains its maximal value $ 2\pi/\beta$ away from $\lambda  = 1$. \la{fig:gamma} } 
\end{center}
\end{figure}

\begin{figure}
\begin{center}
\includegraphics[width=\columnwidth]{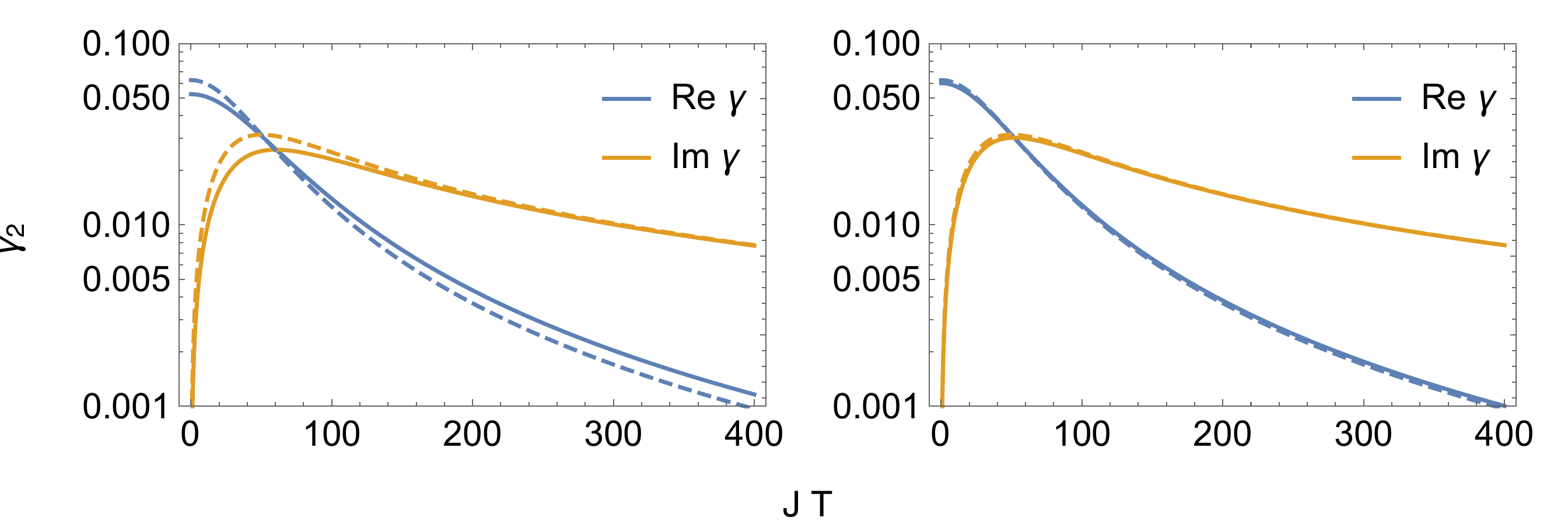}
\caption{ $\gamma_2 = \gamma_1^*$ as a function of $\cj T$ for $\beta\cj = 100$. On the left, $\lambda = 0.9$, whereas on the right $\lambda = 0$. Solid lines are obtained by numerically solving equation \eqref{streich}. The dashed lines are the ``pinching'' approximation derived in \eqref{eq:pinch}. At large $\beta \cj$, pinching is a reasonable approximation even when $1-\lambda^2$ is not small (left). \la{fig:gamma2} } 
\end{center}
\end{figure}

We can use our results for $\gamma_1 = \gamma_2^*$ to compute the 2-pt correlator $e^{g(\beta, \tau)}$ as a function of $\tau$, see figure \ref{fig:correlator}. This will be needed for the evaluation of the on-shell action for the Renyi-2 entropy in the next subsection.

\begin{figure}
\begin{center}
\includegraphics[scale=.6]{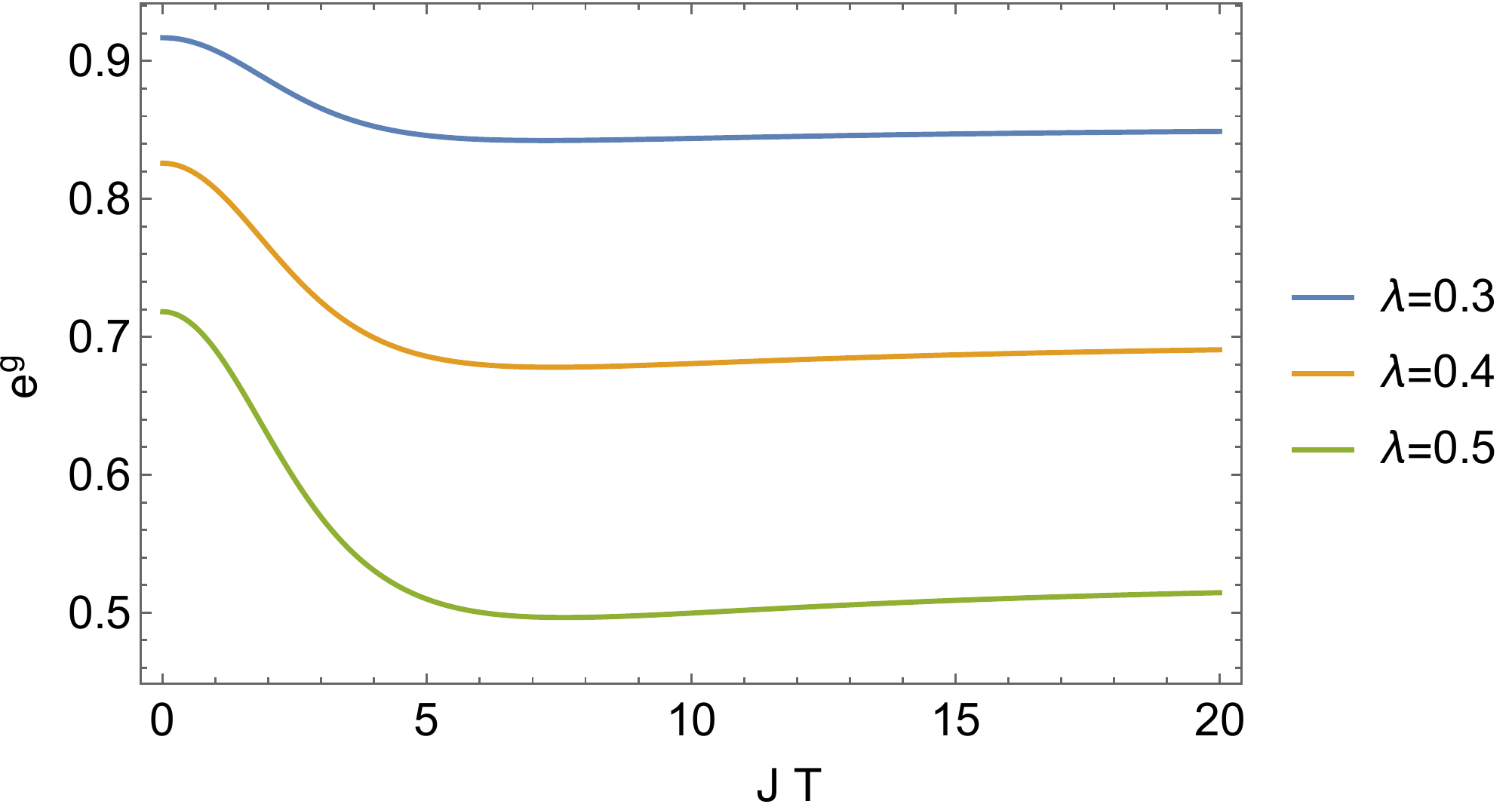}
\caption{ Here we show the disk contribution to the 2-pt correlator $e^{g(t_1,t_2)}$ evaluated at the quench sites ($t_1 = 0$, $t_2 = \beta/2+iT$) for $\lambda = 0.3,0.4,0.5$, and $\beta\cj=5$. Note that at small values of $\lambda$ the correlator stays large, which indicates that the distance between the quench sites is staying relatively small, even at large Lorentzian times. This is qualitatively similar to the Schwarzian behavior where the geometry gets pinched by the brane. \la{fig:correlator} } 
\end{center}
\end{figure}

\subsubsection{Evaluation of the action}
Now we would like to evaluate the on-shell action of the disk. We start with the large-$q$ Liouville action \cite{Cotler:2016fpe}:
\eqn{ S=  {N \over 8q^2} \int dt_1 dt_2 \;  \pd_1 g \pd_2 g - 4 \cj^2 e^{-\mu \Omega}  e^g   }
Then in terms of $\gh$ \eqref{hatg}:
\eqn{ S =  {N \over 8q^2} \int dt_1 dt_2 \;  \pd_1 \hat{g} \pd_2 \hat{g} + \mu^2 \pd_1 \Omega \pd_2 \Omega - \mu \pd_1 \hat{g} \pd_2 \Omega - \mu \pd_1 \Omega \pd_2 \hat{g} - 4 \cj^2  e^{\hat{g}} }
Let us consider $\pd S/\pd \mu$. Since we are evaluating the action on-shell, only the explicit $\mu$ dependence contributes:
    \eqn{ -\pd S/\pd \mu &=  {N \over 8q^2} \int  \pd_1 \hat{g} \pd_2 \Omega + \pd_2 \hat{g} \pd_1 \Omega \\
			     &= -{N \over 8q^2} \int dt_1 \pd_1 \gh (t_1, \tau) \sgn(t_1-\tau) + \int dt_2 \pd_2 \gh(\tau,t_2) \sgn(t_2-\tau)\\
			     &={N \over 2q^2} \lb  \gh(\tau,\tau) - \gh(\beta, \tau)  \rb
		     }
So to compute the action, we may simply integrate the correlator as a function of $\lambda$
\def\lt{\tilde{\lambda}}
\eqn{I(\tau, \taub) = -{N \over  q^2} \int_{\lt = \lambda}^1 2 \lt^3 d\lt   \log \lb  \frac{\alpha_1(\lt) }{\sin (\alpha_1(\lt) \tau + \gamma_1(\lt) )} \rb }
In this expression, we have subtracted the action at $\lambda = 1$, which is required when we normalize the thermofield double.
This expression can be computed numerically. We display the result in Figure \ref{fig:action}. 
We see that the answer grows at late times $T$. This suggests that the disk contribution alone will lead to a violation of unitarity at late times. (It does not prove that there is a violation since in principle quantum corrections could in principle stop the growth, but based on both the quantum Schwarzian computations and the finite $N$ numerics, this seems unlikely.)

\begin{figure}
\begin{center}
\includegraphics[scale=.6]{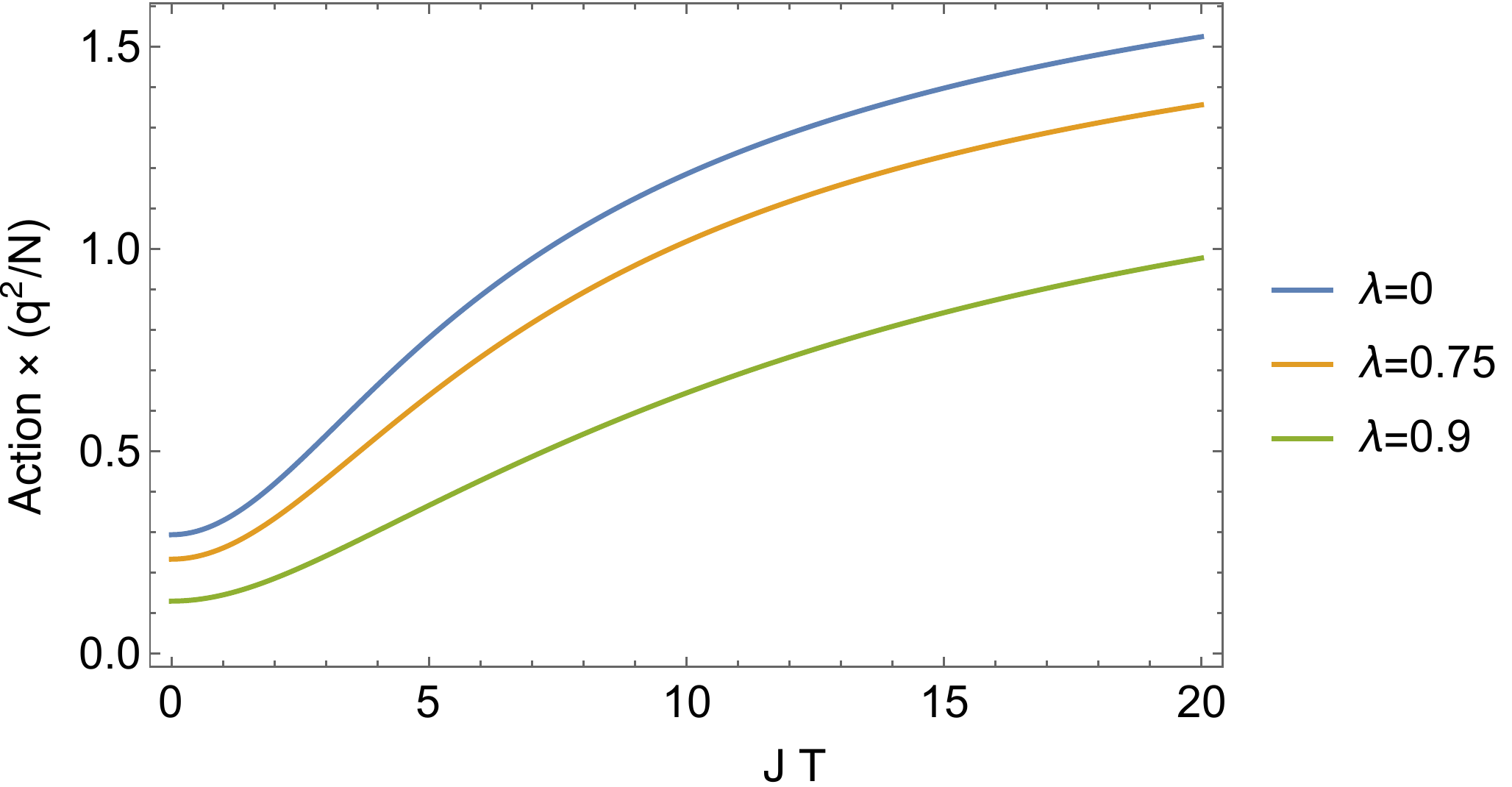}
\caption{ The large $N$ on-shell action of the disk as a function of $\cj T$ for $\beta \cj = 5$ for various values of $\lambda$. The growth of the action implies that the typical overlap $\braket{\tfd,J}{\tfd,J'} \sim e^{-I}$ is shrinking as a function of time. Smaller values of $\lambda$ lead to more decorrelated states, which agrees with the larger values of the action. The growth at late times suggests that there is a unitarity problem if we only include the disk solution.  \la{fig:action} } 
\end{center}
\end{figure}

\pagebreak

\subsection{Wormhole in large q SYK}
\def\lr{{\mathsf{LR}}}
\def\lll{{\mathsf{LL}}}
\def\rr{{\mathsf{RR}}}
\def\rl{{\mathsf{RL}}}
\def\glr{g_{\lr} }
\def\gll{g_{\lll} }
\def\betax{\beta_\mathrm{aux}}

\begin{figure}[H]
\begin{center}
\includegraphics[width=1\columnwidth]{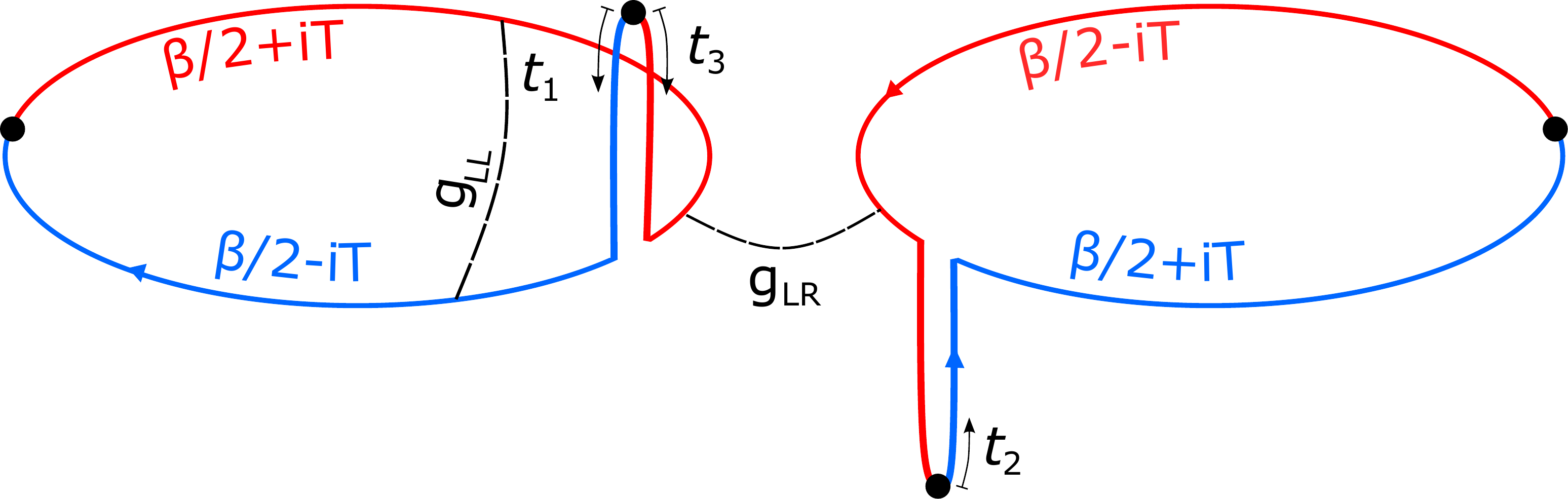}
\caption{ Contour relevant for the Renyi-2 wormhole, with our conventions labeled. The black arrows start at the points where $t_i = 0$ and they point in the direction of positive $t_i$. The correlator between points $t_1$ and $t_2$ behaves as if they are on two sides of a thermofield double.  \la{figc2} }  
\end{center}
\end{figure}

Here we present a preliminary exploration of the wormhole in large $q$ SYK, leaving a more thorough analysis for the future. When we introduce a second side, the large $N$ variables include a $g_\lr$. Remarkably in the standard large $q$ approximation, the $g_\lll$ and $g_\lr$ variables are uncoupled in the equations of motion:
\eqn{
	\pd_1 \pd_2 \gll =- 2 \cj^2 \exp \lp {\gll- \hm\Omega_\lll } \rp , \quad 	\pd_1 \pd_2 \glr = 2 \cj^2 \exp \lp {\glr - \hm\Omega_\lr } \rp. \la{eq:eom}}
Furthermore, the equations of motion just differ by a sign, which can be accounted for by changing the direction of time.
The basic idea is that using the same tricks as above, we will have a $g_{LR}$ that obeys a Liouville like equation. So we expect a solution that is very similar to the Schwarzian wormhole, except that we ``build'' the solution from the large $q$ disk solution. More explicitly, consider an ansatz: 
\eqn{\exp\lp {g_\toc^\lr(u_1,u_2)} \rp  &= \lb {\alpha \over \cj \sin \left[  \alpha \lp u_1 -\tilde{u}_2\rp  + \gamma \right]} \rb^2, \quad \tilde{u}_2 = u_2 - u_b \\
\exp \lp {g_\otoc^\lr(u_3,u_2) }\rp       &= 
	\frac{\alpha^2}{\cj^2 }\left[  { \sin \lp  \alpha \lp u_3-\tilde{u}_2 \rp  + \gamma \rp }   -\frac{ \cj  \left(1 - \lambda^{2} \right) }{\alpha } \sin \left(\alpha u_3  \right) \sin \left(\alpha \tilde{u}_2\right)\right]^{-2 } \la{eq:glr}\\
\exp \lp {g_\otoc^\lll(u_1,u_3) }\rp       &= 
	\frac{\alpha^2}{\cj^2 }\left[  { \sin \lp  \alpha \lp u_1-u_3\rp  + \gamma \rp }   -\frac{ \cj  \left(1 - \lambda^{2} \right) }{\alpha } \sin \left(\alpha u_1 \right) \sin \left(\alpha u_3\right)\right]^{-2 }\\
\exp \lp {g_\toc^\lll(u_1,u_1') }\rp       &= \lb {\alpha \over \cj \sin \left[  \alpha \lp u_1 -u_1' \rp  + \gamma \right]} \rb^2.
}
We are adopting a convention where $u_i = 0$ corresponds to a quench site, and the second quench site on the left (right) side is at $u_1 = \tau$  ($u_2 = \tau$).
Furthermore, $u_1, u_3$ runs clockwise on the LHS, whereas $u_2$ runs counterclockwise on the RHS. So for example, the solution $g_\toc^\lr$ is valid when $u_1 u_2 <0$ before $u_1$ or $u_2$ cross another quench site. This accounts for the sign difference in the equations of motion of \eqref{eq:eom}.
To analytically continue, we set $u_i = it_i$.
This gives a sign convention $t_i$ illustrated in Figure \ref{figc2}.
One can confirm that the equations of motion are satisfied if $\alpha = \sin \gamma$. The remaining task is to determine $\alpha$ and the time shift $u_b$.


As a warm-up, it is useful to reconsider the Schwarzian wormhole from a slightly different perspective. There, too, the equations of motion are locally the same as for the disk solution. The main idea is that we can build the wormhole by considering the disk solution with some auxiliary temperature $\betax$. 
The disk solution in the Schwarzian limit is obtained by taking two disks of circumference $\beta_E$ and joining them together at the quench sites. 
The wormhole is constructed by cutting along the diameter of each partial disk and then gluing to another copy of the solution, see figure \ref{fig:stanfordworm} and Appendix \ref{app:r2wormhole}. Pasting together the two solutions gives a new solution that is a topological cylinder.

Now before cutting, the total length of the doubled disk solution is $2\betax$. When we cut along the diameter of the partial disks and join the two copies, we remove a portion of the boundary that has length $4 \times (\betae/2)$.  Then requiring that the total length of the boundary is $2\beta$ (a factor of $\beta$ for each side) gives
\eqn{ 2\betax - 2\betae = 2\beta\la{eq:trivial} } 
Now for the disk solution, we have from \eqref{eq:tanSch}
\eqn{ -\tan \lp \frac{\betax \pi}{2 \betae}\rp  = \frac{4\pi C}{\delta \betae } .}
Eliminating $\betax$ reproduces \eqref{eq:tanb}, as desired:
\eqn{\tan \left(\frac{\pi \beta}{2 \beta_{E}}\right)=\frac{\delta \beta_{E}}{4 \pi C} \approx \frac{(1-\lambda^2) \beta_E \cj}{2\pi} .\la{eq:tanb2} }
The last equality is based on the approximate relation derived in the previous subsection, around \eqref{eq:tana}.

Now we can follow the same strategy to find $\beta_E$ as a function of $\beta$ in the large $q$ solution. The key point is that although we do not have a clear geometric picture like in the Schwarzian case, nevertheless the $\toc$ correlator is exactly thermal at an effective temperature $1/\betae$, so the procedure is entirely analogous. In particular, just as we cut $\betae/2$ of the boundary particle on the partial disk, we remove $\betae/2$ of the $\toc$ correlator, and paste it to a second copy. If we enforce equation \eqref{eq:trivial}, this pasting procedure gives a smooth correlator.
Indeed, the special thing about removing exactly $\betae/2$ of the solution is that the $\toc$ 2-pt function is minimized at this time. So when we continue the 2-pt function $g_{\lr}$ onto the ``back side'' of the solution, there will be no discontinuity in any of its derivatives. 

Now while \eqref{eq:trivial} holds for both the Schwarzian and the large $q$ theory, the relation between $\betae$ and $\beta$ will be modified. For the $\toc$ correlator with any $\alpha$, we can define an effective inverse temperature $\beta_E$:
\eqn{\frac{\alpha}{\cj} =  \cos \lp \alpha \beta_E \over 2 \rp.}
In general, $\betae$ differs from $\betax$ due to the added energy from the insertion of the operator.
Together with the relation between $\betax$ and $\alpha$ derived in the previous subsection
\eqn{\sin \lp \alpha \betax/2 + 2\gamma\rp  = \lambda^2 \sin \lp \alpha \betax/2\rp,  \alpha/\cj = \sin \alpha}
and \eqref{eq:trivial}, we can eliminate $\betax$ and obtain $\beta_E$ as a function of $\beta$ for the wormhole solution. We show the results for in \ref{fig:betabeta}.

This solution for $\alpha(\betae)$ can be substituted into \eqref{eq:glr} along with
\eqn{
    u_b = \betax/2 = (\beta+\betae)/2.}
The last requirement just says that in the auxiliary disk solution, the second quench site appears after a boundary time $\betax/2$. 
We can check that with the above requirements, all the $\toc$ correlators obey the correct boundary conditions, e.g. $\glr(0,0) = \glr(\tau,\taub), \glr(0,\taub) = \glr(\tau,0)$. 

We expect that the wormhole solution discussed above to be an approximate solution to the large $N$ equations of motion when the Lorentzian time $T$ is large. 
To borrow the Schwarzian language, the reason why we need this condition is that we have essentially ignored ``windings.'' In the Schwarzian description, there are multiple geodesics that connect two points; in general, a correlator will receive contributions from all of these geodesics unless the wormhole throat $b$ is very long (which happens at large $T$, see Appendix \ref{app:matter}).
To state the issue in the large $q$ formalism, note that we have described the solution in patches \eqref{eq:glr}, but actually the patches overlap. While the $\toc$ solutions are valid whenever the two times are on the same side of a quench, the $\otoc$ solutions are supposed to be valid ``near'' the quench site at $u_1 = u_3 = 0$. When the $u_i$ are near the opposite quench site, we should use a similar ansatz:

\eqn{\exp \lp {g_\otoc^\lll(u_1,u_3) }\rp       &= 
	\frac{\alpha^2}{\cj^2 }\Bigg[  { \sin \lp  \alpha \lp u_1-u_3\rp  + \gamma \rp }  
	-\frac{ \cj  \left(1 - \lambda^{2} \right) }{\alpha } \sin \left(\alpha (u_1-\tau) \right) \sin \left(\alpha (u_3-\tau)\right)\Bigg]^{-2 }\\
	\exp \lp {g_\otoc^\lr(u_3,u_2) }\rp       &= 
	\frac{\alpha^2}{\cj^2 }\Bigg[  { \sin \lp  \alpha \lp u_3-\tilde{u}_2 \rp  + \gamma \rp } 
	-\frac{ \cj  \left(1 - \lambda^{2} \right) }{\alpha } \sin \left(\alpha (u_3-\tau)  \right) \sin \left(\alpha (\tilde{u}_2-\tau) \right)\Bigg]^{-2 } \la{eq:glr2}\\
	}
We believe that the full solution to the large $N$ equations of motion is close to a sum of the two ansatzs the large $\alpha T$ limit, with a sign appropriate for the periodicity/anti-periodicity of the correlators. Summing the two ansatz only makes sense if the correlators are small in the overlap region. For fixed $\beta \cj, \lambda<1$ this is satisfied if $\alpha T \gg 1$, since the correlators decay exponentially in the overlapping region $e^g \sim \alpha^2 e^{-2\alpha T}$. A similar issue arises in finding the double cone in SYK \cite{Saad:2018bqo} and also for the finite temperature wormhole in the coupled SYK model see also \cite{Maldacena:2018lmt}. To check this more carefully, one would need to use a different large $q$ approximation, discussed in \cite{Maldacena:2018lmt} that is valid when the correlators $G_\lll$ are small (and therefore $|g_\lll/q| \gg 1$.) It would also be interesting to solve the finite $q$ $G,\Sigma$ equations numerically.



A regime where the solution simplifies is when $\beta \cj  \gg 1$ while keeping $\lambda<1$ fixed. The wormhole correlators become
\eqn{
\exp\lp  g_\otoc^\lll(u_1,u_3)  \rp              &= {1 \over \lp -(1-\lambda^2) \cj^2 u_1 u_3 + \cj u_{13} + 1\rp^2} \\
  \exp \lp {g_\toc^\lr(u_1,u_2) }\rp    &= {1 \over \lp \cj (u_1 - u_2) + 1 \rp^2} \\
        \exp \lp {g_\otoc^\lr(u_3,u_2) }\rp  &= {1 \over \lp -(1-\lambda^2) \cj^2 u_3 u_2 + \cj u_{32} + 1\rp^2} \la{eq:syklowtemp}
}
Notice that the correlator across the wormhole is maximal at $u_1 = u_2 = 0$! This means that the wormhole is basically what is depicted in the right side of Figure \ref{fig:infbeta}. Since both circles intersect the asymptotic boundary, we can interpret the figure as either having a single boundary, or two boundaries that are infinitely long, where the two boundaries are separated at asymptotic infinity.
This wormhole solution can be contrasted with the disk solution in the same limit.
At large $(1-\lambda)^2 \beta \gg 1$, the wormhole relation is $\betae \approx \beta$. On the other hand, the disk in this limit gives $\betae \approx \beta/2$. This means that for the disk, the 1-sided correlator $G_\lll(0,\beta/2+iT)  \approx 1$ is large whereas $G_\lr =0$. But for the wormhole $G_\lll(0,\beta/2+iT)$ is small, but the two-sided correlator $G_\lr(0,0) \approx 1$. Such qualitative behavior of the correlators seems to be roughly in agreement with the $q=4, N=20$ numerics, see Figure \ref{fig:l0c}. 

\begin{figure}
\begin{center}
\includegraphics[scale=.6]{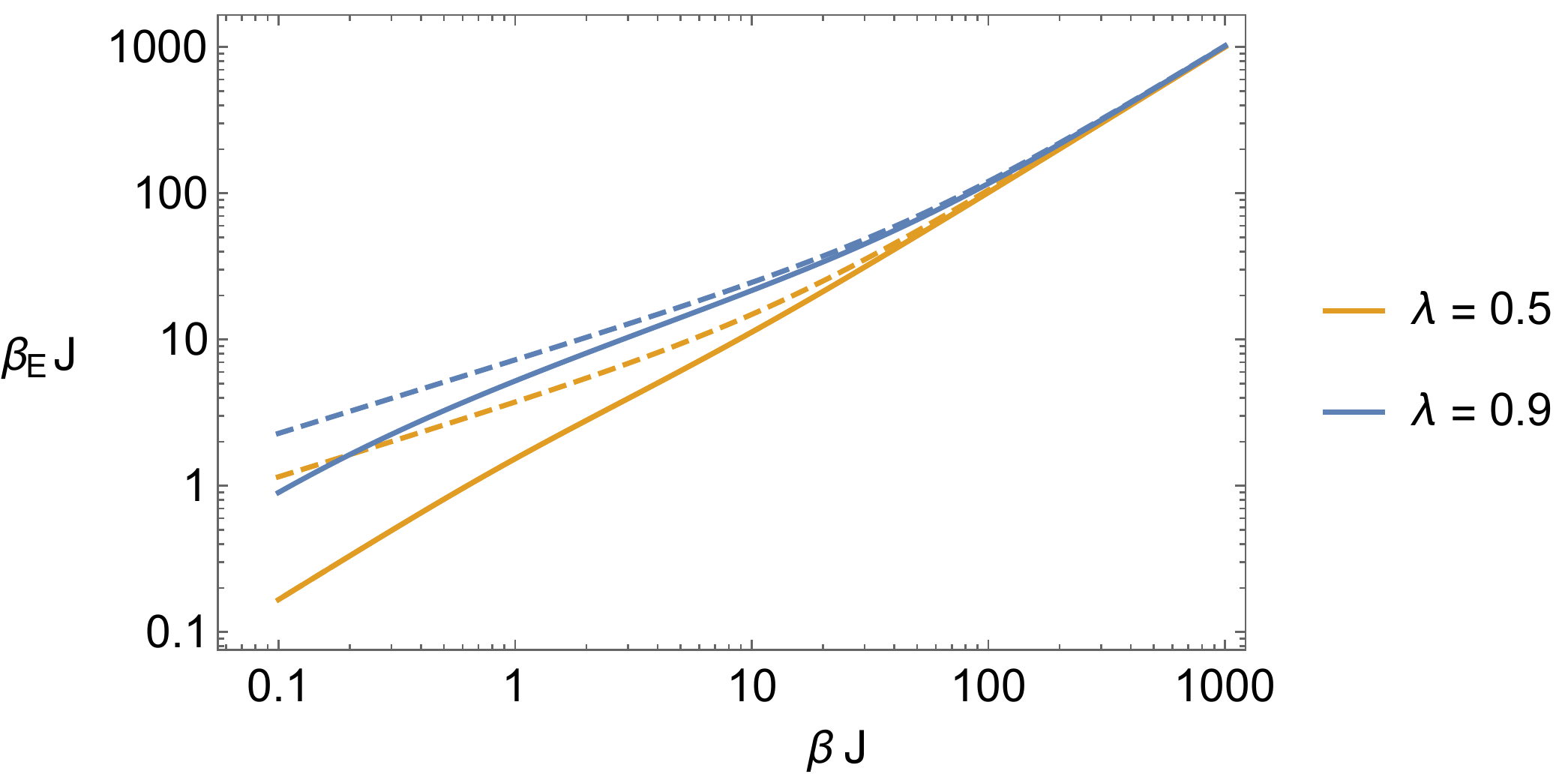}
\caption{ The effective inverse temperature $\beta_E$ as a function of $\beta$ in units of $\cj$. We show the results for $\lambda = 0.9$ and $\lambda=0.5$.  The dashed lines are the Schwarzian predictions given by \eqref{eq:tanb2}. At very large $\beta\cj$, the growth is approximately linear $\betae \approx \beta$.  \la{fig:betabeta} } 
\end{center}
\end{figure}

Note that our setup has various exact discrete global symmetries, which implies that there are multiple wormhole solutions.
These symmetries  are the same as the ones for the ``ramp'' in SYK \cite{Saad:2018bqo}, so we will be brief.
First note that $(-1)^F_L$ and $(-1)^F_R$ generate a $\mathbb{Z}_2 \times \mathbb{Z}_2 $ global symmetry, 
that leaves $G_\lll$ and $G_\rr$ invariant but takes $G_\lr \to -G_\lr$ when only one $(-1)^F$ is applied.
Thus any wormhole solution $G_{LR} \ne 0$ spontaneously breaks one of the two $\mathbb{Z}_2$ symmetries; further, the solutions come in pairs. For the large $q$ solutions, we should write $G_\lr = \pm e^{g_\lr/q}$ 
When $q$ is a multiple of 4, there is the additional symmetry $G_\lr\left(t, t^{\prime}\right) \rightarrow i G_\lr\left(t,-t^{\prime}\right), \quad G_\rl\left(t, t^{\prime}\right) \rightarrow i G_\rl\left(-t, t^{\prime}\right), \quad G_\rr\left(t, t^{\prime}\right) \rightarrow-G_\rr\left(-t,-t^{\prime}\right)$ that should be treated similarly.

Another setup in which one can study the wormhole of the journal analytically is Brownian SYK. There, the journal contains a whole function's worth of couplings $\{ J_{ijkl}(t) \}$. See Appendix \ref{app:brown}.
\pagebreak

\section{Reconstruction with erroneous knowledge of the couplings}

The goal of this section is to illustrate how bulk reconstruction using the boundary system is affected by erroneous or incomplete knowledge of the couplings. The discussion will be mostly  qualitative. We will keep things simple and use the Petz Lite protocol of \cite{westcoast}.

In the simplest setting of bulk reconstruction, one considers a ``code subspace'' of the boundary Hilbert space that share the same bulk background geometry but differ in the state of the matter. In our case, we consider a code subspace composed of perturbations of the time evolved thermofield double state $| \beta + i T; J \rangle$, where $J$ labels the couplings of the Hamiltonian $H_J$ used to prepare the state. We use $|i ; J \rangle$ to denote the different matter states spanning the code subspace. If we like, we can view $i$ as some species index for a particle in AdS. Note that we are not yet averaging over couplings.

We will take implementing successful bulk reconstruction to mean that there's an operator that can transition between any two states of the code subspace; for $i,j,k,l$ in some orthonormal basis, there exists an operator ${\cal O}_{ji}$ such that 
\begin{align}
    \langle k; J | {\cal O}_{ji} |l; J \rangle  \approx \delta_{kj}\delta_{li}.  \label{reccond}
\end{align}
We use the approximate signs $\approx$ to indicate that we are ignoring non-perturbatively small effects from non-factorization wormholes, and also to account for the approximation of using Petz lite. This condition is easily satisfied by constructing the ``global'' $J$ operator, 
\begin{align}
    {\cal O}^J_{ji} \equiv   |j; J \rangle\langle i; J |,
\end{align}
constructed out of the basis states of the code states. This gives the square of the norm of the states:
\begin{align}
     \langle k; J | {\cal O}^J_{ji}|l; J \rangle  =  \eqfig{0.12\columnwidth}{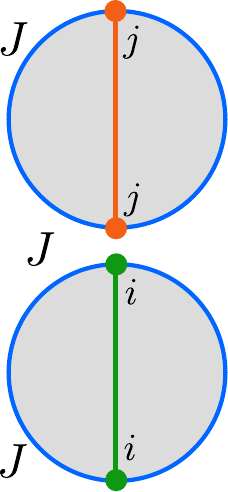} \times \delta_{jk}\delta_{li}
\end{align}
In this equation, the orange and green colors indicate the particle species (or equivalently the index $i$ or $j$).
The setting we want to consider is where we don't know the value of the coupling $J$ in the Hamiltonian that's used in preparing the code subspace, and we are tasked with finding an operator that transitions between two specified code words $i$ and $j$. This is risky because using a wrong value of the coupling, say $J'$ instead of $J$, leads to an eventual breakdown of the reconstruction:
\begin{align}
    \langle j; J | {\cal O}^{J'}_{ji} |i; J \rangle =  \langle j; J |j; J' \rangle \langle i; J' |i; J \rangle = \eqfig{0.14\columnwidth}{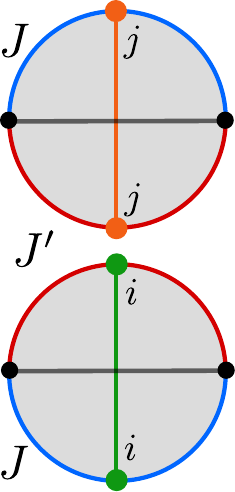}
\end{align}
For $J \neq J'$, the right hand side decays in time due to the backreaction of the shocks (depicted in black) created by the sudden change in the boundary conditions. We showed that in the case of JT + free boson BCFT model of section \ref{sec:renyi}, that this decays exponentially
\begin{align}
    \langle j; J | {\cal O}^{J'}_{ji} |i; J \rangle \approx \left[\pi \epsilon \over \beta \cosh(\pi T/\beta)\right]^{(J-J')^2 \over 2 \alpha' \pi^2 }.
\end{align}
(We are ignoring the gravitational backreaction of the orange and green particles and quantum Schwarzian effects, hence the $\approx$ sign.) The reconstruction therefore fails at late times, since the states prepared with different couplings become more orthogonal under time evolution. This is the same effect that resulted in a unitarity problem, which hinted at the presence of an island.

If we are given the knowledge of the distribution of the original couplings, then we can do better. A natural procedure is to average the couplings in the operator over this distribution. We can estimate the failure on average by also averaging over the couplings of the system:
\begin{align}
    \int dJ dJ' P(J) P(J') \langle j; J | {\cal O}^{J'}_{ji} |i; J \rangle &\approx {1\over  1 + {2 \over \pi^2 \alpha' m^2} \ln\left[ {\beta \over \pi \epsilon} \cosh(\pi T/\beta)\right]} \approx { \beta \pi \alpha' m^2 \over  {2  T } } \label{avgPetz}
\end{align}
At first sight it appears that the reconstruction fails, even on average, albeit more slowly than if we don't average. However, we note that this decay only depends on the statistical properties of the couplings, and also weakly on $i, j$ (not shown here). This means we can improve our reconstruction by simply scaling the operator by a time dependent factor $C_{ij, m}(T)$. We will take this to mean that the reconstruction is successful, at least on average.

However, there's another, more severe way in which this operator can fail: the sum over couplings enhances the contribution of a wormhole that connects the bra/ket of the operator/state its ket/bra. This in particular means that ``flipped'' matrix elements which should vanish actually get a contribution:
\begin{align}
     \int dJ dJ' P(J) P(J') \langle i; J | {\cal O}^{J'}_{ji} |j; J \rangle \approx \eqfig{0.12\columnwidth}{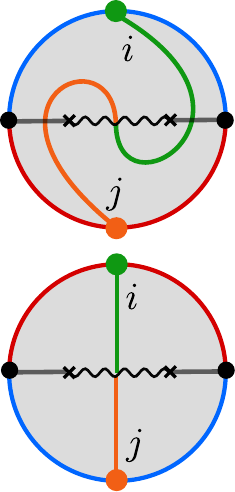} = \ \ \eqfig{0.25\columnwidth}{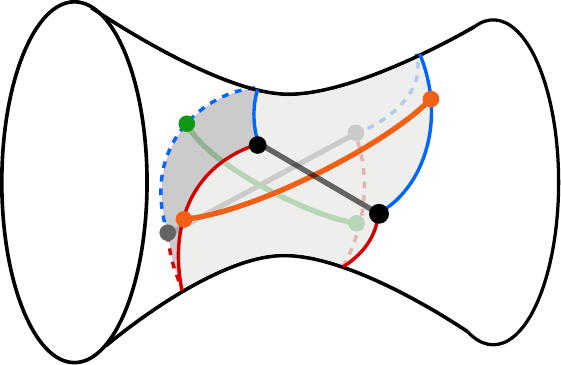}
\end{align}
Averaging over the same value of the boundary coupling is indicated by like colors. At early times, this is a small contribution that only mildly violates our condition \eqref{reccond}, but at large times, when the disk is highly suppressed, it is comparable to the ``good'' matrix elements  $\langle j; J | {\cal O}^{J'}_{ji} |i; J \rangle$ that come from the disk (which are decaying). So, if we rescale the operator to set the ``good'' elements $\approx 1$, we will also rescale these ``flipped'' elements $\langle j; J | {\cal O}^{J'}_{ji} |i; J \rangle \sim 1$. Thus we see that this reconstruction attempt fails badly at late times.

Furthermore, even when the disk contribution to the ``good'' matrix elements is small, there is not a substantial contribution from the wormhole since the bulk states in the throat are orthogonal, and winding contributions decay exponentially in $b \sim T$. 
One can see graphically that the overlap is zero by considering
\begin{align}
     \int dJ dJ' P(J) P(J') &\langle j; J | {\cal O}^{J'}_{ji} |i; J \rangle \approx \eqfig{0.12\columnwidth}{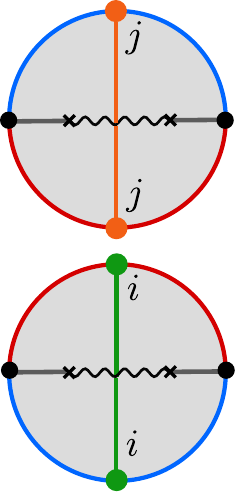}  &= \eqfig{0.25\columnwidth}{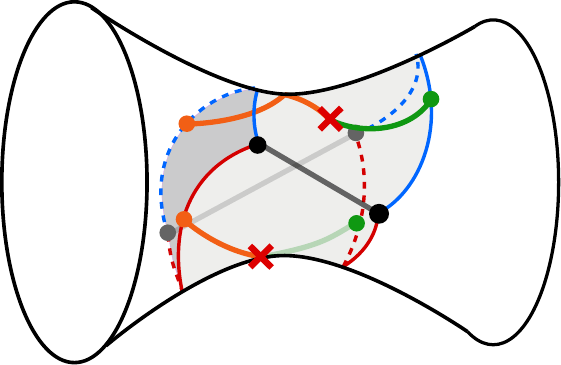} = 0.
\end{align}
The red crosses above indicate the orthogonality of the bulk states. A similar use for wormholes was first pointed out in \cite{westcoast}. However, failure is not guaranteed because the particles can pair up outside the branch cut:
\begin{align}
   \int dJ dJ' P(J) P(J') \langle j; J | {\cal O}^{J'}_{ji} |i; J \rangle \approx \eqfig{0.14\columnwidth}{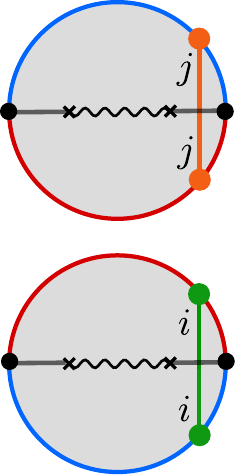}
\end{align}
For such operators outside the horizon, the wormhole and the disk contribution give similar answers, so there are no significant contributions to the ``flipped'' elements and rescaling the operator with time should succeed.




The right interpretation of this is that the entanglement wedge of the boundary develops a blind spot in the bulk; the reconstruction fails because the particle falls into the island. This island is the entanglement wedge of reference purifying the system, namely what we've been calling the journal in the previous sections. With this interpretation, the averaged operator used in \eqref{avgPetz} is nothing but the Petz lite operator, obtained by tracing out the journal,
\begin{align}
    O_{ji}^\mathsf{Petz \, Lite} &= \Tr_\mathrm{journal} \left[ \int \, dJ \sqrt{P(J) P(J')} |j; J \rangle \langle i; J' |_\mathsf{Sys} \otimes | J \rangle \langle J' |_\mathsf{journal} \right] \\
    &= \int \, dJ P(J) \, |j; J \rangle \langle i; J |_\mathsf{Sys}
\end{align}

We can adapt the above discussion to the case of having partial knowledge by defining a reconstruction controlled by the result of measuring the couplings. This is just a controlled Petz Lite,
\begin{align}
    {\cal O}_{ji} = \sum_{\bkap} \underbrace{\sum_{\bmu} P(\bmu, \bkap)\ketbra{j; \bkap, \bmu}{i; \bkap, \bmu}_\sys}_{ O_{ji}^\mathsf{Petz \, Lite}(\bkap)} \otimes \ketbra{\bkap}{\bkap}_\kno 
\end{align}
We note that this can be generalized to the case of full Petz with the replacement
\begin{align}
    O_{ji}^\mathsf{Petz \, Lite}(\bkap) \rightarrow \left[\Pi^\mathsf{code}_\mathsf{Sys}(\bkap)\right]^{-{1\over 2}} O_{ji}^\mathsf{Petz \, Lite}(\bkap) \left[\Pi_\mathsf{Sys}^\mathsf{code}(\bkap)\right]^{-{1\over 2} }
\end{align}
where
\begin{align}
     \Pi^\mathsf{code}_\mathsf{Sys}(\bkap) \equiv \Tr_{\bmu, \bkap'} \left[ \Pi_{\bkap'} \Pi^\mathsf{code} \Pi_{\bkap'} \right]
\end{align}
is the trace over the couplings of the projector onto the code subspace when projected onto a given value of the known couplings.



It is interesting to speculate how the failure of reconstruction due to islands manifests itself in a theory with fixed couplings. Given that it arises from the emergence of an island and wormholes (as a result of averaging) it suggests that something like half wormholes \cite{Saad:2021rcu} are important for reconstructing the black hole interior. This suggests that we need to go beyond semi-classical gravity to address the firewall question \cite{Almheiri:2012rt,Almheiri:2013hfa}.

Besides the Petz map, one can also try to use simpler methods of bulk reconstruction. One option is to use the Maldacena-Qi Hamiltonian \cite{Maldacena:2018lmt} which directly couples the left and right sides\footnote{We thank Juan Maldacena for suggesting this direction.}. In \cite{Maldacena:2018lmt}, they showed that the wormhole does not require perfect correlation between couplings. This suggests that at early times, as long as the couplings are not too uncertain, evolution with the coupled Hamiltonian will still lead to an eternal traversable wormhole, which implies that the entanglement wedge of the two sides includes all of AdS. At late times, the wormhole is very long, so the correlations between the two sides is not strong enough to prevent the wormhole from growing. It would be interesting to understand this more quantitatively.

\pagebreak
\section{Discussion}

\begin{figure}[H]
\begin{center}
\includegraphics[width=1\columnwidth]{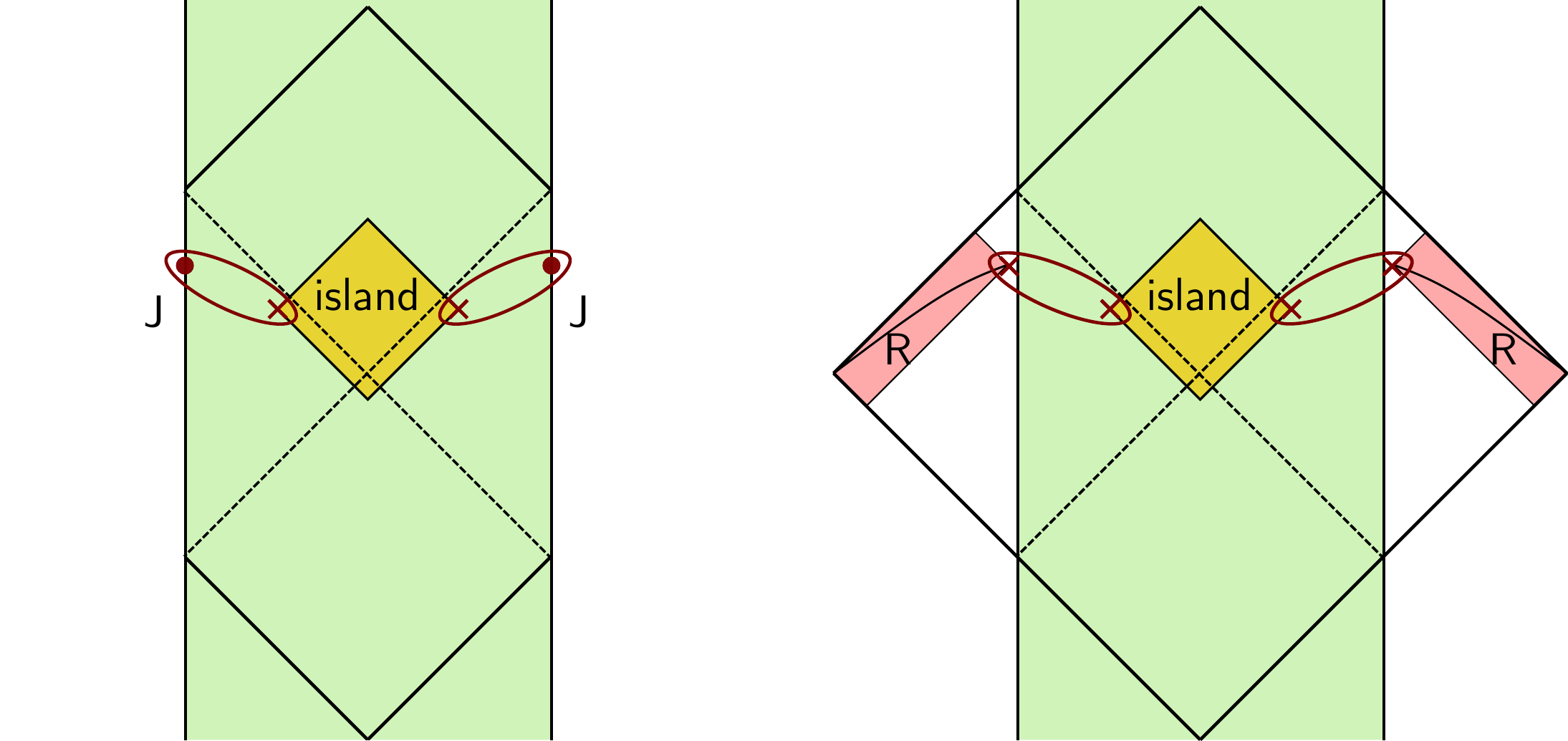}
\caption{ Island of the journal versus the island in the East coast setup. At late times, the separation between the twists from the island are nearly null separated from the boundary twists. In this OPE limit, one can approximate the entropy as the sum of two semi-infinite intervals discussed in section \ref{sec:peninsula}. \label{fig:EC}  } 
\end{center}
\end{figure}

\subsection{Bathing in the unknown \la{bathanalogy}}

In the introduction of this paper we motivated the entanglement entropy between the system and the journal as a measure of the uncertainty in the couplings, and we studied its  growth under controlled time evolution and its effects on the entanglement wedge of the system. 

Here we briefly motivate a different (but physically identical) picture of thinking of the journal as a bath with which the black hole system interacts. This makes the problem more analogous to the East Coast model \cite{eastcoast}, where the black hole interacts with a bath. Just as an example, we can consider the SYK system. The Hamiltonian \footnote{To reproduce a probability distribution over the couplings that is independent of $\beta$, we will generally need to include an explicit $\beta$ dependence in the Hamiltonian.} on the combined SYK and journal systems is
\eqn{H = \sum_{i < j < k < l } \hat{J}_{ijkl} \psi_i \psi_j \psi_k \psi_l + \hat{J}_{ijkl}^2 \frac{N^3}{6 J^2 }}

Here we are using the undergraduate quantum mechanics ``hat'' notation to emphasize that $\hat{J}$ is a quantum mechanical operator. The interaction is quite simple in a sense because there is no conjugate momentum to $\hat{J}$ in the Hamiltonian; however, it is sufficient to transfer quantum information to the journal. 
Then the formation of the island for the journal is qualitatively similar to the formation of the island for the radiation, see Figure \ref{fig:EC}.

\subsection{Evaporating BH $+$ journal}
It would be interesting to study how the state of the Hawking radiation of an evaporating black hole is affected by the uncertainty in the couplings. A concrete case would be to consider the ``East coast'' model where we have JT gravity $+$ CFT, joined to a flat space region where the CFT continues. In addition to $\cft_1$ in the bulk, we could also have a second $\cft_2$ that lives only in the black hole region. In other words, we have a boundary condition $J$ for $\cft_2$ that prevents any transmission into the flat space region. Furthermore, we assume that in the semi-classical picture there is no interaction between $\cft_1$ and $\cft_2$.

We get information about the state of the radiation by looking for the entanglement wedge of the radiation. In this setup there are two systems, the radiation and the journal, vying for ownership of the island. We can guess at the winner by comparing the rate of entropy growth of the two systems. The first to $~2 S_{BH}$ wins. We found in section \ref{trivialsurface} that the entropy of the journal grows only logarithmically, while it was shown in \cite{Almheiri:2019yqk} that the entropy of the radiation grows linearly time. Hence we believe that the radiation wins the race, at least at first.

An open question is whether ownership of the island is ever transferred to the journal. If it does get transferred, then one interpretation is that large uncertainty in the couplings reduces the system to the semi-classical state; this would support the recent ideas connecting premature ensemble averaging to Hawking's calculation \cite{Marolf:2020xie, Bousso:2019ykv, Bousso:2020kmy}. In the case that the island doesn't get transferred, then a possible interpretation is that the radiation alone purifies the journal. This translates physically to the statement that the couplings can be determined by the measurements on the Hawking radiation.




\subsection{Spectral form factor} 
One kind of coupling that always exists is the overall normalization of the Hamiltonian $\lambda$. This case is interesting because we can make contact with the spectral form factor.
If the overall coupling is uncertain, time evolution with $H_{\lambda} = \lambda H$ and inverse time evolution with $H_{\lambda'} = \lambda' H$ will not cancel by an amount $T (\lambda-\lambda')$.
The Renyi-2 entropy of the journal is therefore non-trivial:
\eqn{\tr \rho^2 &= \int p(\lambda) p(\lambda') \left|\tr \lp e^{i (\lambda-\lambda') H T - \hf (\lambda+\lambda') \beta H} \rp\right|^2 \\
&= \int p(\lambda) p(\lambda')  \left |Z\lp \hf (\lambda+\lambda') \beta + i (\lambda-\lambda') T \rp\right |^2 }
Here we see that the Renyi-2 is essentially the time-averaged spectral form factor. The specific kind of time-averaging depends on $p(\lambda)$.
The fact that the spectral form factor cannot decay to zero is just a consequence of the unitarity bound on this Renyi entropy. Hence the ``journal'' perspective conceptually unifies Maldacena's information paradox \cite{Maldacena:2001kr} with the information paradox of evaporating black holes.
Given that the plateau of the spectral form factor is not explained by a single wormhole  but doubly non-perturbative effects \cite{Saad:2018bqo, Saad:2019lba}, one could wonder if such effects play a role in the entropy of the journal.

\subsection{Chaos and the Loschmidt Echo}
The lack of our ability to do bulk reconstruction even when there is a small uncertainty in couplings is closely related to quantum chaos.
In holography, one usually diagnoses chaos using the $\otoc$, e.g, we create a perturbation at some time in the past $W(-T)$, and watch it grow. 
The perturbation is localized at some time $-T$.

An alternative diagnostic of chaos is the
\href{http://www.scholarpedia.org/article/Loschmidt_echo}{Loschmidt echo}. 
Both diagonistics involve (backwards) time evolution $\exp( i H_1 T)$, but in the Loschmidt echo, one changes the Hamiltonian going forward $\exp( i H_2 T)$ by a small amount $H_2 = H_1 + \epsilon W$. 
So whereas the $\otoc$ is a localized pertubation in time, the Loschmidt echo is completely de-localized. 
The simplest diagnostic of chaos is the decay of the inner product
\eqn{ \bra{\psi'} \exp( -i H_2 T) \exp(i H_1 T) \ket{\psi} } 
Clearly if $H_2 = H_1$ this product does not decay, but in general it should decay and then exhibit some erratic oscillations of order $e^{-S}$. 
What we showed was that the disk contribution leads to a decay, but wormholes are needed to explain the erratic oscillations. We showed this for $\ket{\psi} = \ket{\beta, J_1}$, $\ket{\psi'}= \ket{\beta, J_2}$ in SYK and in JT gravity, but we believe that the lessons should be fairly general.

One can also study the size of the Loschmidt operator $\exp( -i H_2 T)\exp(i H_1 T) $. At infinite temperature, this is given by sandwiching the size operator $\sim \psi_L \psi_R$ with the state $\exp( -i H_2 T)\exp(i H_1 T) \ket{0}$. To define a finite temperature version of size, we would need to specify whether the thermal ensemble should correspond to $H_1$ or $H_2$, but for small perturbations this is a minor detail. For large $q$ SYK, it follows from the disk solution at $\lambda \approx 1$ in \ref{sec:largeq} that the size grows exponentially with Loschmidt Lyapunov exponent 
\eqn{\lambda_L = 2 \alpha,}
e.g, the same Lyapunov exponent as the thermofield double. At low temperatures, the maximal chaos exponent follows from the bulk picture and the relation between size and the symmetry generators \cite{Lin:2019qwu}. Indeed, in the bulk, a change in the couplings at time $-T$ inserts a matter shockwave; the gravitational backreaction is responsible for the exponential growth in size. But we have calculated the exponent at finite temperature in the large $q$ theory, where SYK is not simply described by the Schwarzian mode.

An interesting question is whether one can argue for a bound on the Loschmidt Lypaunov exponent $\lambda_{L} \le 2\pi/\beta$ along the lines of \cite{Maldacena:2015waa}.


\subsection{Singularity?}
Our work shows that the black hole interior is quite sensitive to the precise values of the couplings of the theory. Even ignorance about the irrelevant couplings appears to be enough to prevent the interior from being reconstructed at late times. 
This is surprising if we believe that irrelevant operators have small effects in the bulk. On the other hand, irrelevant operators can have a large effect near the black hole singularity. In JT gravity, one can show that the profile of a free scalar field with mass determined by $\Delta$ will diverge near the inner horizon except for integer values of $\Delta$ \cite{Lin:2019qwu}. It is tempting to speculate that the island associated to the journal that forms at late times is trying to censor us from accessing the region near the singularity. Adopting the spirit of Penrose's cosmic censorship conjecture \cite{penrose1969gravitational}, if we do not have access to the UV couplings of the boundary theory, it seems reasonable that we cannot reconstruct the region near the black hole singularity that is sensitive to those couplings. This obviously deserves more study.

\subsection{Janus's Journal}
It would be interesting to consider the journal entangled to a ``conventional'' holographic theory which has no disorder average.
We could consider a situation the journal records the value of $\lambda$ in $\en$ SYM. An off diagonal element of the journal density matrix would involve half of the thermal cylinder with one value of the gauge coupling $g$ and the other half with a different value $g'$. 
For the vacuum state, this problem is rather well-studied: one considers the Euclidean path integral on a sphere with gauge coupling $g$ on half of the sphere and $g'$ on the other half \cite{Bak:2003jk,Clark:2004sb,Clark:2005te,Gaiotto:2008sd}.
The problem at finite temperature would involve a path integral on a torus, with top and bottom halves of the torus differing \cite{Bak:2007jm}. 
It would be interesting to find wormholes in such a setup (or in other holographic setups with well-defined string duals), as this would pose a sharp factorization problem in $\en$.
More generally, one can consider a holographic CFT$_1$ joined to a CFT$_2$ at some interface, see \cite{Bachas:2021fqo, Simidzija:2020ukv, May:2021xhz} for some recent discussions. Our paper suggests that the square of such quantites could have wormholes.

\subsection{Conclusion}

We are usually taught that gravity is a universal force, e.g., that spacetime is sourced by energy-momentum. What we seem to be learning here is that the spacetime in the interior of the black hole is in some sense highly non-universal.
This seems related to the following equation:
\eqn{P(\mathsf{firewall}) =\eqfig{0.6\columnwidth}{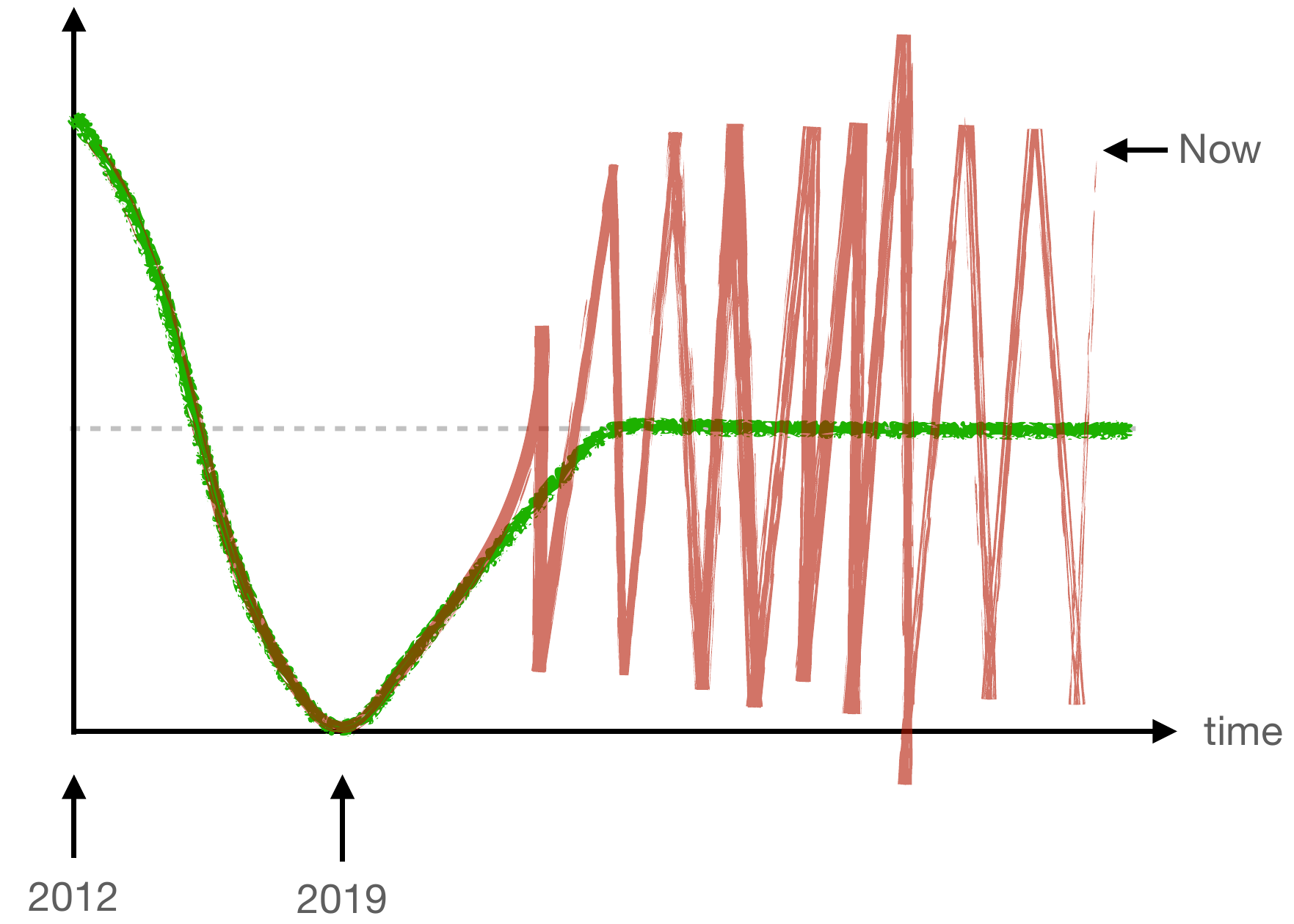} } 
\begin{center} {\it What gravity doesn't know, neither do we...}
\end{center}
\section*{Acknowledgments} 

We thank Alexey Milekhin, Adam Levine, Venkatesh Chandrasekar, Alex Streicher, Yiming Chen, Juan Maldacena, Edgar Shaghoulian, Douglas Stanford, Edward Witten, and Ying Zhao for discussions.
We thank Douglas Stanford for permission to reproduce his figure as our Figure \ref{fig:stanfordworm}, and Juan Maldacena for repeatedly threatening us to finish the draft.

\appendix

\section{Compact free boson \la{windings}}

In section \ref{matterentropy}, we considered the matter entropy of a non-compact boson, with Gaussian measure over the Dirichlet boundary conditions. 
Another case we can consider is the compact free boson $X \equiv X + 2 \pi r$ with the same action as in \eqref{eq:freebosonaction}.
The boundary conditions we consider are again Dirichlet conditions on the free field, labeled by $X$. 
(For a review of boundary states in the compact boson, see \cite{Oshikawa:1996dj}.)
The boundary condition changing operator has dimension
\eqn{\Delta_b = {1 \over \alpha'}  \lp{X_1 - X_2 + 2\pi r w \over 2\pi}\rp^2, \quad w \in \mathbb{Z} \la{eq:bccCompact} .}
In the compact case, there are actually infinitely many boundary condition changing operators that change $\ket{X_1}$ to $\ket{X_2}$; in string theory this is the fact that an open string ending on 2 D-branes can wind around the compact dimension $w$ times.
Then \eqref{eq:renyi} becomes
\eqn{ Z_n = {1 \over (2 \pi r)^n} \int  \prod_{i=1}^n dX_i e^{-A(X_i - X_{i+1})^2/2 }\la{eq:freebosonrenyi}}
We can evaluate this integral à la Fadeev-Popov by first fixing $X_1 = 0$, performing $n-1$ integrals, and then integrating over $X_1$. This gives
\eqn{Z_n ={1 \over (2\pi r)^{n-1}} \sqrt{\frac{(2 \pi)^{n-1}}{\det \mathbf{M}} } =  \sqrt{\frac{1}{ n (2 \pi A r^2)^{n-1}  } } .}
In writing the above equations, we implicitly summed over windings. This seems to be a choice that in principle might depend on UV regulator. 
The replica trick
\eqn{S=-\left.\partial_{n}\left(\frac{\log Z_{n}}{n}\right)\right|_{n=1} = \hf \log 2 \pi A r^2  - c'(1) }
At large $T$, $A r^2 \approx \frac{T r^2}{2\pi\beta \alpha'}$. $S \sim  \hf c \log (T/\beta)(r^2/\alpha')$ for $c$ free bosons. When this entropy exceeds the entropy of the quantum dots $S \sim 2S_0$ there is a unitarity paradox. This happens at
\eqn{T \sim \beta e^{2S_0/c} \alpha'/r^2.}

\section{von Neumann entropy for 2 boundary states \la{app:2bd}}
Consider a general BCFT with two boundary states, each with equal probability. We can label the boundary conditions $\sigma = 1$ or $\sigma = -1$. Then $\Delta_i$ will be some value $\delta$ if the boundary conditions change, or else it will be 0. We can write this as $\Delta_i = \frac{\delta}{2} (\sigma_i \sigma_{i+1} - 1) $. Now we are instructed to compute
\eqn{Z_n = c_n \sum_{\sigma_i} \exp \lp - \frac{J}{2} \sum_{i=1}^n  \sigma_i \sigma_{i+1} - 1 \rp}
This is an Ising model on a periodic lattice with $n$ sites. Using the standard transfer matrix, 
\eqn{Z_n(J) = \lp 1-e^{-J}\rp ^n + \lp 1 + e^{-J} \rp^n,}
\eqn{S_m = (1- \pd_n) \log Z_n = -\frac{1}{2} \log \left(1-e^{-2 J }\right)-e^{-J } \coth ^{-1}\left(e^{J }\right)+\log (2) \la{ising} }
This expression is real when $e^J > 1$.
Notice that $J \to \infty$, we get $S_m = \log 2$. So if we have a very long interval, the matter entropy just reflects the uncertainty in the boundary condition. (The entropy of the Journal is saturated.)
In writing this expression, we assumed that $c_1 =  1$ and that $\pd_n c|_{n=1} = 0$. The first condition is easy to justify but the second one is a bit mysterious.

More generally, we could choose the boundary conditions $\sigma_i$ to appear with different probabilities. Then we would give an Ising model in a magnetic field.

\section{Zero temperature island}

A case where we do not need to take the OPE limit is the zero-temperature or extremal limit. The geometry in this limit becomes the \pcr{} plane:
\eqn{ds^2 = (dx^2+dy^2)/y^2, \quad y \in (\epsilon,\infty)}
The Schwarzian boundary is located at $y = \epsilon$ and the \pcr{} horizon is at $y \to \infty$.
The BCFT bulk-boundary 2-pt function in these coordinates is
\eqn{
	\ev{O_{(h,0)}(y) O_{(0,h)}(\bar{y}) O_{(h_b,h_b)}(x') }\sim {y^\Delta \epsilon^{\dbd}\over (2z)^{\Delta - \Delta_\partial} (y^2 + x^2)^{\Delta_\partial} }  
}
Here we have included the warp factor $\Omega \sim 1/y$ for AdS. Notice that when the boundary dimension vanishes $\Delta_\partial = 0$, this correlator is independent of $y$, as required by symmetry. For our application, we take the bulk operator to be a twist and the boundary operator to be a boundary condition changing operator.
When the separation $x =  0$, we get $c_n (\epsilon/y)^{\dbd}$. 

For the case of 2 boundary states, we have $e^J = y e^\delta $. So the limit $J \to \infty$ corresponds to an island that is very close to the \pcr{} horizon. For a free boson, 
\eqn{ S_m &= \frac{\log (\ta)}{2}+\sqrt{4 \ta+1} \coth ^{-1}\left(\sqrt{4 \ta+1}\right)+1- 2 c_1' \\
\tilde{a} &\sim \log (y/\epsilon).  }


\section{Renyi-2 entropy for a marginal journal\la{app:matter}}

In this subsection we consider the problem of evaluating the integral \eqref{eq:matter} for a marginal deformation $\Delta =1$:
\eqn{ -I_m = {\chi_0^2 \over 2\pi} \int_0^{u_*} du_1 \int_0^{u_*} du_2 \frac{t'(u_1) t'(u_2)}{2 \sin^2 \lp \frac{t(u_1) - t(u_2)}{2}\rp  }   \la{eq:matterr2} }
To evaluate this integral, it is convenient to work in Poincare coordinates $f(u) = \tan t(u)/2$ instead of Rindler coordinates $t(u)$:
\eqn{ - \tilde{I}_m = \int du_1 \, du_2 \lb \frac{f'(u_1) f'(u_2)}{ \lp f(u_1) - f(u_2)\rp^2 } \rb   \la{eq:matterf2} }
One subtlety about this integral is that we need to regulate the divergence coming from $t_1 \to t_2$. A simple regulator is to restrict the integration domain to the region $|u_1 - u_2| > \epsilon$: 
\eqn{-I &= 2 \int_{0}^{u_*- \epsilon} du_2 \int_{u_2 + \epsilon} du_1  \lb{f'(u_1)  f'(u_2)  \over (f(u_1) -f(u_2))^2}\rb \\
&= 2\int_{0}^{u_* - \epsilon} f'(u_2) du_2  \lb {}{1 \over f_2-f(u_*)}-{1 \over f_2-f(u_2+\epsilon)} \rb \\
&\approx 2\int_{0}^{u_* - \epsilon}  du_2  \lb {}{f'(u_2) \over f_2-f(u_*)}+\epsilon\inv - \frac{f''(u_2)}{2t'(u_2)}   \rb \\
&\approx 2 \lb \log(f(u_*-\epsilon)-f(u_*))- \log(f(0)-f(u_*)) - \hf \log(f'(u_*)) + \hf \log (f'(0)) + u_*\epsilon\inv \rb\\
&\approx 2 \lb \hf \log(\epsilon^2 f'(u_*) f'(0) )- \log(f(0)-f(u_*)) + u_* \epsilon\inv \rb \\
&=   (2u_*/\epsilon) + \log{ \lb  \frac{\epsilon^2 f'(u_*) f'(0) }{ \lp f(u_*) - f(0) \rp^2 }   \rb }\\
}
Mapping back to Rindler coordinates,
\eqn{e^{-I} = e^{-\tilde{I} \delta} = e^{2u_*/\epsilon} \lb  \frac{\epsilon^2 t'(u_*) t'(0) }{ 2\sin^2 \lp \frac{t(u_*) - t(0)}{2} \rp }   \rb^{\delta}, \quad  \delta= \chi_0^2/(2\pi) \la{eq:matterint}}
In SYK, we expect $\epsilon \sim 1/\cj$, since the thermal two-pt function goes like $(\beta \cj)^{-\Delta}$.
In computing the overlap between states, we should normalize the states, e.g., $\frac{|\braket{\chi_0}{0}|^2}{\braket{0}{0}\braket{\chi_0}{\chi_0}}$.
The denominator $\braket{\chi_0}{\chi_0}$ gives a similar factor, except that $u_* \to  2 u_*$ and $\beta = 2 u_*$ since we turn on the source over the entire thermal circle (both the bra and the ket). Using the same regulator as above, the denominator gives
\eqn{\braket{\chi_0}{\chi_0} \sim \lb  e^{4u_* \delta/\epsilon}\frac{\epsilon^2 t'(u_*) t'(0) }{ (\epsilon t'(0) )^2 }   \rb^{\delta} \sim e^{4u_* \delta/\epsilon},  }
So the effect of the denominator is to remove the $1/\epsilon$ divergence to the matter action.
We conclude that the normalized overlap is

\eqn{\frac{\braket{\chi_0}{0}}{\sqrt{\braket{0}{0}\braket{\chi_0}{\chi_0}}} = \frac{1}{\sqrt{Z(\beta) Z(2u_*) }} \int \frac{Dt}{\slt} \, e^{-\sch(t)} \lb  \frac{\epsilon^2 t'(u_*) t'(0) }{ 2\sin^2 \lp \frac{t(u_*) - t(0)}{2} \rp }   \rb^{\delta} \la{overlap:final},}
where $Z(\beta)$ is the Schwarzian thermal partition function, see \cite{Stanford:2017thb}.
For the SYK model, we expect that the UV regulator $\epsilon \sim 1/\cj$ since the thermal two-pt function goes like $(\beta \cj)^{-\Delta}$. For JT gravity at finite cutoff, $\epsilon$ should be related to the JT cutoff $(ds/du)^2 =1/\epsilon_{JT}^{2}$.
Here we are imagining evaluating \eqref{eq:matterint} with the classical thermal solution $t(u)$. However, our entire derivation also applies for off-shell $t(u)$. Hence a more accurate answer would be obtained by integrating \eqref{overlap:final} over all paths $t(u)$ with the Schwarzian action \eqref{overlap:final}. Such an integral can be performed exactly, see e.g. \cite{Mertens:2017mtv, Lam:2018pvp, Saad:2019pqd, Yang:2018gdb}. 

As a warmup for the wormhole, we can consider the thermal circle and turn on the source in two intervals, $[u_1,u_2]$ and $[u_3,u_4]$. This will clearly give a 2-pt function $\ev{O(u_1) O(u_2)}\ev{O(u_3) O(u_4)}$ when $u$ and $u'$ are both in the same interval. But we will also get a contribution when $u,u'$ are in opposite intervals. There will be no divergences from this region of integration. We get
\eqn{- \tilde{I}_m &\supset 2 \int_{u_1}^{u_2} du \, \int_{u_3}^{u_4} du' \lb \frac{f'(u) f'(u')}{ \lp f(u) - f(u')\rp^2 } \rb\\
&= 2 \log  \lp \lb \frac{f(u_2) - f(u_4)}{f(u_2)-f(u_3)} \rb \lb \frac{f(u_1) - f(u_3)}{f(u_1)-f(u_4)} \rb \rp \\
e^{-I} &\sim \lp  \frac{f(u_2) - f(u_4)}{f'(u_2) f'(u_4)}\frac{f'(u_2)f'(u_3)}{f(u_2)-f(u_3)} \frac{f(u_1) - f(u_3)}{f'(u_1)f'(u_3)}\frac{f'(u_1) f'(u_4)}{f(u_1)-f(u_4)} \rp^{\delta}
}
Including the contribution from the two intervals, we get

\def\ccm{\mathcal{C}_{-\delta}}
\eqn{e^{-I_m} &\sim \cc(u_1,u_2)\cc(u_3,u_4) \lp \frac{\cc(u_2,u_3) \cc(u_1,u_4)}{ \cc(u_2,u_4) \cc(u_1,u_3)} \rp }
Here $\cc(u_i,u_j)$ denotes the 2-pt function of a boundary primary of dimension $\delta$:
    \eqn{\cc(u_i,u_j) = \lb \frac{f'(u_i) f'(u_j)}{ \lp f(u_i) - f(u_j)\rp^2 } \rb^{\delta} }
Let us consider the OPE limit $u_2 \to u_3$ (on a circle, this is also equivalent to $u_1 \to u_4$). Then we get the simplification
\eqn{e^{-I_m} &\sim  \cc(u_2,u_3) \cc(u_1,u_4)    }
Similarly in the OPE limit for the opposite channel $u_1 \to u_2$ or $u_3 \to u_4$:
\eqn{e^{-I_m} &\sim  \cc(u_1,u_2) \cc(u_3,u_4)   }
So we see that in the OPE limit, the effect of the couplings is to insert a product of conformal 2-pt functions. We can think of this as inserting a pair of ``domain walls'' in the bulk. In the extreme limits, the domain walls will prefer to link up to their nearest partner.

More precisely, note that $\cc_{12} \propto \exp \lp {-\delta d_{12}}\rp$ where $d$ is the renormalized geodesic distance in the bulk between $u_1$ and $u_2$. So when $d_{12} \ll d_{13}, d_{14}$ we get the first OPE limit, and similarly for the other limits.

Finally, note that one can generalize this computation easily to $n_I > 2$ intervals. In particular, since the matter action is bilocal in times, we will need to just consider each pair of intervals:
\eqn{e^{-I_m} &\sim \prod_{A=1}^{n_I} \cc(u_{A1}, u_{A2}) \times \prod_{B<C} \lp \frac{\cc(u_{B2}, u_{C1}) \cc(u_{B1}, u_{C2}) }{\cc(u_{B2}, u_{C2} )\cc(u_{B1}, u_{C1}) }\rp .\la{eq:manyint} }
In this expression, $A,B,C$ index the intervals; the endpoints of the interval are denoted by $A = [A1, A2]$.

\subsection{Renyi-2 wormhole} \la{app:r2wormhole}

First consider the double trumpet geometry:
\eqn{ds^2 = d\rho^2 + \lp \frac{b}{2\pi}\rp^2 \cosh^2 \rho \, d\theta^2, \quad \rho \in (-\infty,\infty), \quad \theta \in [0,2\pi] \la{dtrump} }
This is conformally related to a cylinder of length $\pi$ and radius $b$:
\eqn{ds^2 = \frac{d\sigma^2 + \lp \frac{b}{2\pi}\rp^2 \, d\theta^2}{\sin^2 \sigma}, \quad \sigma \in (0,\pi), \quad \theta \in [0,2\pi] }
To specify a cutout shape, one needs to specify some $\theta_L(u)$ and $\theta_R(u)$. We are interested in a setup in which the left circle has boundary conditions $\chi(u_L)$ and the right circle has boundary conditions $\chi(u_R)$. More specifically, $\chi(u_L)$ is constant and non-zero in some time interval $[0, u]$ and similarly for the right hand side.
Now we can view the double trumpet \eqref{dtrump} as a quotient of AdS$_2$ by a global time translation $e^{i b E}$.
If we consider the universal cover of the double trumpet, we get the picture:
\begin{figure}[H]
\begin{center}
\includegraphics[width=0.5\columnwidth]{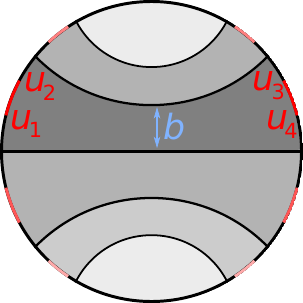}
\caption{ Here we display the fundamental domain of the double trumpet geometry (dark gray), as well as some of its images (lighter gray). We also show the parts of the boundary where a source is turned on ({red}) and its images (lighter pink). \la{fig:quotient}  }
\end{center}
\end{figure}
We can evaluate the matter partition function by going to the universal cover. What started out as a 2-interval computation on the cylinder becomes a computation involving infinitely many intervals. However, we can still apply \eqref{eq:manyint}. However, we claim that in the large Lorentzian time limit, we are in the OPE regime where \eqref{eq:manyint} will just give 
\eqn{e^{-I_m} \sim \cc^{LR}(u_1, u_4) \cc^{LR}(u_2, u_3).}
We will assume that this approximation holds; calculate the backreaction, and then check that our approximation is self-consistent.

To understand why this is the case, consider the expression for the matter action in these coordinates:
\eqn{ -I_m^\trum &= {1 \over 2\pi} \int du_L \, du_R \lb \frac{\theta_L'(u_L) \theta_R'(u_R)}{2 \cosh^2 \lp \frac{b}{2\pi} \frac{\theta_L(u_L) - \theta_R(u_R)}{2}\rp } \rb \chi_L(u_L) \chi_R(u_R)\\
&+ {1 \over 2\pi}\int du \, du' \lb \frac{\theta_L'(u) \theta_L'(u')}{2\sinh^2 \lp \frac{b}{2\pi}\frac{\theta_L(u) - \theta_L(u')}{2}\rp } \rb\chi_L(u) \chi_L(u')\\
&+ {1 \over 2\pi}\int du \, du' \lb \frac{\theta_R'(u) \theta_R'(u')}{2\sinh^2 \lp \frac{b}{2\pi}\frac{\theta_R(u) - \theta_R(u')}{2}\rp } \rb\chi_R(u) \chi_R(u')\\}
The above formula is slightly imprecise since $\theta$ is a periodic variable. In principle we should sum over all windings $\theta \to \theta + 2\pi n$. However, we see that if $b$ is very large, only a single winding will dominate.


\begin{figure}[H]
\begin{center}
\includegraphics[width=\columnwidth]{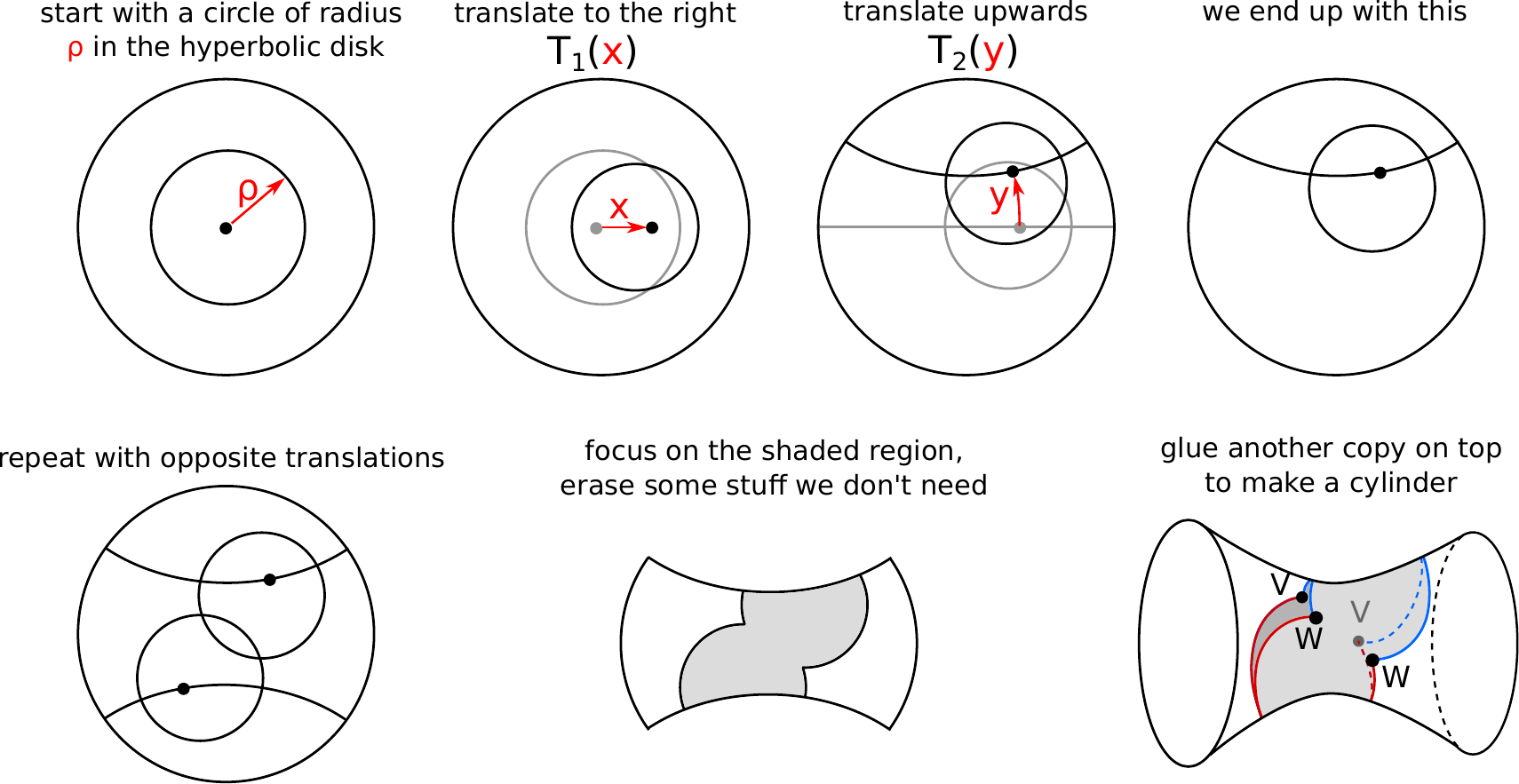}
\caption{ (Adapted from \cite{StanfordMore} with permission).  This gives the cut-and-paste construction of the wormhole \cite{StanfordMore} that contributes to the Renyi-2 entropy. In Stanford's context, the wormhole arises from inserting 2 pairs of operators; in our context, these operators are boundary condition changing operators that arise from turning on a marginal source. At large Lorentzian times, we end up with essentially the same wormhole. If we already know the 2-operator disk solution, a ``shortcut'' procedure is to start with the solution on the bottom left, with some auxiliary inverse temperature $\betax$. When we ``erase some stuff we don't need,'' we are left with a boundary of length $\betax-\betae$. After gluing to a second copy, we get the constraint $\beta = \betax-\betae$. } \la{fig:stanfordworm}
\end{center}
\end{figure}
Here $\rho$ is the radius of the circle; the circumference of the circle defines an effective inverse temperature $\beta_E$,
\eqn{2\pi \sinh \rho =\beta_E/\epsilon}
where $\epsilon$ is the JT regulator.
The relationship between the parameters $\{\beta_E, x,y\}$ and the physical inputs $\{ \beta, \tau, \delta\}$ is given in \cite{StanfordMore}, equation (B.14-16):
\begin{align}
&\tan \left(\frac{\pi \beta}{2 \beta_{E}}\right)=\frac{\delta \beta_{E}}{2 \pi} \la{eq:berel} \\
&\sinh ^{2}(y)=\frac{\left(\frac{\beta_{E} \delta}{2 \pi}\right)^{2}}{1+\tan ^{2}\left(\frac{\pi i t}{\beta_{E}}\right)} \\
&\cosh ^{2}(x)=\frac{\left(1+\left(\frac{\beta_{E} \delta}{2 \pi}\right)^{2}\right)\left(1+\tan ^{2}\left(\frac{\pi i \tau}{\beta_{E}}\right)\right)}{1+\left(\frac{\beta_{E} \delta}{2 \pi}\right)^{2}+\tan ^{2}\left(\frac{\pi i \tau}{\beta_{E}}\right)} .
\end{align}
\la{eq:param}
In the small $\delta \beta$ limit, $\beta_E^2 = \pi^2 \beta/\delta$.
Now we can check that our approximation is self-consistent. At large $\tau$,
\begin{equation}
b=4 y \approx \frac{4 \pi}{\beta_{E}}\left( \tau- 2\tau_*\right), \quad \tau_*=\frac{\beta_{E}}{2 \pi} \log \left(\frac{2 \pi}{\beta_{E} \delta}\right).
\end{equation}
Here $\tau_*$ can be interpreted as the scrambling time $\tau_* \sim \beta_E/2\pi \log \frac{S-S_0}{\Delta S}$.
So at large times the wormhole is large and we expect the approximation to be valid. 

Another regime we can explore is the extremal limit $\beta \to \infty, \tau =0$. We also need to decide how $\delta$ scales in this limit. The simplest case is $\delta$ held fixed. Then $\pi \beta/(2 \beta_E) \approx \pi/2$ so in this limit $\beta_E \approx \beta$. Furthermore, $y \sim \log \beta$, so at least classically the wormhole is getting large. 

A useful parameter is the amount $\tau_a$ of the $\beta_E$ circles that overlap, which governs the distance between $W_L$ and $W_R$, or equivalently $V_L$ and $V_R$. Consider the case $\tau=0$ for simplicity. The disks of circumference $\beta_E$ that are used to build the solutions are always cut along a geodesic that passes through the center of the disk, such that there is no discontinuinty in the bulk geometry. So requiring that the total boundary length is $2 \beta$ gives
\eqn{\tau_a = (\beta_E - \beta)/2. \la{ta} }
\eqn{ \ev{V_L V_R} \sim \ev{W_L W_R} \sim \lp  \frac{\pi}{\beta_E \sin \lp \pi \tau_a/\beta_E \rp  }\rp^{2\delta}   }

To evaluate the action more carefully, we can follow \cite{Goel:2018ubv}.

Furthermore, the distances 
    \eqn{
\begin{aligned}
e^{D_{12}} & \approx \lp \frac{\beta_E}{2\pi \epsilon} \rp^2 \sinh ^{2}\left(\frac{\pi}{\beta_{E}} \lp -i \tau_a +  t_{12}\rp \right) \\
e^{D_{13}} & \approx \lp \frac{\beta_E}{2\pi \epsilon}\rp^2  \left[i \sinh \left(\frac{\pi}{\beta_{E}} t_{13}\right)+ \frac{\delta \beta_{E}}{\pi} \sinh \left(\frac{\pi}{\beta_{E}} t_{1}\right) \sinh \left(\frac{\pi}{\beta_{E}} t_{3}\right)\right]^{2}\\
e^{D_{23}} &\approx \lp \frac{\pi}{\beta_{E}} \rp\left[ \sinh \left(\frac{\pi}{\beta_{E}} \tau_{13}\right)+m \frac{\beta_{E}}{\pi} \sin \left(\frac{\pi}{\beta_{E}} (\beta - \tau_2) \right) \sin \left(\frac{\pi}{\beta_{E}} \tau_{3}\right)\right]^{2}
\end{aligned}
}

An interesting limit of these formula is when $\beta \to \infty$ with $\delta$ held fixed. Then $\tau_a \to 0$ and $\beta_E \to \infty$.
\eqn{
\begin{aligned}
e^{D_{12}} & \approx   \lp t_{12}/ 2\epsilon \rp ^2 \\
e^{D_{13}} & \approx \lp i \frac{ t_{13}}{2 \epsilon} + \frac{\delta}{2\epsilon} t_1 t_3 \rp^2
\end{aligned}
}
This agrees with the large $q$ wormhole in the appropriate limit, see \eqref{eq:syklowtemp}.

\section{Brownian SYK \la{app:brown}}
in this section, we compute the Renyi-2 entropy of the journal, where the journal records the entire history of couplings $\{ J_{ijkl}(t)\}$ for Brownian SYK.
So the Hilbert space of the journal is that of a ${N \choose q}$ ``fields'' and not just that of a collection of point particles.

\subsection{Decorrelated spectral form factor}
Let us start by considering a simple generalization of the spectral form factor as a warmup. Usually in Brownian SYK, we write
$\ev{|Z(T)|^2}_J = \ev{\tr U_J(T) \tr U_J(T)\inv}_J$ where $\ev{\cdots}_J$ means to disorder average over $J$ with the appropriate Gaussian measure. Here we consider the quantity $\ev{\tr U_J(T) U_{J'}(T)\inv  }_{J,J'}$ where we disorder average over $J,J'$ with a Gaussian measure that correlates $J$ and $J'$ by an amount $r = \ev{J J'}$.
\begin{equation}
\begin{aligned}
	\ev{\tr U_J(T) U_{J'}(T)\inv } & \approx \int \mathcal{D} G \mathcal{D} \Sigma \exp \left\{-\frac{N}{2} \int_{0}^{T} d t\left[\frac{2 J}{q}\left(\frac{1}{2^{q}}-i^{q} G(t)^{q}\right)+\Sigma(t) G(t)\right]\right\} \\
& \times \int \mathcal{D} \psi_{a}^{(L)} \mathcal{D} \psi_{a}^{(R)} \exp \left\{-\frac{1}{2} \int_{0}^{T} d t\left[\psi_{a}^{(A)} \partial_{t} \psi_{a}^{(A)}-\psi_{a}^{(L)}(t) \psi_{a}^{(R)}(t) \Sigma(t)\right]\right\} \la{e2}
\end{aligned}
\end{equation}
Here we are thinking of the $L$ system as one of the $U(T)$'s and the $R$ system as the other $U(-T)$.
THe second line computes the trace over an $L \cup R$ of $2N$ fermions.

If we now consider time-independent solutions, we get a simple action
\begin{equation}
\exp \left\{N\left[\log \left(2 \cos \frac{T \Sigma_\lr}{4}\right)-\frac{J T}{q 2^{q}}+r i^{q} \frac{J T}{q} G_\lr^{q}-\frac{T}{2} \Sigma_\lr G_\lr\right]\right\}
\end{equation}
Let's set $J = 1$ from now on. We get the equations of motion
\eqn{G  = -(1/2) \tan (\Sigma T/4) , \quad i^q G^{q-1} r  =  \Sigma/2 }
This gives
\eqn{\Sigma/2 = r i^q  \lp -1/2 \rp^{q-1} \tan \lp \Sigma T/4 \rp^{q-1}, r i \sigma 2^{-q}=  \Sigma  \\
	\sigma =   \tanh^{q-1} \lp \sigma   {rT   2^{-q}}  \rp
}
There can be at most $q+1$ solutions to the above equation.
Let us consider the case $q=2$ for simplicity.  We get three solutions when $rT/4 > 1$. We get only 1 solution, $G=0$ when $r T/4 < 1$. The solution $G=0$ gives the exponential decay.
\eqn{\ev{Z_1(T) Z_2(-T)}_\text{dip} \sim \exp \lp  N\log 2 - {J N T \over q 2^q } \rp }
For large values of $r T$, we get a different solution $\sigma = \mp 1$ or
\begin{equation}
G_\lr=\pm \frac{i}{2}, \quad \Sigma_\lr=\mp \frac{r i }{2^{q-2}}
\end{equation}
For the usual spectral form factor, these solutions give us an action that is independent of $T$ (for large $T$). However, $r<1$ these solutions predict an exponential decay:
\eqn{\ev{Z_1(T) Z_2(-T)}_\text{ramp} \sim \exp \lb -  N T  (1-r)\over q 2^q \rb }
So we get an exponentially decaying term. Note that this formula is valid when $r T \gg 1$.

\subsection{Journal disk}
Now consider the problem we are actually interested in, the Renyi entropy of the journal.
Let us start by finding the disk solution. Then we just need to consider $\ev{ \tr U_J(T) U_{J'}(T)}$.
The difference between this setup and \eqref{e2} is the just the choice of boundary conditions. Instead of the second line in \eqref{e2} computing a trace, we should write
\eqn{ 2^{N/2} \bra{0} \exp \lp  \int H dt\rp\ket{0} = 2^{N/2} \lp e^{\pm {i \over 4} T \Sigma }\rp^{N}
}
Note the normalization $2^{N/2} = \tr 1$.
We are left with an action
\begin{equation}
	\exp \left\{N\left[ \frac{\pm i T \Sigma}{4} -\frac{J T}{q 2^{q}}+i^{q} r \frac{J T}{q} G^{q}-\frac{T}{2} \Sigma G\right]\right\}
\end{equation}
Extremizing the action gives
\eqn{G = \pm i/2, \quad \Sigma = i^q r J G^{q-1} = \pm rJ }
The on-shell action is
\eqn{Z \sim \exp \lb \frac{-J T(1-r)}{q2^{q} } \rb}
We see that this disk solution will actually conflict with unitarity at late times $T$.
Note that when $r=0$, the top and bottom disk are independent. This is equivalent to the usual Brownian disk with $2T$. Indeed, from SSS we see that the disk has an action
\eqn{Z_\text{disk} \sim 2^{N/2} \exp {- \frac{JT N}{2 q 2^q}}}

To be a bit more explicit, consider a two-point function 
\eqn{G(T',T)&=\tr \lp U(T') \psi_j U(T) \psi_j\rp  = \bra{1} U(T')^L \psi_j^L U(T)^L \psi_j^L \ket{1} \\
	    & = - i\bra{1}  U(T')^L \psi^L_j \psi^R_j U(T)^L \ket{1}\\
	    & = -i G_{LR}.
}
The first line is true for any maximally entangled state $\ket{1}$. In the second line, we specialize to the choice $J_- \ket{1} = \lp \psi^L + i \psi^R \rp \ket{1} = 0$.

\subsection{Wormhole}
At long last, we find the wormhole. Actually for $r=0$, we have already found the wormhole in the form of the Brownian ``ramp'' solution. More generally, we need to introduce 4 sides $L_1, L_2$, $R_1, R_2$ such that $L_1$ and $L_2$ are perfectly correlated but $L_1 R_1$ are correlated by an amount $r$.


We will have a quadratic fermion action
\eqn{& \int \mathcal{D} \psi_{a}^{(L)} \mathcal{D} \psi_{a}^{(R)} \exp \left\{-\frac{1}{2} \int_{0}^{T} d t\left[\psi_{a}^{(A)} \partial_{t} \psi_{a}^{(A)}-\psi_{a}^{(L)}(t) \psi_{a}^{(R)}(t) \Sigma(t)\right]\right\} \la{e1}}

It is possible to diagonalize the anti symmetric $\Sigma$ matrix, so we could try to find constant solutions. To reduce the number of variables, we could try an ansatz
\eqn{\Sigma_{L1R2} &= \Sigma_{R1L2}, \\ \Sigma_{L1L2} &=  -\Sigma_{R1R2},\\ \Sigma_{L1R1} &=  \Sigma_{L2R2} }
This is motivated by the UV property $\psi_L \ket{1} = -i\psi_R \ket{1}, \psi_R \ket{1} = i \psi_L \ket{1}$.

At large times $T$, we expect that only the smallest eigenvalue will matter.

\eqn{ -S = N \frac{ T}{2}  \sqrt{s^2_{1L2R} + s_{1L2L}^2-s^2_{1L1R} } - \frac{2 J T}{q 2^q} + i^q  \frac{2 J T }{q} \lp r G_{1L2R}^q + r G_{1L1R}^q + G_{1L2L}^q \rp - {T \Sigma_A G_A} }
We find two solutions of this action $G_{1L2L} = \pm 1/2$ which leads to an action that is independent of time. For this solution $|G_{1R2R}| = 1/2$ but all other correlators are zero. So it looks the wormhole is ``made'' of disks, just like the large $q$ SYK wormhole.



\mciteSetMidEndSepPunct{}{\ifmciteBstWouldAddEndPunct.\else\fi}{\relax}
\bibliographystyle{utphys}
\bibliography{BigReferencesFile.bib}{}

\end{document}